\documentclass{article}

\usepackage{arxiv}

\usepackage[utf8]{inputenc} % allow utf-8 input
\usepackage[T1]{fontenc}    % use 8-bit T1 fonts
\usepackage{hyperref}       % hyperlinks
\usepackage{url}            % simple URL typesetting
\usepackage{booktabs}       % professional-quality tables
\usepackage{amsfonts}       % blackboard math symbols
\usepackage{nicefrac}       % compact symbols for 1/2, etc.
\usepackage{microtype}      % microtypography
\usepackage{lipsum}

\usepackage{amsmath}
\usepackage{bm}
\usepackage{tabularx}
\usepackage{physics}
\usepackage{graphicx}
\usepackage{listings}
\usepackage[round]{natbib}
\usepackage{multicol}
\usepackage{amssymb}
\usepackage{setspace}
\usepackage{caption}
\usepackage{color}
\usepackage[font=small]{subcaption}
\title{AI-driven Bayesian inference of statistical microstructure descriptors from finite-frequency waves}

\author{
  Wouter Klessens\ \\
  Department of Earth Sciences\\
  Utrecht University\\
  Utrecht, the Netherlands \\
  \texttt{w.h.a.klessens@students.uu.nl} \\
   \And
 Ivan Vasconcelos\ \\
  Department of Earth Sciences\\
  Utrecht University\\
  Utrecht, the Netherlands \\
  \texttt{i.vasconcelos@uu.nl} \\
  \And
 Yang Jiao\ \\
  Materials Science and Engineering\\
  Arizona State University\\
  Tempe, AZ, USA \\
  \texttt{yang.jiao.2@asu.edu} \\
}

\begin{document}
\maketitle

\rhead{\scshape AI-inferred microstructures from waves}

\begin{abstract}
    Most geological processes {\color{black} are tied to} heterogeneous microstructures, e.g. contaminated groundwater flow through porous media, production-induced compaction and/or microfracturing of rocks {\color{black} under stress } leading to large-scale earthquakes. Still, the ability to reliably image materials at the microscale from {\color{black} long-wavelength} wave data remains a major challenge to the geophysical, engineering and medical fields. {\color{black}  At present, when it comes to describing microstructures,} we generally rely on tomographic images of specific in-situ-acquired samples; a technique that is reliable and {\color{black} can capture} realistic {\color{black}  structures}. {\color{black}  This digital sample approach, however, makes the generalisation of microstructure complexity from single samples difficult to abstract and generalise.} Here, we present a novel framework to constrain microstructure geometry and properties from long-scale waves, which {\color{black}  are known} to be sensitive to microheterogeneities. To quantify microstructures while maintaining generality we use two-point statistics, from which we can analytically compute scale-dependent effective wave {\color{black} properties - such as wavespeeds and attenuation - } building on recent strong-contrast expansion (SCE) theory {\color{black} for (visco)elastic wavefields}. {\color{black} A key aspect of this SCE approach is that it yields wave-equation-based effective properties at finite propagation frequencies, overcoming limitations of static-elasticity approaches in capturing scale-dependent wave behaviour.} We apply the theory to acoustic and elastic waves in the long-wavelength regime and additionally to higher-frequency {\color{black} fields at scales close to the coherent} scattering regime. By evaluating both analytical two-point correlation functions and that of image samples we observe that both effective wavespeeds and attenuation of long-scale waves predominantly depend on volume fraction and phase properties, and that especially attenuation at small scales is highly sensitive to {\color{black}   the geometry of }microstructure heterogeneity (e.g. {\color{black}geometric } hyperuniformity) due to {\color{black} incoherent inference of sub-wavelenght} multiple scattering. {\color{black}   Our main goal is not only to discuss effective wave behaviour, but primarily to investigate inferring microstructure properties from observed effective wave parameters.} To both solve this highly nonlinear inversion problem and take uncertainty into account, we use the supervised machine learning method of Random Forests (RF) to construct a Bayesian inference approach. Assuming fixed medium contrasts, we can accurately resolve two-point correlation functions sampled from {\color{black}  various microstructural configurations, including: }a bead pack, Berea sandstone and Ketton limestone samples. {\color{black}  Rather importantly, we show that} inversion of small-scale{\color{black}-induced effective} elastic wave data yields the best results{\color{black}, particularly compared to single-wave-mode (e.g., acoustic only) information}. Additionally, we show that the retrieval of {\color{black}  microscale} medium contrasts is more difficult {\color{black} - as it is highly ill-posed -} and can only be achieved with specific a priori knowledge. {\color{black}   Our results, being the first fully nonlinear inference  models of complex micro-descriptors from finite-frequency elastic wave properties, are promising for many imaging applications, such as earthquake hazard monitoring, non-destructive testing, imaging fluid flow properties in porous media, quantifying sub-wavelength tissue properties in medical ultrasound, or designing materials with tailor-made wave properties.}    
    
\end{abstract}

% keywords can be removed

\section{Introduction}

The causal relation between microstructure heterogeneity and macroscopic properties and phenomena {\color{black} - such as wave propagation -} is of great importance in nature, however {\color{black} this relation} is also often very complex due high degrees of non-linearity {\color{black} between microstructure and long-scale properties}. {\color{black} In the case of geophysics, for example, we} would benefit enormously from an improved ability to {\color{black} remotely } map subsurface microstructures since e.g. earthquake {\color{black} nucleation} mechanisms - either anthropogenic or natural - are controlled by them. The former can be caused by either injection or extraction of material from the subsurface, affecting the material microstructure and eventually resulting in rock failure after the formation of microfractures. {\color{black} As a well-known case study}, gas extraction from the Slochteren sandstone in the Groningen field for over fifty years now has led to significant damage due to production-induced earthquakes. In this scenario, the removal of gas from the sandstone microstructure causes a decrease in pore fluid pressure, resulting in a larger vertical effective stress and compaction (\cite{spiers2017new, van2018reservoir, van2015induced}). In addition, many potential measures to mitigate climate change require human management of the subsurface, these could equally lead to detrimental side effects {\color{black} similar to those seen at Groningen}. Fluid injection in geothermal spots aims {\color{black} at maximising } permeability {\color{black} - e.g., through natural fracturing - } to stimulate circulation for economically viable energy production. The resulting increase in pore pressure can however cause shear failure especially along pre-existing faults where material cohesion is close to zero (\cite{gaucher2015induced, jha2014coupled}). A similar process occurs with CO$_2$ sequestration (CCS) that alterates both physical and chemical properties of the subsurface microstructure (\cite{white2016assessing}). This introduces the additional risk of groundwater contamination, controlled by microstructure characteristics like permeability, which in turn depends on porosity and distribution of the pore space (\cite{kang2010pore}). In addition countless more examples of microstructure importance in geophysics, there is an equal interest from engineering and medical disciplines (e.g. \cite{dawson2003enhanced, latch2006microheterogeneity, sadeghi2017chemotherapy}).   \\ 

Up until now, characterisation and quantification of in-situ microstructures relies on risky and expensive sample acquisition {\color{black} (e.g., through borehole coring)}, {\color{black} followed by laboratory }tomographic methods. More specifically, X-ray micro computed tomography ($\mu$-CT) is a well-established non-destructive imaging technique capable of monitoring 4D microstructure evolution with up to sub-micrometer resolution (\cite{cnudde2013high}). After imaging, there is a need for quantitative approaches to {\color{black} appropriately} capture {\color{black}, abstract and realistically describe the widely-varying} orders of complexity of materials. {\color{black} Though much progress has been achieved, the task of microstructure quantification remains} a major challenge. {\color{black} To this end, }statistical microstructure descriptors (SMDs) are widely used solutions. SMD examples are the n-point correlation functions $S_n$ (\cite{torquato1982microstructure}), defined as the probability that all n points in a certain random configuration are found in a phase of interest. One problem, however, is the large computational demand to sample statistical descriptors with high order correlations that are necessary to describe high order material complexity. More specifically, when it comes to calculating correlations for $n\geq3$, it becomes computationally inefficient to track all possible configurations. Recently \cite{chen2019hierarchical} introduced n-point polytope functions $P_n(r)$, giving the probability that for random regular polytopes the vertices separated by $r$ fall into a certain phase. This approach can easily succesfully describe correlations up to  high orders, e.g. $n=8$, since the polytope regularity allows for an efficient numerical sampling. The functions of both examples are equal to the two-point correlation function $S_2$ when $n=2$, which albeit non-unique (e.g. \cite{yeong1998reconstructing, cule1999generating}) is an accessible descriptor for applications ranging from microstructure reconstruction (\cite{yeong1998reconstructing, jiao2007modeling, jiao2008modeling}) to prediction of macroscopic properties (\cite{rechtsman2008effective, kim2020multifunctional, kim2020effective}). In some applications, a combination of two-point correlation functions with lineal-path functions ($L_2$), or two-point cluster functions ($C_2$) was used to account for the non-uniqueness in reconstruction problems (e.g. \cite{yeong1998reconstructing, jiao2009superior}). \\

Next to microstructure quantification, there is the need to determine effective physical properties of these micro-heterogeneous media using upscaling methods. One such approach is to use closed-form approximation formulas where geometrical - thus mathematically tractable - inclusions are used to model microstructures (\cite{gaunaurd1983resonance, kachanov2005quantitative}). The disadvantage of inclusion-based approaches, however, is that these models are often too simplistic and too many competing models are available for describing for microstructure-induced effective properties, such as wave properties. In contrast, without committing to geometrical models, exact Strong-Contrast Expansion (SCE) formulations are useful, as they rely on the above-mentioned n-point correlation functions to describe micro-scale heterogeneity (\cite{torquato1997effective, rechtsman2008effective, kim2020multifunctional, kim2020effective}). In general, these SCE compute effective wave characteristics (wavespeeds and attenuation) as a function of frequency by taking ensemble averages over the microstructure descriptors. This procedure allows for an accurate description of microstructure influence on finite-frequency wave propagation at long wavelength scales, as scattering is negligible with respect to incident wave propagation. However, near the scattering regime (smaller wavelengths) SCE theory requires an additional correction approximately accounting for coherent scattering (\cite{kim2020effective, kim2020multifunctional}). 

Imaging techniques such as seismic and ultrasound imaging in geophysics (e.g. \cite{biondi20063d, virieux2009overview, stahler2011monitoring, niederleithinger2015embedded, ravasi205d}) and in medical applications (e.g. \cite{gennisson2013ultrasound, sigrist2017ultrasound, guasch2020full}) can remotely characterise heterogeneous materials, however, they are limited to the resolution of their wavelength which is often larger than the desired microscales. Still, both in geophysics ad medicine, studies show that these techniques are indeed sensitive to heterogeneity at the microscopic scale (\cite{sayers2005seismic, hatchell2005rocks, vasconcelos2007seismic, gennisson2003transient,  tanter2008quantitative, lu2016label,  sadeghi2017chemotherapy, komar2019advancing}). The goal of this study, therefore, is to make a first step toward remotely imaging microstructures by inferring quantitative information in the form of SMDs  - and possibly microstructure phase properties - from effective finite-frequency wave data. More specifically, the main goal is to extract two-point statistics ($S_2$) from effective wavespeed and attenuation computed using SCE theory which we extended to visco-acoustic and -elastic media. Here, we compute acoustic and elastic wave parameters from long-wavelength theories and additionally elastic wave properties near the coherent scattering regime and invert these in three separate inference setups.   \\

In general, the microstructure model space can have many degrees of freedom - including physical phase contrasts and microstructure geometry - thus the inversion is likely to be highly ill-posed. Therefore, it is important that we have access to the prediction uncertainty as well as a method that can deal with the full nonlinear nature of the forward calculation. For these reasons, we use a fully probabilistic approach to retrieve complete posterior distributions.
More specifically, we rely on the machine learning algorithm random forests (RF, \cite{breiman2001random}), a random ensemble of decision trees (\cite{breiman1984classification}) to unravel the non-linear relation between in- and output data, which has shown to be useful in geophysical contexts (\cite{reading2015combining, rouet2017machine}). The Scikit-learn library in Python (\cite{pedregosa2011scikit}) is a very accessible and powerful tool to this end. The data in this study are synthetic, and consist of a large collection of two-point medium statistics with corresponding analytically computed effective wavespeed and attenuation per propagating frequency. \\

Finally, one recently-discovered, interesting measurable state of soft matter is the so-called hyperuniformity, which is characterised by slow decay of volume fraction fluctuations at large scales with respect to the volume of a certain expanding observation window (\cite{zachary2009hyperuniformity}). Hyperuniform materials occur in various forms in nature, e.g. as crystals, and can also occur in the geophysical field as two-phase heterogeneous media (\cite{xu2017microstructure}).  \cite{kim2020effective} showed that hyperuniformity can have signatures in effective wave properties; it namely suppresses attenuation due to scattering. This leads us to the hypothesis that when we are able to get accurate estimates of SMDs we may be able to detect hyperuniformity from effective wave data.

Here, we begin by briefly reviewing low-order statistical microstructure descriptors (SMDs) in the form of two-point correlation functions, and on how we sample them from digital microstructure images. Next, we cover the theory of strong-contrast expansions (SCEs), which allow us to directly tie effective wave properties at finite frequencies to SMDs - discussing the cases of acoustic, and the so-called local and nonlocal elastic SCEs. Using this framework connecting SMDs to wave properties, we then present the inverse problem of inferring microstructures in a Bayesian sense, and describe our supervised-learning Random-Forest (RF) approach to solving the inference problem. Finally, we present inference results for three benchmark microstructures with different degrees of complexity, discussing the role of features and priors in the estimation, together with physical insights based on the different SCE theories used. 

\section{Statistical microstructure descriptors}

\subsection{Two-point statistics}
Here we consider microstructures of two phase random media - examples are porous media, composites and polycrystals - and generalise them using two-point correlation functions~\citep[e.g.,][]{torquato1988two}. Throughout this work we consider binary microstructure images, where in case of a porous medium the black and white regions represent pores (phase 2) and the solid (phase 1) respectively. We introduce the indicator function for phase i as: 
\begin{equation}
\label{eqn:indic}
    \mathcal{I}^{(i)}(\boldsymbol{x}) = 
    \left\{
    \begin{array}{ll}
    1, \; \boldsymbol{x} \in  \mbox{phase i}\\
    0, \; \mbox{otherwise} 
    
    \end{array}
    , \; \mbox{for} \; i = 1, 2
    \right.
\end{equation}
The volume fraction of each phase is expressed by the quantity $\phi_i \equiv \langle \mathcal{I}^{(i)}(\boldsymbol{x}) \rangle$, i.e. the ensemble average. From this it follows that the two-point correlation function reads 
\begin{equation}
S_2^{(i)}(\boldsymbol{x}_1, \boldsymbol{x}_2) \equiv \langle \mathcal{I}^{(i)}(\boldsymbol{x}_1) \mathcal{I}^{(i)}(\boldsymbol{x}_2) \rangle. 
\end{equation}
For statistical homogeneous and isotropic media which we consider here and $\boldsymbol{r} = \boldsymbol{x}_2 - \boldsymbol{x}_1$, we have:
\begin{equation}
    S_2^{(i)}(\boldsymbol{x}_1, \boldsymbol{x}_2) = S_2^{(i)}(\norm{\boldsymbol{r}}) = S_2^{(i)}(r).
\end{equation}
Logically, for $r=0$ we have
\begin{equation}
    S_2^{(i)}(0) = \phi_i,
\end{equation}
and when there are no long range orders,
\begin{equation}
    \lim_{r \to \infty} S_2^{(i)}(r) = \phi_i^2.
\end{equation}
We can analytically estimate the average pore or grain size with the derivative of $S_2^{(i)}(r)$ at $r=0$:
\begin{equation}
\label{eqn:poresize}
    \bar{l}^{(i)} = \abs{\frac{\phi_i}{S_2^{(i) \prime } (0)}},
\end{equation}
Which we can directly determine from the intersection of that slope with the $x$-axis (\cite{mosser2017reconstruction}). 
The scaled autocovariance function, which is independent of volume fraction and the choice of phase $(i)$, equals
\begin{equation}
    f(r) = \frac{\chi(r)}{\phi_1 \phi_2} = \frac{S_2^{(i)}(r) - \phi_i^2}{\phi_1\phi_2},
\end{equation}
with $\chi(r)$ the autocovariance function, and is useful for comparing curve shapes. All necessary conditions for $S_2$ and $f$ can be found in \cite{torquato2006necessary}. \\

%Spectraldensity - scattering measurements: problem with finite resolution images and no analytical representation as in \cite{kim2020multifunctional}/HYPERUNIFORMITY?! \\

In the special case of hyperuniformity, random media possess unique geometric regularity properties and can be mathematically detected when the autocovariance spectral density has a zero infinite wavelength component (\cite{zachary2009hyperuniformity}):  \\
\begin{equation}
\label{eqn:hyper_fourier}
    A_\tau = \lim_{k \to 0} \Tilde{\chi}(k) = 0.
\end{equation}
Therefore, using the convention $\Tilde{\chi}(\boldsymbol{k}) = \int_{\mathbb{R}^d} \chi (\boldsymbol{r}) \exp (- i \boldsymbol{k} \boldsymbol{r}) d\boldsymbol{r}$, the hyperuniform condition in the spatial domain is (e.g. \cite{xu2017microstructure}):
\begin{equation}
\label{eqn:hyper_int}
    A_\tau = \int_r \Big[ S_2 (r) - \phi^2 \Big] \Omega_d (r) dr = 0,
\end{equation}
where $\Omega_d (r)$ is the surface area of a sphere with radius $r$ in $d$ dimensions. Here, we mainly use the spatial formulation for image samples because numerical fast Fourier transform computations are bound by finite $r$ samples from their resolutions - making the evaluation of Fourier coefficients near zero frequency fairly unstable. In addition, instead of evaluating absolute numbers, we compare the relative degree of hyperuniformity between samples and eventually relate this to effective wave properties. \\

\begin{figure}[t!]
	\centering
	\begin{subfigure}[t!]{0.48\textwidth}
		\centering
		\includegraphics[width=\textwidth]{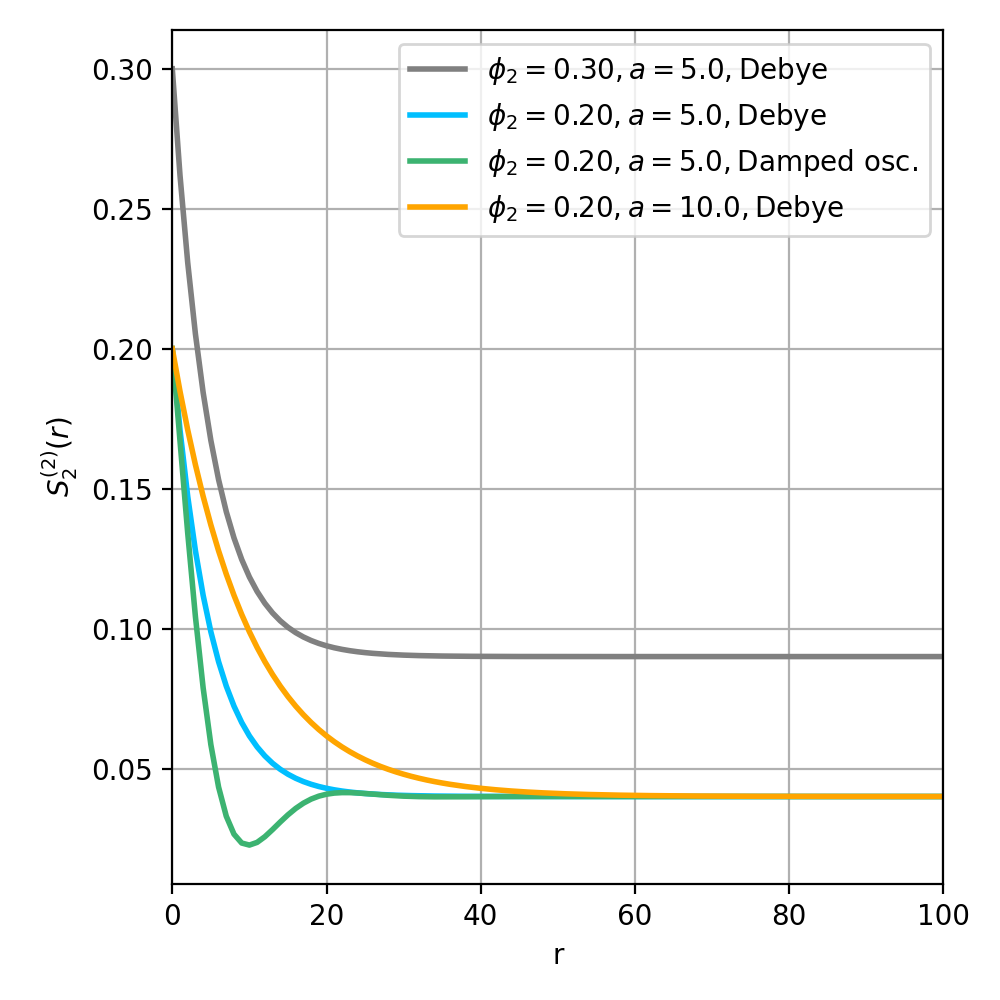}
		\caption{\label{fig:S2s}}
	\end{subfigure}%
	\quad
	\begin{subfigure}[t!]{0.48\textwidth}
		\centering
		\includegraphics[width=\textwidth]{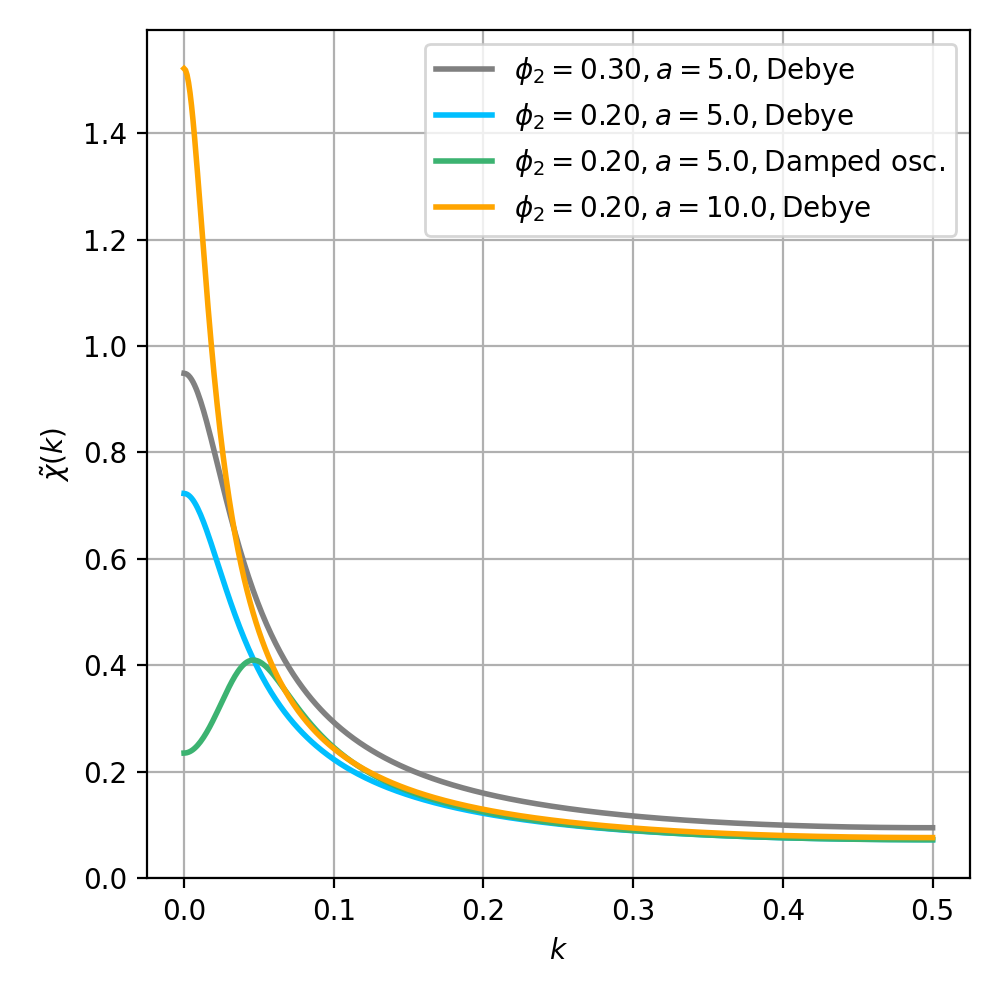}
		\caption{\label{fig:chi_of_k}}
	\end{subfigure}
	\caption{\small (a) Examples of analytical two-point correlation functions with varying medium length scales (here a is either $a_D$ or $a_O$), medium type (8) and (9) and volume fraction $\phi_2$. For the damped oscillating function: $q = 1/4$ and $\psi = 0$. In (b) the corresponding spectral densities are shown to compare the degree of hyperuniformity (\ref{eqn:hyper_fourier}).}
	\label{fig:analytical_S2}
\end{figure}

When aiming at studying the geometry of micro-heterogeneous materials, it is useful to introduce a few analytical examples of scaled autocovariance functions of statistical homogeneous and isotropic media, since wisely chosen combinations have shown to closely fit those of certain real microstructure samples (\cite{jiao2007modeling, jiao2008modeling}). Arguably, the simplest example of scaled autocovariance functions is that of a Debye random medium (\cite{debye1957scattering}), which is a monotonically decaying function (figure \ref{fig:S2s}):
\begin{equation}
\label{eqn:f_Debye}
    f_D(r) = \exp(r/a_D),
\end{equation}
where $a_D$ is a characteristic length scale of particles with random shapes and sizes. This scaled autocovariance function has frequently shown to be realizable for reconstruction (e.g. \cite{yeong1998reconstructing}). Damped oscillations are an originally hypothetical - though realizable - family of functions, given by (\cite{yeong1998reconstructing}):
\begin{equation}
\label{eqn:f_DO}
    f_O(r) = \exp(r/a_O) \cos(q r + \psi),
\end{equation}

characterised by anti-correlations controlled by the wavenumber q (figure \ref{fig:S2s}). The exponent with length scale $a_O$ determines the oscillation amplitude and with it the long scale correlations. Note that when $q, \psi \to 0$, the function equals that of Debye media. The oscillation troughs represent `repulsions' of particles due to the first phase separating them - in other words, geometric anti-correlations for a single material phase. Penetrable spheres, as an opposing example to impenetrable spheres, are allowed to connect and therefore the oscillating nature in the scaled autocovariance function diminishes. Note that the damped oscillating autocovariance function is significantly more hyperuniform compared to that of Debye random media (figure \ref{fig:chi_of_k}). Other examples are power-law correlated materials and cubic lattice packings (e.g. \cite{rechtsman2008effective}). In general two point correlation function shapes of real samples can be approximated using complex maps of convex combinations and products of basis functions (for examples see \cite{jiao2007modeling}). Here we will use only convex combinations of $m$ functions giving the effective scaled autocovariance function $f^e$:
\begin{equation}
\label{eqn:autocov_sum}
    f^e(r) = \sum_{i=1}^{m} \alpha_i f_i (r) \, ,
\end{equation}
\noindent when choosing to represent families of scaled autocovariance functions for the purpose of training in machine learning.

\subsection{From microstructure images to SMDs}
Next to analytical correlation functions we will evaluate real microstructure data for benchmarking. To this end, we built a fit-for-purpose Python class including a method that takes two-dimensional tomographic images into two-phase binary microstructures using image processing tools from scikit-image (\cite{van2014scikit}). Subsequently, this microstructure is quantified into two-point correlation functions. We will use the orthogonal
sampling method of \cite{yeong1998reconstructing}, which has proven to be accurate for statistical isotropic media. 

\begin{figure}[t!]
	\centering
	\begin{subfigure}[b]{0.32\textwidth}
		\centering
		\includegraphics[width=\textwidth, trim={75 0 75 0},clip]{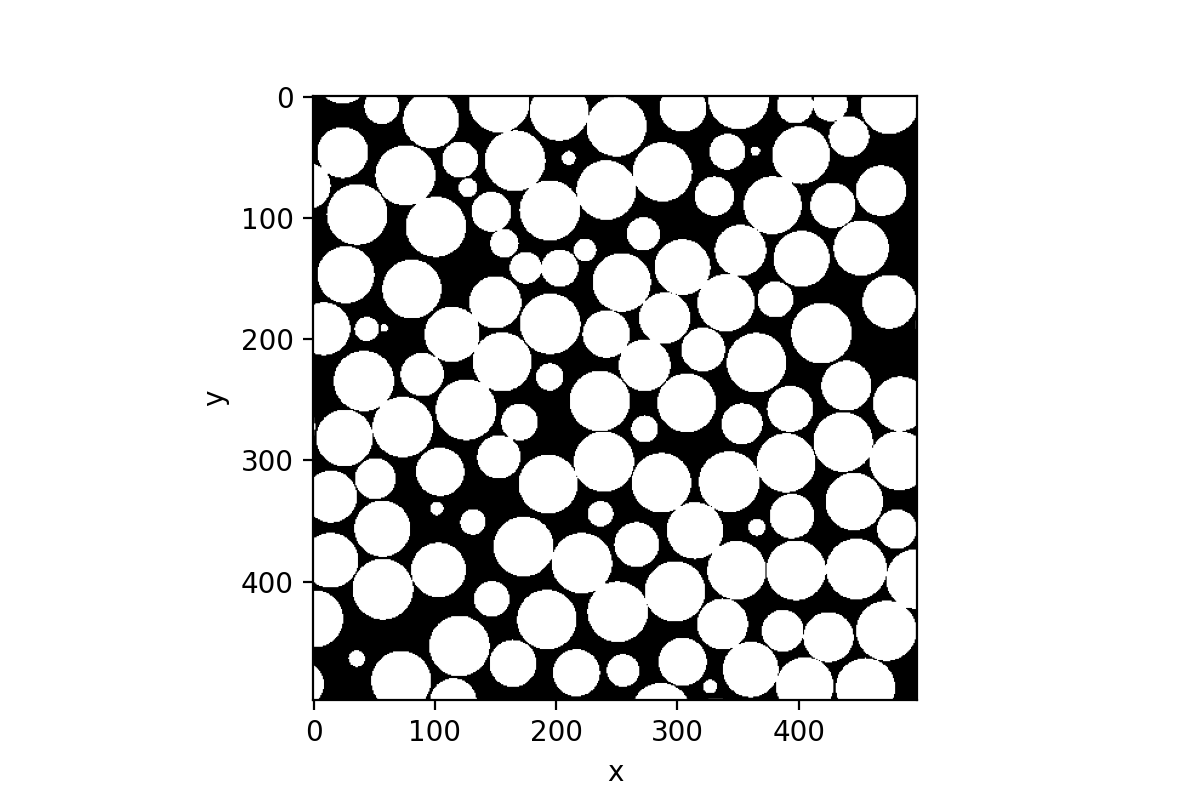}
		\caption{\label{fig:beadim}}
	\end{subfigure}%
	\,
	\begin{subfigure}[b]{0.32\textwidth}
		\centering
		\includegraphics[width=\textwidth, trim={75 0 75 0},clip]{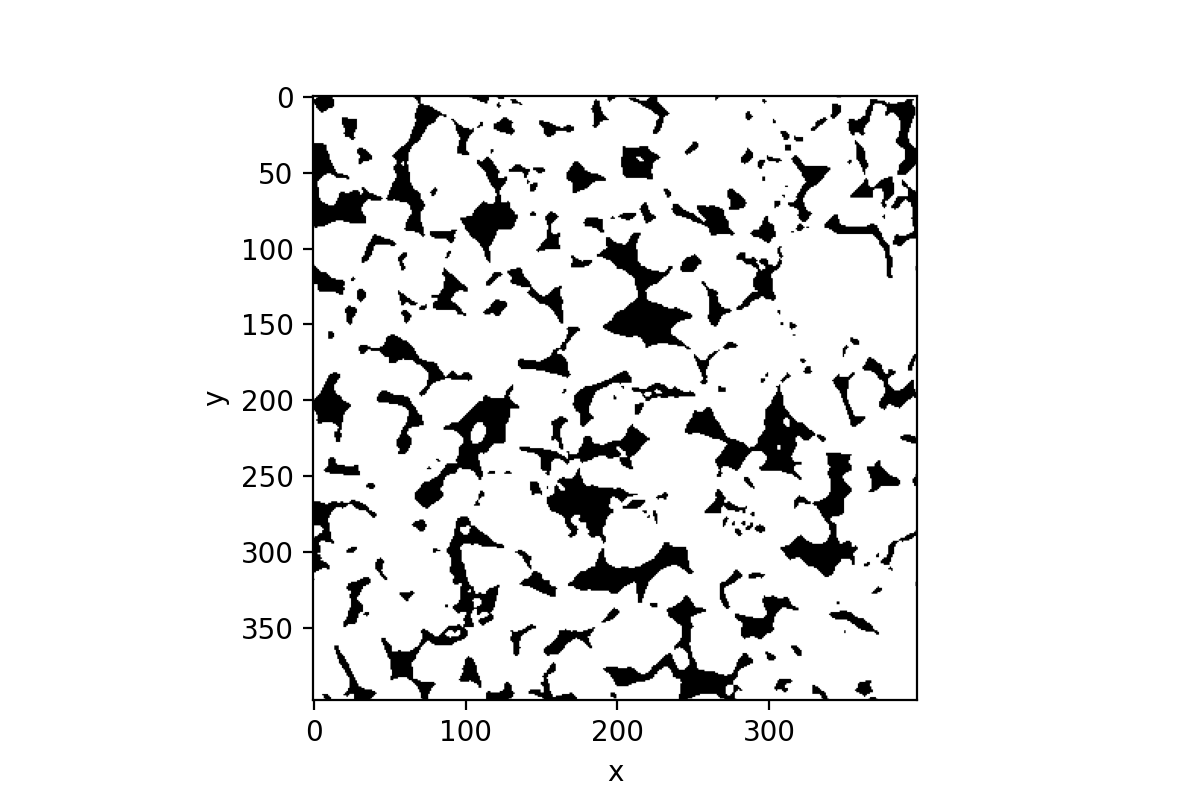}
		\caption{\label{fig:bereaim}}
	\end{subfigure}
	\, %vspace\baselineskip
	\begin{subfigure}[b]{0.32\textwidth}
		\centering
		\includegraphics[width=\textwidth, trim={75 0 75 0},clip]{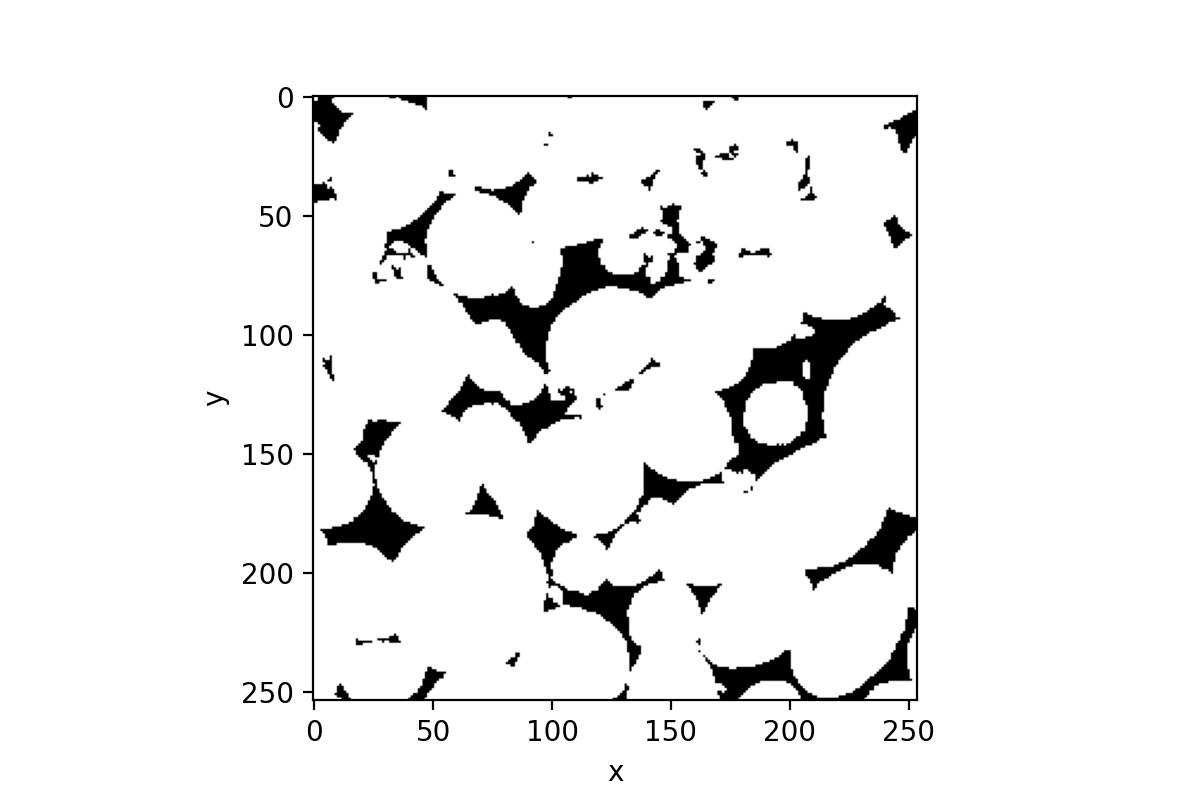}
		\caption{\label{fig:kettonim}}
	\end{subfigure}%
	%\,
	%\quad
	%\begin{subfigure}[b]{0.47\textwidth}
	%	\centering
	%	\includegraphics[width=\textwidth, trim={75 0 75 0},clip]{asphalt_microstructure_structure.png}
	%	\caption{\label{fig:asphaltim}}
	%\end{subfigure}
	\caption{\small{Two-dimensional representations of microstructure samples used in this study. The white phase ($i=1$) contains the particle phase and the black phase represents the pores. We refer to these samples as (a) bead pack, (b) Berea sandstone and (c) Ketton limestone (\cite{mosser2017reconstruction}).}}
	\label{fig:Mossersamples}
\end{figure}

\begin{figure}[t!]
    \centering
    \begin{subfigure}[b]{0.4\textwidth}
		\centering
		\includegraphics[width=\textwidth, trim={0 0 0 0},clip]{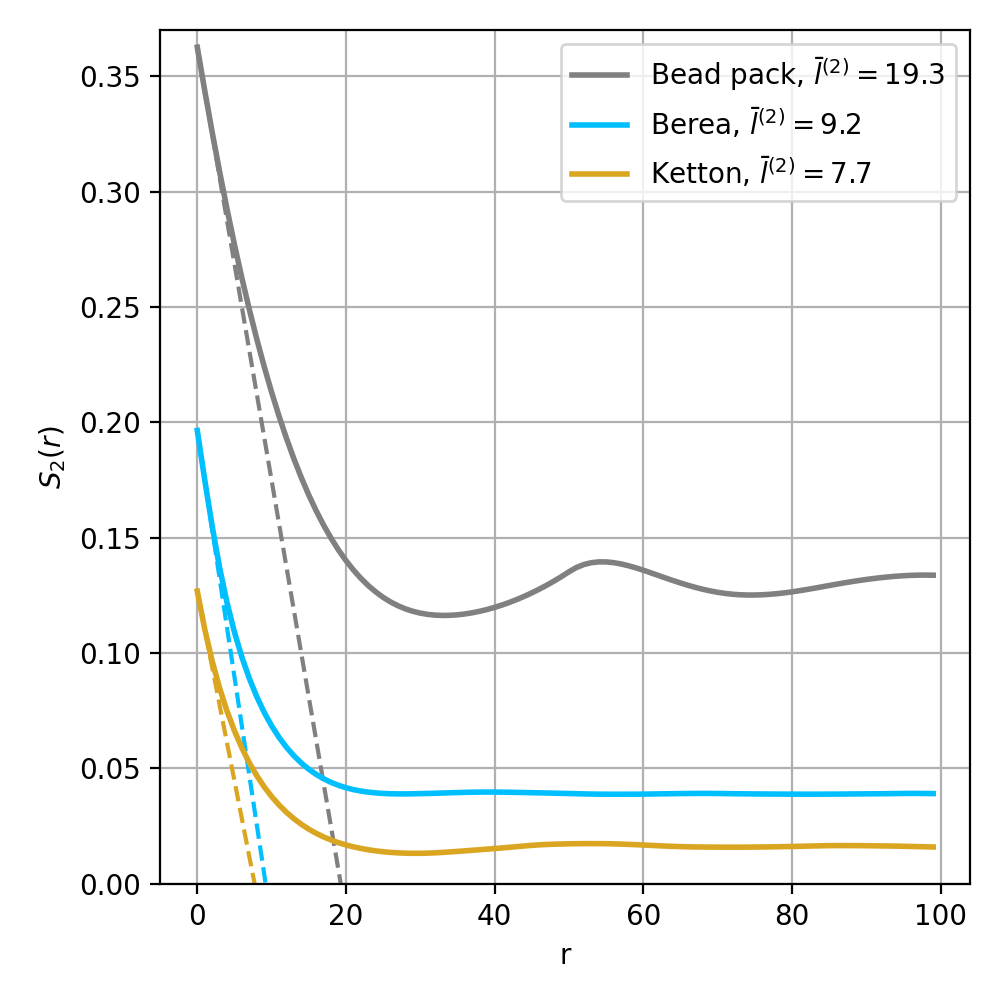}
		\caption{\label{fig:MosserS2}}
	\end{subfigure}%
	\quad
	\begin{subfigure}[b]{0.4\textwidth}
		\centering
		\includegraphics[width=\textwidth, trim={0 0 0 0},clip]{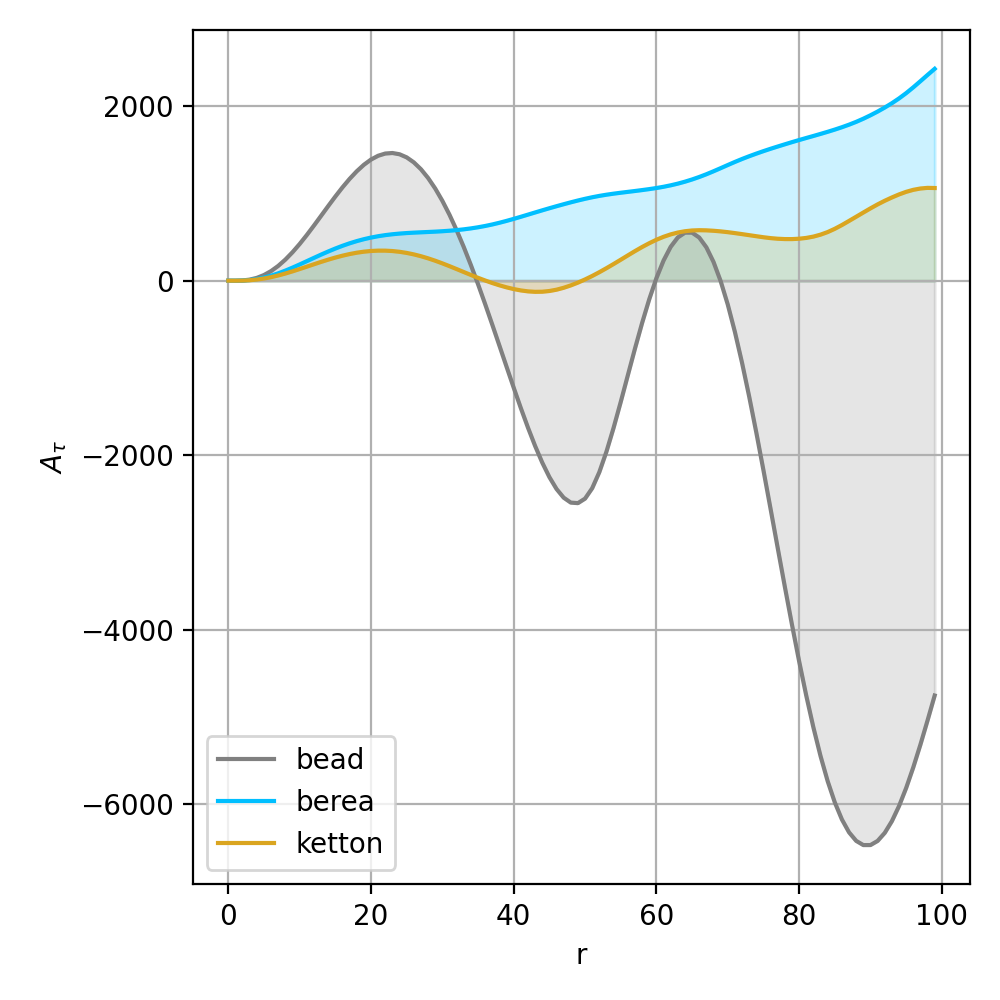}
		\caption{\label{fig:Mosserhyper}}
	\end{subfigure}
	\caption{\small{SMDs for samples in figure \ref{fig:Mossersamples}, with the two-point correlation functions in (a) and the integrals of equation (\ref{eqn:hyper_int}) in (b). The dashed line in (a) is provided to give an estimate of the average pore size per sample. The sum of the marked areas in (b) yields the degree of hyperuniformity.}}
	\label{fig:MosserSMD}
\end{figure}

The binary samples in figure \ref{fig:Mossersamples} are from the open data repository of \cite{mosser2017reconstruction}. The bead pack sample consists of equally sized spheres with 50 pixels diameter and a porosity of $\phi_2 = 0.36$. It has a total one-dimensional size of 500 pixels and a pixel size of $3 \mu$m. The well-known berea sandstone in figure \ref{fig:bereaim} contains angular grains of medium to fine size with a pore space fraction of $\phi_2 = 0.20$ (\cite{pepper1954geology}) and extents 400 pixels in each direction with a pixel size of 3 $\mu$m. The Ketton sample is less porous with relatively large grains, as is characteristic for oolitic limestones (\cite{emery1988origin}). The image size is 256 pixels per dimension and the pixel size is 15.2 $\mu$m. 
%The asphalt sample is taken from \cite{valenta2010macroscopic} and is characterised by complex structures of different scales. The image size is 500 pixels per dimension and the pixel size is estimated at $\sim$ 100 $\mu$m.
Figure \ref{fig:MosserSMD} shows the radial averaged two-point correlation functions for these samples. In general the length of these curves equal half the total sample size. However to generalise we truncate each curve at $r = 100$ pixels.
The oscillating nature that arises in the bead pack curve is expected since the sample consists of impenetrable spheres, as discussed in the previous section. On the other hand, the Ketton sample contains grains that can be seen as penetrable spheres resulting in little to no repulsion in the $S_2$ curve. A low amplitude oscillation arises in the $S_2$ of the Berea sample. 
%The asphalt sample has multiple scales of heterogeneities while the average pore size is small. The $S_2$ curve captures this in the large slope at $r = 0$ and the relatively slow decay for larger $r$ values. 
To verify statistical isotropy we compared this radial average with the two-point correlation function in all directions. With the exception of the Ketton sample, which shows significant anisotropy, all samples have no directional variance and can be treated as isotropic (Appendix \ref{sec:AppA},  \cite{mosser2017reconstruction}). In this work we only work with the radial average correlation functions and as such treat the Ketton sample as statistically isotropic. This suffices for this work as we are only interested in the connection between $S_2$ and effective wave properties, however for potential reconstruction purposes in real case scenarios, it is recommended not to use the radial average for such types of anisotropic samples. Although samples here are shown in two dimensions, the calculations in the following section are based on three-dimensional media which is justified by the assumption of statistical isotropy. 

Figure \ref{fig:Mosserhyper} shows the integral in (\ref{eqn:hyper_int}) as a function of cut-off values $r$. For a hyperuniform material this curve would be converging to approximately zero. These samples are thus not hyperuniform but we can distinguish relative differences. Namely, the increasing order of degree of hyperuniformity is bead pack, Berea and then Ketton. Here we clearly see the effect of resolution limitations, which the spectral density would not reveal. Moreover, if this curve were truncated at $r \approx 70$, we would conclude that the bead pack was completely hyperuniform, leading to misinterpretations. As an additional side-note, if the two-point correlation functions would have been sampled from two-dimensional images the surface area $\Omega_d(r)$ would have been proportional to $r$ instead of $r^2$, resulting in a relatively lower weight on the higher radial distances. In that scenario the bead pack curve in figure \ref{fig:Mosserhyper} might not have descended as strongly causing different degrees of hyperuniformity - here we maintain 3D weighting in the hyperuniformity measure to represent geometric uniformity of corresponding 3D microstructures.

\section{Effective wave properties at finite frequencies}

Together with phase properties we can analytically compute frequency-dependent effective wave properties from two-point correlation functions using SCE theory (\cite{kim2020effective}). More specifically, the aim is to compute the effective longitudinal (P-wave) and transverse (S-wave) wavespeeds ($\rm{Re} \lbrace c_e^L \rbrace$ and $\rm{Re} \lbrace c_e^T \rbrace$, respectively) and attenuation coefficients ($\rm{Im} \lbrace c_e^L \rbrace$ and $\rm{Im} \lbrace c_e^T \rbrace$, respectively) as a function of frequency as if the medium were described by a homogeneous continuum. The input properties are both wavespeeds per phase ($c_i^L$, $c_i^T$), which can either be real- or complex-valued - allowing us to account for intrinsic attenuation at the microscale. \cite{kim2020effective} only used real values, such that each phase on their own is dissipationless. In contrast, we will here extend the theory to viscoelastic media where the individual wavespeeds are complex but constant with frequency - this assumption is reasonable because single-phase particle sizes are already considered much smaller than the wavelength, and as such sub-microscale attenuation is likely frequency-independent regardless of absoprtion mechanism. Here, we describe the direct analytical calculation of effective properties and leave the derivation details to the Appendix. Throughout this work, phase $q$ is the reference phase and phase $p$ the `polarized' or `perturbation' phase ($p \neq q$). Both $p$ and $q$ can be phase 1 or 2 in practice. The mass density $\rho$ is assumed constant in the following derivations and the media are three-dimensional.  
Since we have assumed all media to be statistically isotropic - and thus also macroscopically isotropic - we retrieve isotropic-elastic solutions, where each wave mode is described a single scalar wavespeed. Rather, importantly, here we note that these isotropy assumptions are not a requirement for validity of our inference methods, nor are they a limitation of the SCE theory - we make these assumptions to begin with the simpler, more physically and numerically straightforward cases. Both the SCE theory and our approach extend, in principle, to statistically anisotropic materials, and to phases with arbitrarily complex stiffness tensors - though these more complicated cases are beyond the scope of this study.  \\

 In general, the procedure is to formulate stress perturbations from medium heterogeneities that act locally as point scatterers, resulting in long-wavelength energy loss due to incoherent scattering at small scales. Thus, in the case where the wavelength dominates over the microstructure scale (the long-wavelength regime) the attenuation mechanism is sub-wavelength Rayleigh scattering. In that case we formulate a homogenised (spatially averaged) constitutive relation, where the contrast source is related to the resulting wavefield through a constant tensor. This relation is local in space, meaning that the resulting field at position $\boldsymbol{x}$ depends on the effective medium contrast at $\boldsymbol{x}$. Incident waves with smaller wavelengths, however, cause many scattered waves with more coherent interference resulting in a non-trivial homogenisation of the constitutive relation. The approach of \cite{kim2020multifunctional} and \cite{kim2020effective} for this nonlocal case is to assume that the similar homogenisation is valid when a correction term for the scattered waves is added, which we follow herein. In this manner we approximately capture the shift in physics from the  long-wavelength effective regime to the semi-coherent scattering regime - and can investigate their relative contributions to the inversion problem. 
 
 %The range of wavelengths we include is based on the resolution of wave-based imaging techniques e.g. under the Born approximation, which . \\ 

When it comes to accuracy related to their convergence behaviour, we note that strong contrast expansions can theoretically evaluate infinite sets of n-point microstructure correlation functions. More pragmatically, it has been shown that these expansions converge rapidly (\cite{torquato1997effective}), enabling reasonably accurate approximations when truncated at the two- and three-point level. Here, we rely solely on two-point-truncated SCEs as we only rely on two-point statistics for SMDs. The role of higher-order descriptors and their corresponding SCE correction terms may be important, but is the subject of future research.   

\subsection{Pure-mode longitudinal waves}
Here, we consider the case where longitudinal and transverse waves are decoupled, reducing our problem to pure longitudinal (acoustic) waves. This is the simplest elastic wave setting, and thus the starting point for building an inference method for microstructure. This choice results in scalar wave equations for the two elastic wave modes, each similar in form to that in \cite{rechtsman2008effective}. We follow their derivation as the interference between propagation modes is neglected. For the purpose of analysing frequency-dependent effects, we derive the effective acoustic wavespeed as a function of frequency:
\begin{equation}
\label{eqn:ac_ceff}
    c_e^L(\omega) \equiv \sqrt{\frac{\omega^2}{\sigma_e(\omega)}},
\end{equation}
where $\omega$, the $L$ superscript denotes longitudinal waves, and $k_e^L = \sqrt{\sigma_e}$ are the angular frequency and effective wavenumber, respectively. The latter is defined by:
\begin{equation}
\label{eqn:ac_sigma_eff}
    \frac{\sigma_e + 2\sigma_q}{\sigma_e - \sigma_q} = \frac{\phi_p - A_2(k_q) \, \beta_{pq}}{\phi_p^2 \, \beta_{pq}},
\end{equation}
with 
\begin{equation}
\label{eqn:ac_betapq}
    \beta_{pq} \equiv \frac{\sigma_p - \sigma_q}{\sigma_p + 2 \sigma_q},
\end{equation}
and
\begin{equation}
\label{eqn:A2int}
    A_2(Q) = \frac{Q^2}{2\pi}\int \frac{e^{iQr}}{r} \, \chi(r) dr,
\end{equation}
where $Q$ is an arbitrary wavenumber. This approximation is only valid in the long-wavelength regime. Here, $e^{iQr}/(4 \pi r)$ is the scalar Green's function in the reference phase obeying the Helmholtz equation. See Appendix \ref{subsec:AppB_1} and \cite{rechtsman2008effective} for the derivation. 

\begin{figure}[t!]
    \centering
    \begin{subfigure}[b]{0.48\textwidth}
		\includegraphics[width=\textwidth, trim={0 0 30 30},clip]{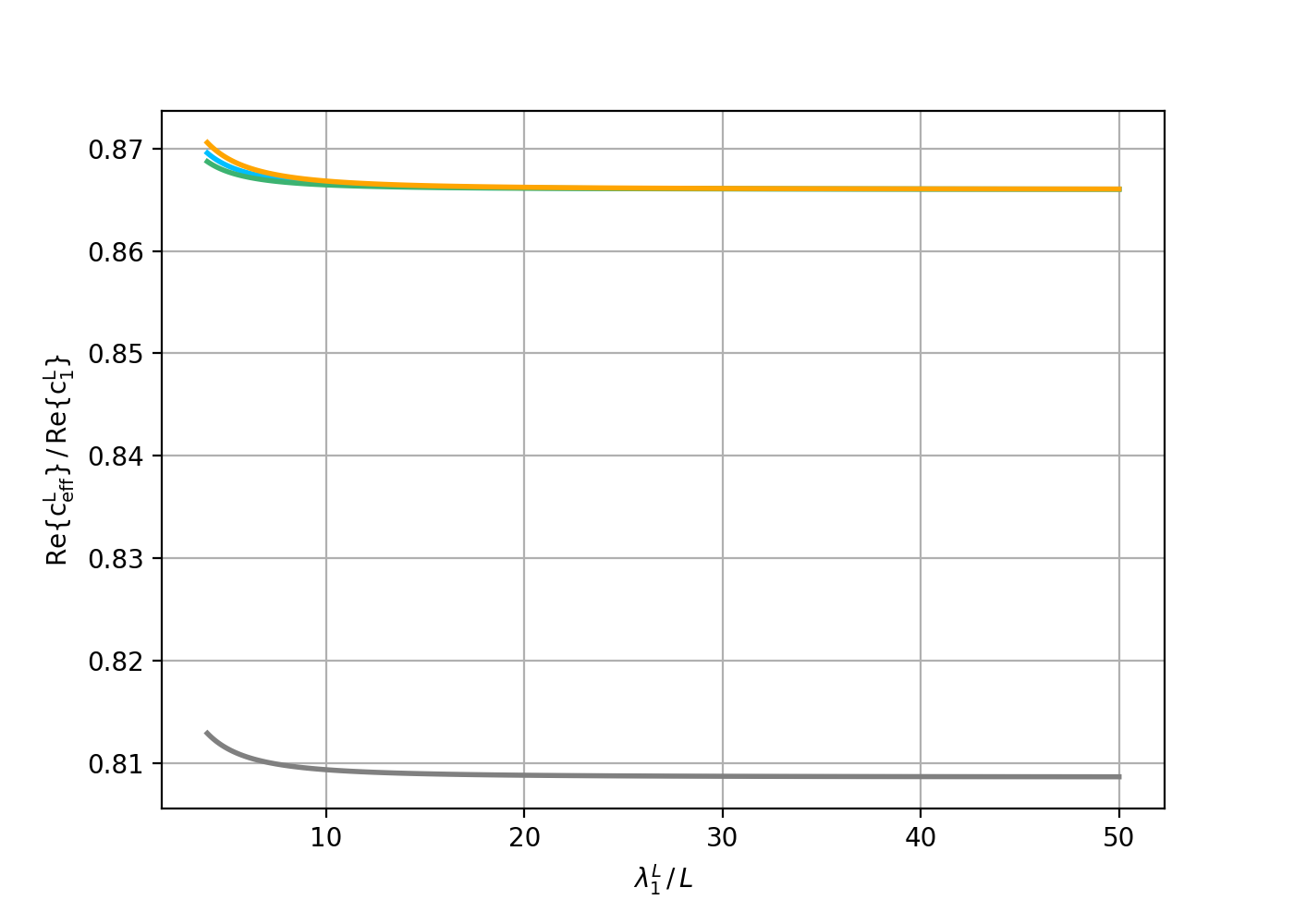}
		\caption{\label{fig:AnalyticalEffCeff}}
	\end{subfigure}%
	\quad
	\begin{subfigure}[b]{0.48\textwidth}
		\includegraphics[width=\textwidth, trim={0 0 30 30},clip]{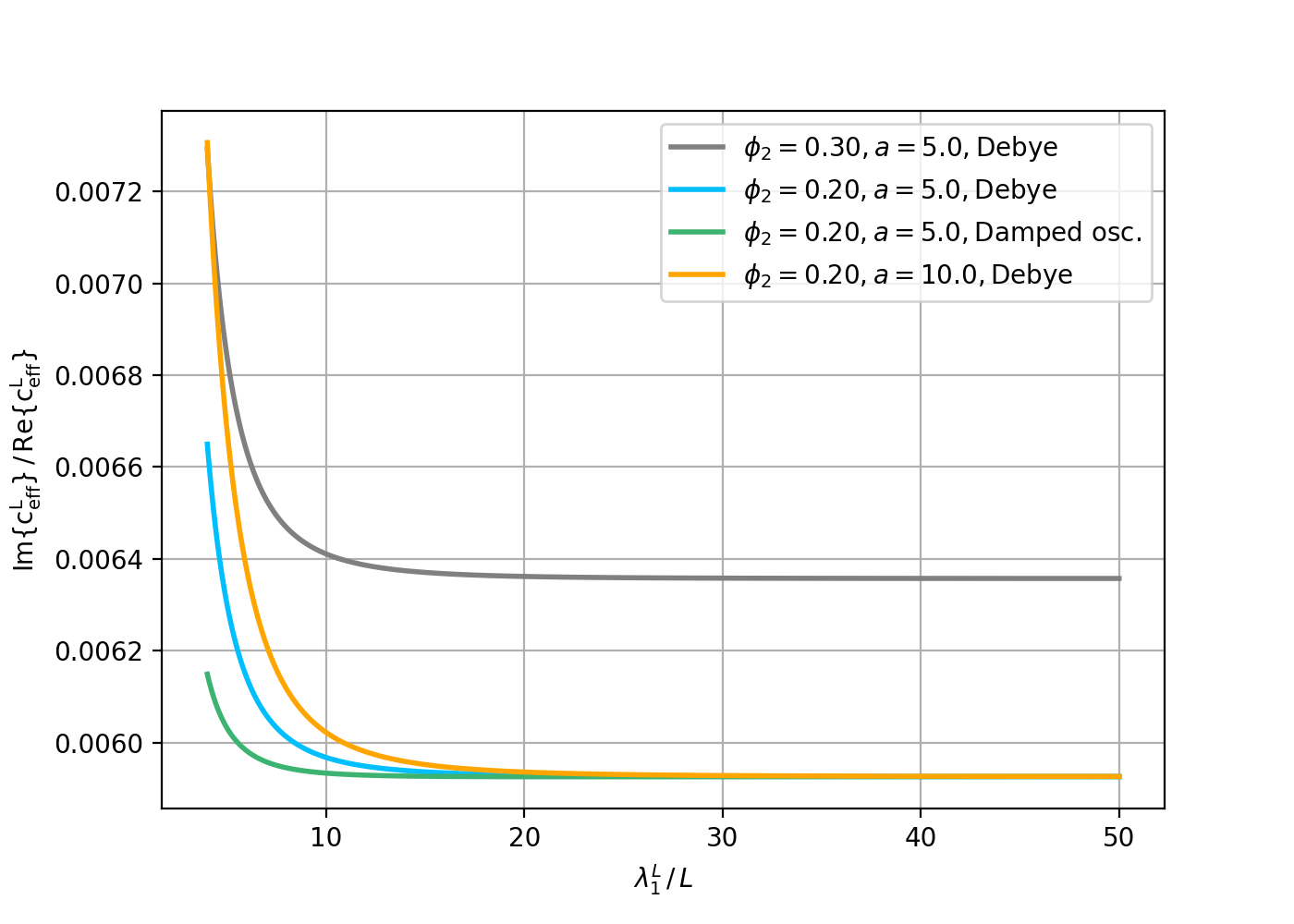}
		\caption{\label{fig:AnalyticalEffAtt}}
	\end{subfigure}
	\caption{\small{Scale-dependent effective acoustic wavespeed (a) and attenuation (b) for the two-point correlation functions in figure \ref{fig:S2s}. Note that the absolute values are predominantly determined by the volume fraction, and the curvature by the shape of $f(r)$. Here we use $c_1^L/c_2^L = A_2^L/A_1^L = 2$.}}
	\label{fig:AnalyticalEffprop}
\end{figure}

Before discussing elastic theory we briefly  discuss effective acoustic properties of the analytical media shown in figure \ref{fig:S2s} to describe the general characteristic behaviour of the SCE calculations; see figure \ref{fig:AnalyticalEffprop}. The frequency range is chosen from $\lambda / L \geq 4$, with $L$ a chosen characteristic microstructure length scale, not larger since most far-field imaging techniques do not have higher resolution due to the diffraction limit (\cite{virieux2009overview, huang2014resolution}) - and not smaller because of the local approach that can not see beyond such small scales. First of all, the absolute wavespeed and attenuation at large scales is predominantly dependent on volume fraction; this is in line with electromagnetic observations of \cite{rechtsman2008effective} in the quasistatic regime. Effective wavespeeds are nearly constant with scale except for wavelengths near the scattering regime where some dispersion occurs. Attenuation increases much more strongly towards the coherent scattering regime, which is expected. The magnitude of coherent versus incoherent scattering at any given frequency clearly on medium geometry - captured by the scaled autocovariance functions. For instance, Debye random media with a larger characteristic length scale cause larger amounts of incoherent scattering. This is in comparison to more `ordered' media, e.g., the damped oscillating function being more hyperuniform than the Debye random media with a similar volume fraction (figure \ref{fig:chi_of_k}) is less sensitive to incoherent scattering, which has been observed before by \cite{kim2020effective} who even found materials that are completely transparent at long wave scales. The fact that this measurable state of matter (i.e., hyperuniformity) shines through in effective wave observations may be of large importance in geophysics and enginering, since hyperuniform materials have shown to possess relatively large brittle strength compared to nonhyperuniform counterparts (\cite{xu2017microstructure}). 

\subsection{Frequency-dependent effective stiffness and wave properties}
Next, we follow the derivation of \cite{kim2020effective} in which we include transverse waves and the interaction between both modes. Now the goal is to determine the effective elastic moduli in the effective stiffness tensor, which in this case are formulated as a function of the reference longitudinal wavenumber (\cite{kim2020multifunctional}):
\begin{align}
\label{eqn:elastic_ceff}
    c_e^L(k_q^L) &\equiv \sqrt{\frac{K_e(k_q^L) + 4/3 \, G_e(k_q^L)}{\rho}}, \\
    c_e^T(k_q^L) &\equiv \sqrt{\frac{G_e(k_q^L)}{\rho}},
\end{align}
where $K_e(k_q^L)$ is the effective bulk modulus and $G_e(k_q^L)$ the effective shear modulus, which are computed through:
\begin{equation}
\label{eqn:K_eff}
    \frac{K_e(k_q^L)}{K_q} = 1 - \frac{\kappa_{pq} \, \phi_p^2 \, (1 + 4G_q / 3/K_q)}{C_2 (k_q^L) + \phi_p \, (\kappa_{pq} \, \phi_p - 1)}, 
\end{equation}
\begin{equation}
\label{eqn:G_eff}
    \frac{G_e (k_q^L)}{G_q} = 1 - \frac{5 \, \mu_{pq} \, \phi_p^2 \, (1 + 4G_q/3/K_q)}{2(1+2G_q/K_q) \, (D_2(k_q^L) + \phi_p \, (\mu_{pq} \, \phi_p - 1))},
\end{equation}
with
\begin{equation}
    \kappa_{pq} \equiv \frac{K_p - K_q}{K_p + 4 G_q/3}
\end{equation}
and
\begin{equation}
\mu_{pq} \equiv \frac{G_p - G_q}{G_p + (3K_q / 2 + 4G_q/3)G_q/(K_q + 2G_q)},
\end{equation}
which are the strong-contrast parameters for the bulk and shear moduli, respectively. In addition:
\begin{equation}
    C_2 (k_q^L) = \sqrt{\frac{\pi}{2}} \, \mathcal{F}(k_q^L) \, \kappa_{pq}
\end{equation}
\begin{equation}
\label{eqn:D_2}
    D_2(k_q^L) = \sqrt{\frac{\pi}{2}} \, \frac{3 \, c_q^L \, \mathcal{F}(k_q^T) + 2 \, c_q^T \, \mathcal{F}(k_q^L)}{3 \, {c_q^L}^2 + 2 \, {c_q^T}^2} \, \mu_{pq}.
\end{equation}
Here, $\mathcal{F}(Q) = 1 / \sqrt{2 \pi} A_2(Q)$ is the so-called attenuation function, valid in only the long-wavelength regime. To extend this theory to the intermediate wavelength regime (i.e., higher frequencies close to the coherent diffraction limit), we add a term to the attenuation function that accounts for incident scattered waves, resulting in the following attenuation function. 
\begin{equation}
\label{eqn:F_nonlocal}
    F(Q) \equiv i \frac{Q^{(3/2)}}{\sqrt{\pi}} \int \sqrt{r} \, e^{iQr} \, \sin (Qr) \, \chi(r) \, dr
\end{equation}
The derivation for this corrext version of $f$ can be found in Appendix \ref{subsec:AppB_2} and in \cite{kim2020effective}.

\subsection{Finite-frequency effective properties from image samples}
To treat the problem of inverting for microstructure descriptors from wave properties, we note that the forward operator in question is non-linear - due to wavefield interactions with the complex microstructure heterogeneity. This forward relation is generally given by 
\begin{equation}
\label{eqn:fwd}
    \boldsymbol{d} = \boldsymbol{G} (\boldsymbol{m}),
\end{equation}
where the data $\boldsymbol{d}$ consist of effective wave properties and the model space ($\boldsymbol{m}$) includes SMDs and elastic properties of each phase. For the porous media we use as an example, the particles (white phase, i = 1) are the stiffer phase. We therefore in general expect the effective stiffness to be reduced by influence of the more compliant pores, as well as scattering-induced dissipation to occur by virtue on introducing microheterogeneity. In addition, the effective attenuation is expected to depend on wavenumber, due to the aforementioned scale-dependent multiple scattering mechanism for energy dissipation.

We use equal physical medium parameters for each sample - i.e., white- and black-phase elastic constants are the same for all case -  even though in reality these can differ significantly based on the materials themselves, and what the choice for the controlling phases are in terms of microstructure features (e.g., pores, mineral phases, etc.). For example, the stiffness of a limestone matrix in general differs from that of sandstones; the stiffness of the pore phase depends on what it stores - e.g. gas, water or oil. In addition, we assume that the samples have equal scales, i.e., equal pixel dimensions - even though in reality the Ketton sample we use here has a larger pixel width than the other samples. This will however not affect our results significantly, since the SCE theory is valid for all microstructure scales - thus we use these samples for method validation and illustration. Here, we choose parameters loosely based on brine-filled sandstone . That is, $c_1^L = 4500$ m/s, $c_2^L = 1500$ m/s, $c_1^L / c_1^T = 1.6$; since shear moduli for fluids is zero, we set this to a relatively small value to avoid numerical issues ($c_2^T = 1$ m/s) and the attenuation, i.e. $Q^{-1}$ where $Q$ is the quality factor, is chosen as $Q_1^L = 250$, $Q_2^L = 50$, $Q_1^T = 150$ and $Q_2^T = 1$ (\cite{anderson1966sound,  toksoz1979attenuation, han1986effects}). 

\begin{figure}[t!]
    \centering
    \begin{subfigure}[b]{0.48\textwidth}
		\includegraphics[width=\textwidth, trim={0 0 0 7},clip]{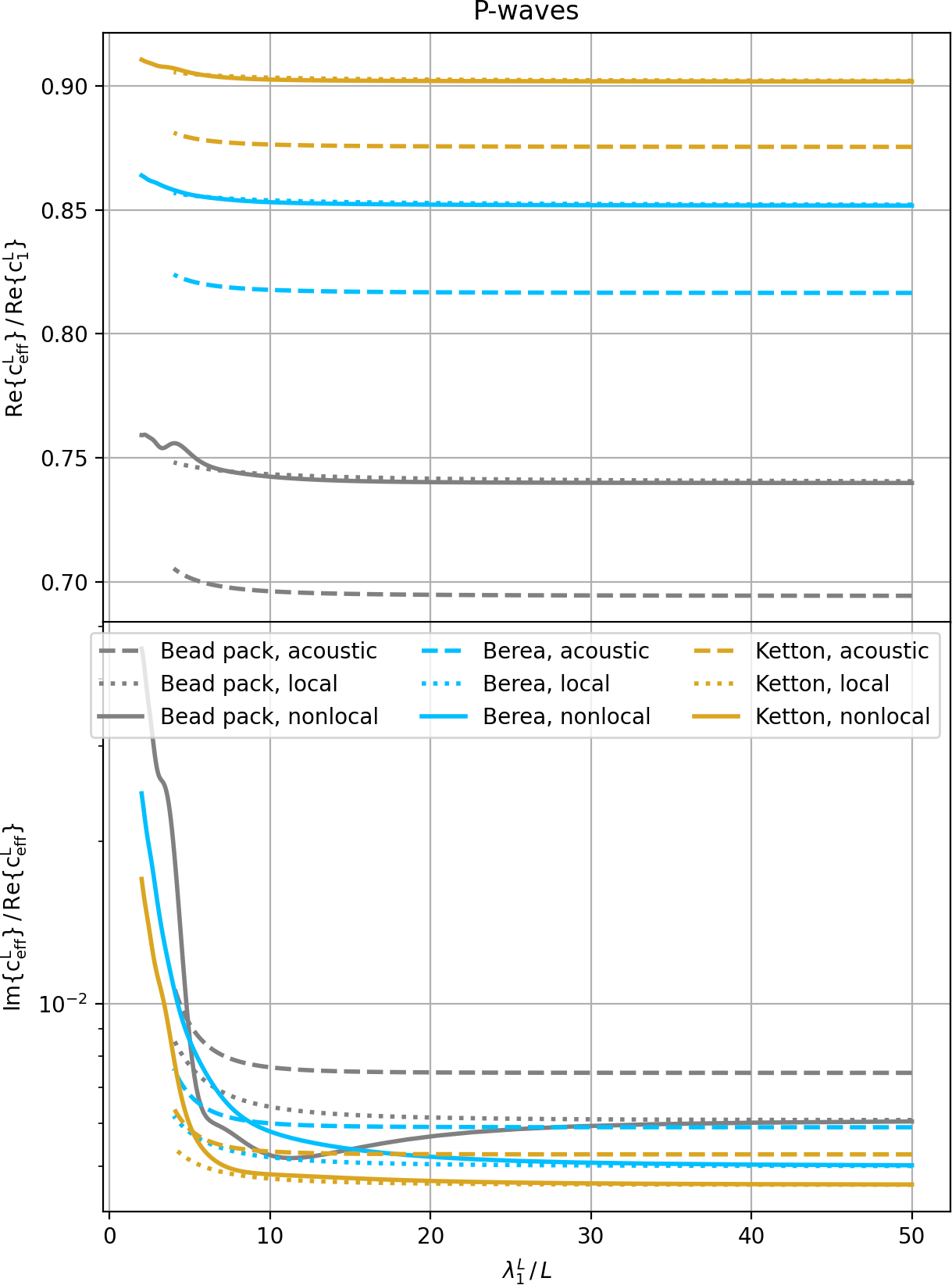}
		\caption{Longitudinal waves\label{fig:MosserEff_Long}}
	\end{subfigure}%
	\quad
	\begin{subfigure}[b]{0.48\textwidth}
		\includegraphics[width=\textwidth, trim={0 0 0 10},clip]{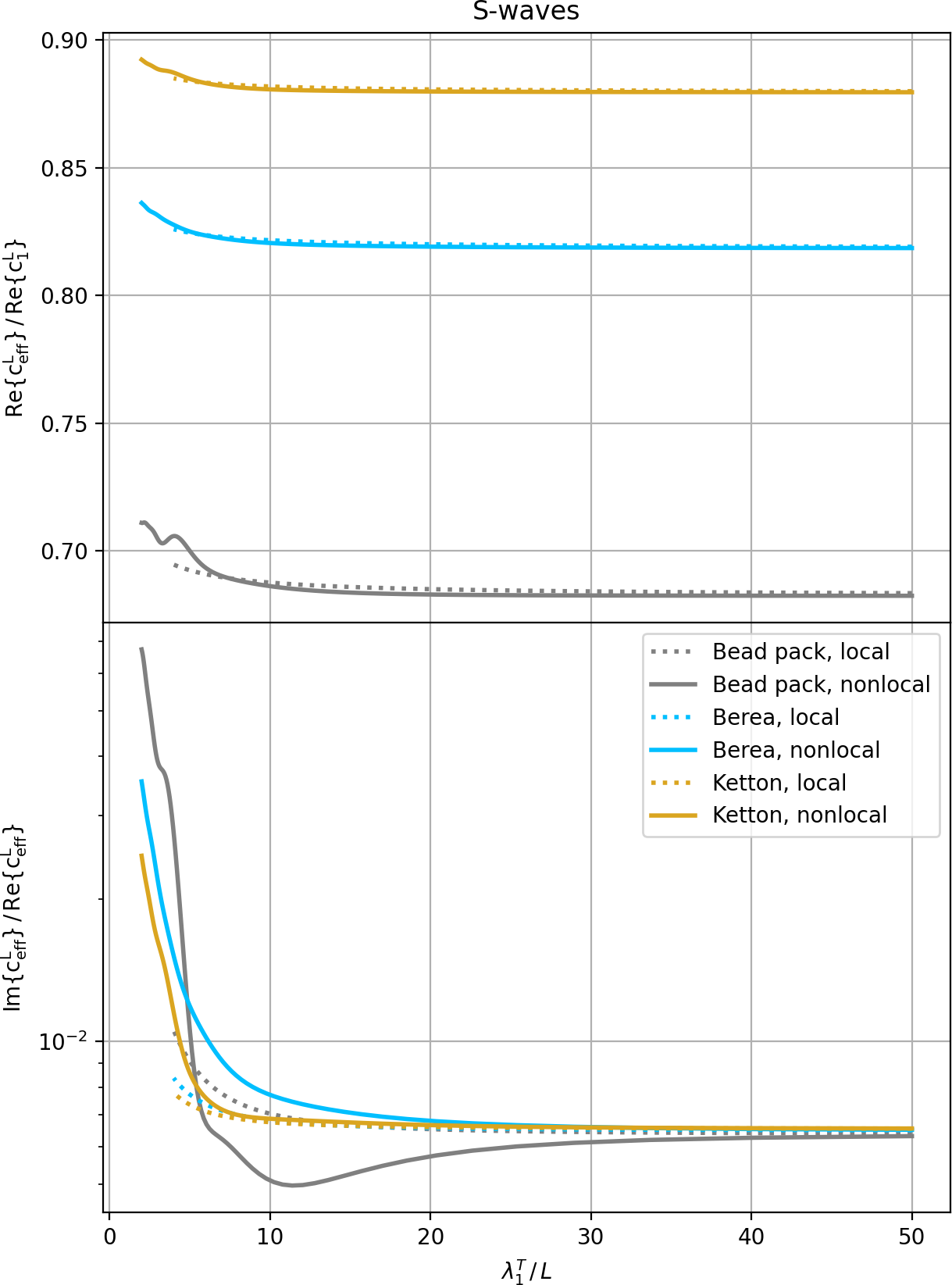}
		\caption{Transverse waves\label{fig:MosserEff_trans}}
	\end{subfigure}
	\caption{\small{Effective (a) longitudinal and (b) transverse wave properties computed from the acoustic (equations (\ref{eqn:ac_ceff}) - (\ref{eqn:A2int})), local elastic (equations (\ref{eqn:elastic_ceff}) - (\ref{eqn:D_2})) and nonlocal elastic (equations (\ref{eqn:elastic_ceff}) - (\ref{eqn:F_nonlocal})) theories. Physical phase quantities used for the samples in figure \ref{fig:Mossersamples} are stated in the main text.}}
	\label{fig:MosserEffprop}
\end{figure}

Figure \ref{fig:MosserEffprop} displays both effective longitudinal and transverse wavespeed and attenuation for the samples used in this study. We extend the frequency range for the nonlocal SCE case from $\lambda / L = 4$ to $\lambda / L = 2$ since to approximately account for coherent scattered waves at the small wavenumbers.  Again, we notice effective wavespeed variation only at small wavelength scales (tied to wave dispersion in coherent scattering) and the effective attenuation increases more strongly at small scales, in agreement with scattering theory. At larger scales attenuation stagnates and approaches the assigned viscous dissipation values. In case of purely elastic media this asymptotic behaviour would disappear and attenuation at long scales would be zero (\cite{kim2020multifunctional, kim2020effective}). Due to the near-zero dependence on frequency, at longer wavelengths, the absolute attenuation values are mainly determined by medium properties and volume fraction. When looking at the attenuation at smaller wavelengths (e.g., $\lambda / L < 10$), we can conclude that there is stronger incoherent scattering in the bead pack sample than in the other samples. This can be caused by several aspects shown in figure \ref{fig:MosserSMD}. For instance, this sample has larger average particle size causing stronger incoherent multiple scattering at the shown scales. More importantly, it appears that the bead pack has the lowest degree of hyperuniformity (figure \ref{fig:Mosserhyper}) - indicating its greater degree of geometric irregularity leads to more pronounced incoherent scattering. Although our samples are all strictly nonhyperuniform, we see relative differences which are likely to have an expression on the effective wave properties. Moreover, the amount of attenuation due to incoherent scattering - next to the difference due to volume fractions - is higher in Berea than in the Ketton sample, which is reflected in their relative hyperuniformity. 

Both the acoustic and local elastic cases have similar trends, however, there are absolute differences between the two; the wavespeed of purely acoustic waves is lower and attenuation higher. This is likely caused by elastic energy transfer - due to sub-wavelength mode conversions - from scattered or incident longitudinal to transverse energy and vice versa, in the case of elastic waves. In contrast, the nonlocal approximation converges to the local theory at large scales - indicating the scales where purely incoherent scattering controls effective wave behaviour. At short scales, it exceeds (in term of absolute magnitudes) both other SCEs, as is expected from the form of correction term for coherent scattering present in the nonlocal SCE. Noteworthy is the observation that the bead pack attenuation has a minimum at intermediate scales. In fact, it seems that the oscillatory behaviour of the sample $S_2$s is captured in the effective properties under the nonlocal approximation, opposite to the local cases, indicating a more unique relation between the data $\boldsymbol{d}$ and model $\boldsymbol{m}$. 

\section{Inverse problem}
\subsection{Bayesian inference}
 A wide variety of approaches is used in imaging (e.g., geophysical, medical) to solve non-linear inverse problems, ranging from deterministic gradient-based least-squares optimization and Gauss-Newton iterative algorithms (e.g. \cite{scales1987tomographic, pratt1998gauss}) to probabilistic approaches aiming to find  a maximum posterior solution by brute force sampling algorithms (e.g. \cite{geyer1992practical}). Instead of providing single best-fit models these methods yield full posterior distributions thus giving uncertainty estimates. Deterministic inversion relies on physical understanding of the forward operator $\boldsymbol{G}$ - which is highly nonlinear in our problem - and often requires linearisation around some chosen prior model. In addition, in most cases it yields but a single solution. Bayesian inference is an alternative approach that provides the posterior probability solution $p(\boldsymbol{m} | \boldsymbol{d})$ obeying Bayes' theorem (\cite{scales1994introductory}): 
 \begin{equation}
 \label{eqn:bayes}
     p(\boldsymbol{m} | \boldsymbol{d}) = \frac{f(\boldsymbol{d}|\boldsymbol{m}) \, \rho(\boldsymbol{m})}{h(\boldsymbol{d})},
 \end{equation}
 which is the likelihood multiplied by the prior beliefs about the model before the data are observed, scaled by the evidence. For a given dataset where the evidence (i.e., data variance) is constant, the posterior distribution directly depends on the prior and likelihood terms. 
 Explicitly calculating the posterior probability using Bayes' theorem is typically difficult. In practice this is only done when the (often naive) assumption of feature independence is applied (\cite{rish2001empirical}). More frequently, the posterior is approximated using sampling methods. Markov Chain Monte Carlo methods like the Metropolis algorithm are a well known example that repeatedly  sample the likelihood term starting from a first model estimate (\cite{tarantola2005inverse}). These methods are however expensive as the forward model needs to be called many times to yield an acceptable posterior distribution including uncertainties. Supervised machine learning (ML) algorithms sample the posterior probability - directly or indirectly - for a given set of $\boldsymbol{d}$ and $\boldsymbol{m}$, i.e. the training data. The posterior solution of the training set therefore completely depends on the prior ($\rho(\boldsymbol{m})$) and on ML model parameters that determine how the in- and output relation is constructed ($\theta$). When these parameters are set satisfactory and the model is trained to map data into model parameters based on the training data, we can insert new data to search for maximum posterior solutions. \\ 
 
 In fact, single ML predictions ($\hat{\boldsymbol{m}} = \hat{\boldsymbol{G}}^{-1} (\boldsymbol{d}$)) are function estimators that map new data into a model and as such approximate the inverse of $\boldsymbol{G}$ in (\ref{eqn:fwd}). Taking the average of our posterior distribution in (\ref{eqn:bayes}) yields this estimator. The misfit between each estimator and actual model is here formulated by the mean squared error (MSE), which can be decomposed into variance and bias (\cite{goodfellow2016deep}):
\begin{align}
\label{eqn:MSE}
    \text{MSE} & = \mathbb{E}[\norm{\hat{\boldsymbol{m}} - \boldsymbol{m}}^2  \\
        & = \sum_{i=0}^{N} \text{Bias}(\hat{m_i}, m_i)^2 + \text{Var}(\hat{m_i}), 
\end{align}
where N is the size of the discrete model vector.
This decomposition will be of practical use when analysing future inferences. Moreover, additional parameters $\theta$ capture the model capacity, high if the training data is optimally fit and lower when the ML model is constrained. This has the direct consequence to the posterior solution that its bias (underfitting) decreases and its variance (overfitting) increases with capacity. Our goal is to find the optimal capacity point where the sum of both contributions is minimal. The advantage of the probabilistic approach is that prediction uncertainties are captured in the posterior. The degree of non-linearity of $\boldsymbol{G}$ on the model parameters determines the shape of the posterior probability, being Gaussian for linear models (if prior distributions are also Gaussian) and deviating from Gaussian with stronger non-linearity (\cite{tarantola2005inverse}).
In the case of Gaussian distributions we can fit an analytical curve to compute standard deviation for uncertainty quantification.

\subsection{Random forests}
Random forests are collections of independent decision trees - each of them being an individual ML model - and as such are deemed as an ensemble training method (\cite{ breiman2001random}). A decision tree is a flowchart-like structure consisting of nodes that split input data samples in the most `efficient’ manner to categorise them, and extends to a specified limited depth or to a depth when no splits are longer possible. Non-linearity therefore arises from a combination of multiple linear splits. For each split the MSE, which is here the distance between the actual target and where it would be after a split, is computed to maximally reduce variance. After the final split a leaf node gives a prediction in the form of either classification or regression where the latter is the case for inversion problems. In general the output is an average over all decision trees, thus providing a single model in terms of expected value. However, when we have access to the predictions of each individual decision tree we can draw posterior distributions that obey Bayes theorem (\ref{eqn:bayes}). 
%Although the posterior probability is then strictly speaking constructed with frequentist statistics, the model follows the principle of Bayesian inference. 
Moreover, we infer posterior beliefs from the prior that the ML model constructed. We will consider an array of size ($N_{samples}, N_{features}$) for the training data where $N_{samples}$ is the amount of data samples and $N_{features}$ is the number of features per sample (here the components of $\boldsymbol{d}$). For each sample we have a corresponding target (here model $\boldsymbol{m}$), being either a scalar or vector, thus giving a total target array of size ($N_{samples}, N_{targets}$). Single decision trees are prone to overfitting - they fit noise - being of low predictive capacity to new data. Bagging (bootstrap aggregating) is therefore applied to reduce variance without raising bias. It randomly selects training samples to construct individual trees allowing multiple sample occurrences (i.e. with replacement), creating a new dataset with size equal to the training set. This is a crucial step, because trees would be correlated when composed from the complete original training set, inhibiting variance reduction. In this way we additionally have access to the generalization error (measure of prediction misfit to unseen data) in the form of out-of-bag error (OOB error), i.e. the mean prediction error of each training sample using only trees that were constructed without that particular sample. The last step that distinguishes the RF from other decision tree algorithms (\cite{breiman1984classification}) is feature bagging. Here the model only considers random feature subsets per split. The rationale is that allowing each tree node to evaluate the entire feature space would lead to many trees making splits based on certain important features, again causing correlation among trees. By default an RF  regression tree considers $N_{features}/3$ per split when feature bagging is applied (\cite{james2013introduction}), however optimal values for this differ per problem and can be found by cross-validation or OOB error evaluation. Similarly we can fine-tune hyperparameters like number of trees, maximum tree depth, the minimum required amount of samples arriving at a node to make a split, the minimum required amount of samples arriving at a leaf node and others in the RandomForestRegressor class of Scikit-learn (\cite{pedregosa2011scikit}). Limiting the tree sizes via the latter three mentioned examples might have the risk to cause a bias increase, whereas a limit on the amount of trees might cause a large variance. In this way RF models accommodate capacity adjustments to find an optimal bias-variance combination. 

\subsection{Inference model setup and feature choice}
In an ideal world we would like to be able to solve for medium contrast and microstructure geometry in the form of $S_2(r)$ - including volume fraction. However, this problem suffers from ill-posedness, since there logically is a trade-off between medium contrast and volume fraction mainly at longer wavelength scales where coherent scattering is negligible (figure \ref{fig:MosserEffprop}). For this reason we will mainly focus on the case where medium contrasts are known a priori, and hence on the retrieval of medium geometry captured in the $S_2$ functions. In one example we instead assume known volume fraction and infer phase contrast and microstructure geometry.

\begin{figure}[t!]
    \centering
    \begin{subfigure}[b]{0.4\textwidth}
		\centering
		\includegraphics[width=\textwidth, trim={0 0 0 0},clip]{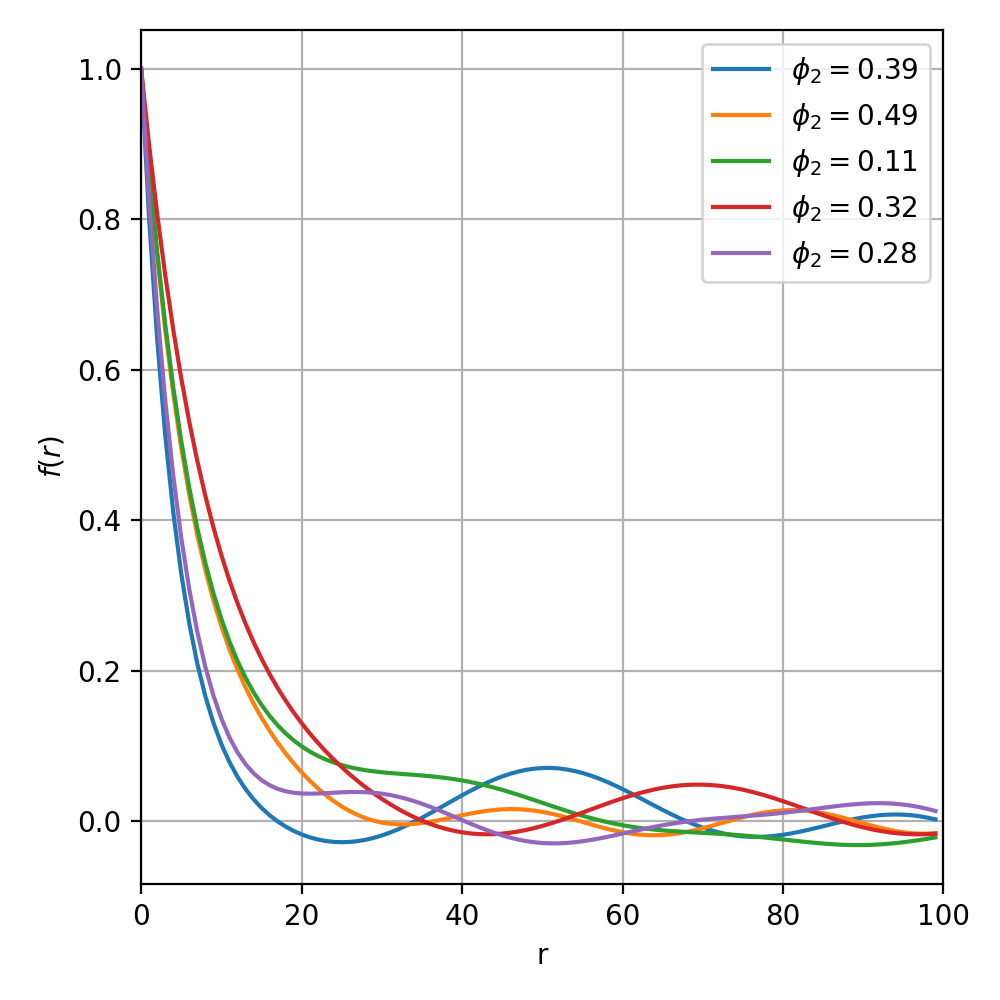}
		\caption{\label{fig:Broadprior}}
	\end{subfigure}%
	\quad
	\begin{subfigure}[b]{0.4\textwidth}
		\centering
		\includegraphics[width=\textwidth, trim={0 0 0 0},clip]{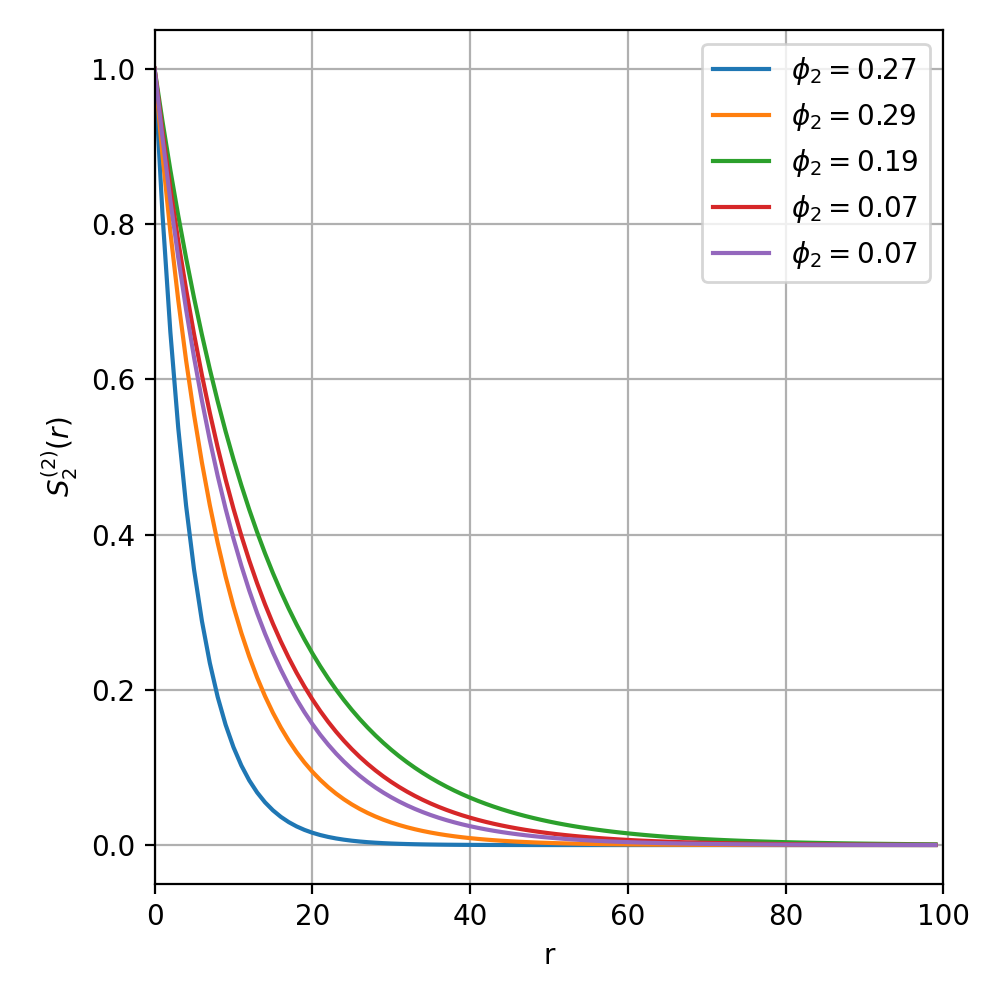}
		\caption{\label{fig:Debyeprior}}
	\end{subfigure}
	\caption{\small{Two different prior $S_2$ distributions, where in (a) we used a combination of Debye (\ref{eqn:f_Debye}) and Damped oscillating (\ref{eqn:f_DO}) basis functions with an additional oscillating perturbation. In (b) we used only Debye functions with different characteristic length scales.}}
	\label{fig:Priors}
\end{figure}

Creating training data for our RF model involves generating either random synthetic $S_2$ models, collecting an array of real image sample data or both, from which we analytically compute effective wave properties. This is directly correlated to our prior $\rho(\boldsymbol{m})$ - limiting variation of included $S_2$ functions restricts the prior and vice versa. Due to relatively large geometry differences of the samples in figures \ref{fig:Mossersamples} and \ref{fig:MosserSMD} we will use a relatively broad prior to include as many $S_2$ curves as possible. To this end we use equation (\ref{eqn:autocov_sum}) to combine scaled autocovariance functions of Debye random media (\ref{eqn:f_Debye}) and damped oscillations (\ref{eqn:f_DO}). This, however, mainly accommodates oscillations at small $r$ values instead of at every radial distance. For that reason, we add an additional perturbation to each scaled autocovariance function, namely summed cosines with random wavenumbers, amplitude and phase (figure \ref{fig:Broadprior}). All variables involved when building the $S_2$ data (volume fraction, characteristic length, etc.) are sampled from uniform distributions. Even though we can not ensure each function's realizability (in terms of yielding microstructures), being only interested in recovering correlation functions of actual samples justifies using potentially non-realizable functions in our prior. In other words, having potentially non-realisable $S_2$ functions in our training serves the purpose of building a wide basis of a priori SMDs and associated wave signatures. 

\begin{figure}[t!]
	\centering
	\begin{subfigure}[t!]{0.48\textwidth}
		\centering
		\includegraphics[width=\textwidth]{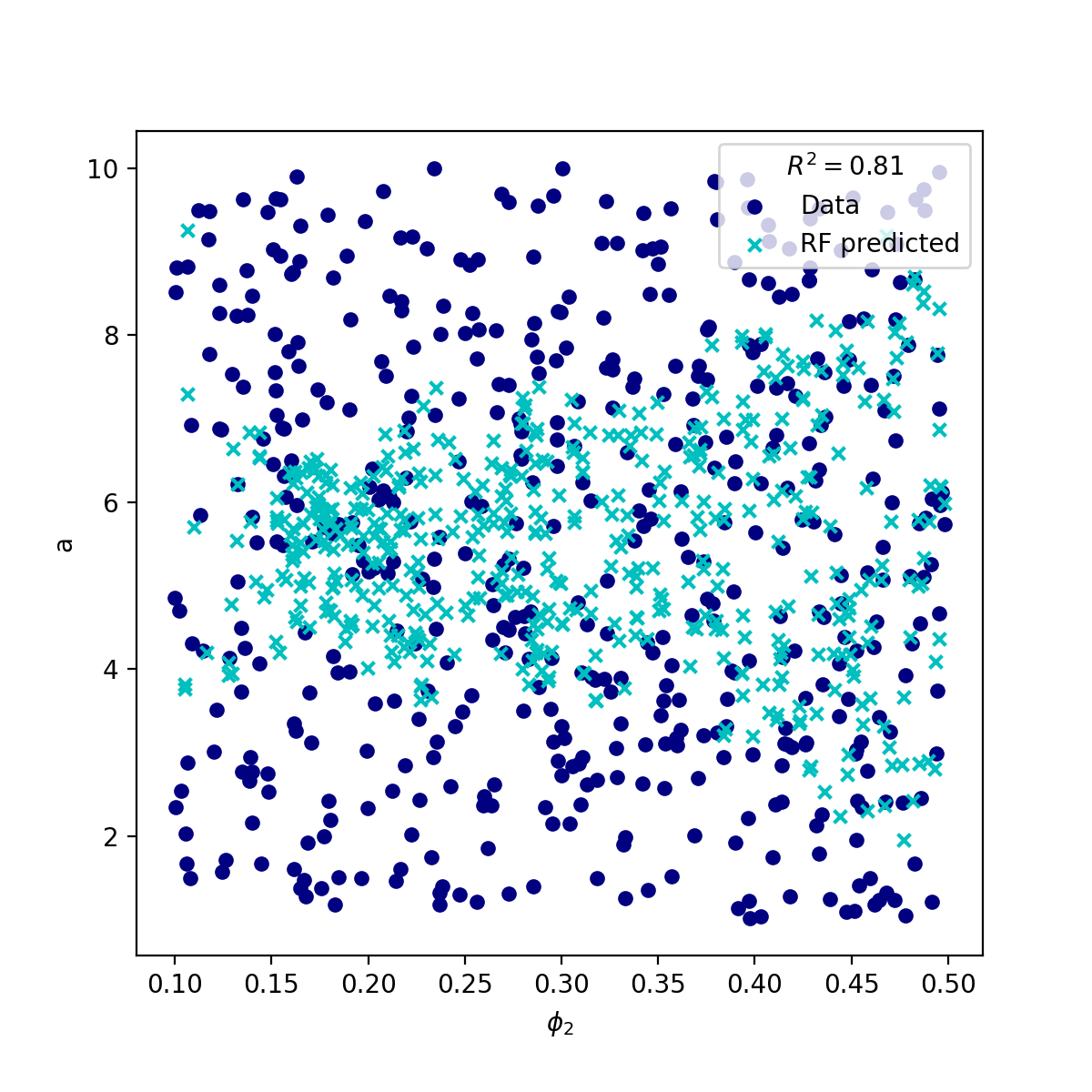}
		\caption{\label{fig:noderivs}}
	\end{subfigure}%
	\quad
	\begin{subfigure}[t!]{0.48\textwidth}
		\centering
		\includegraphics[width=\textwidth]{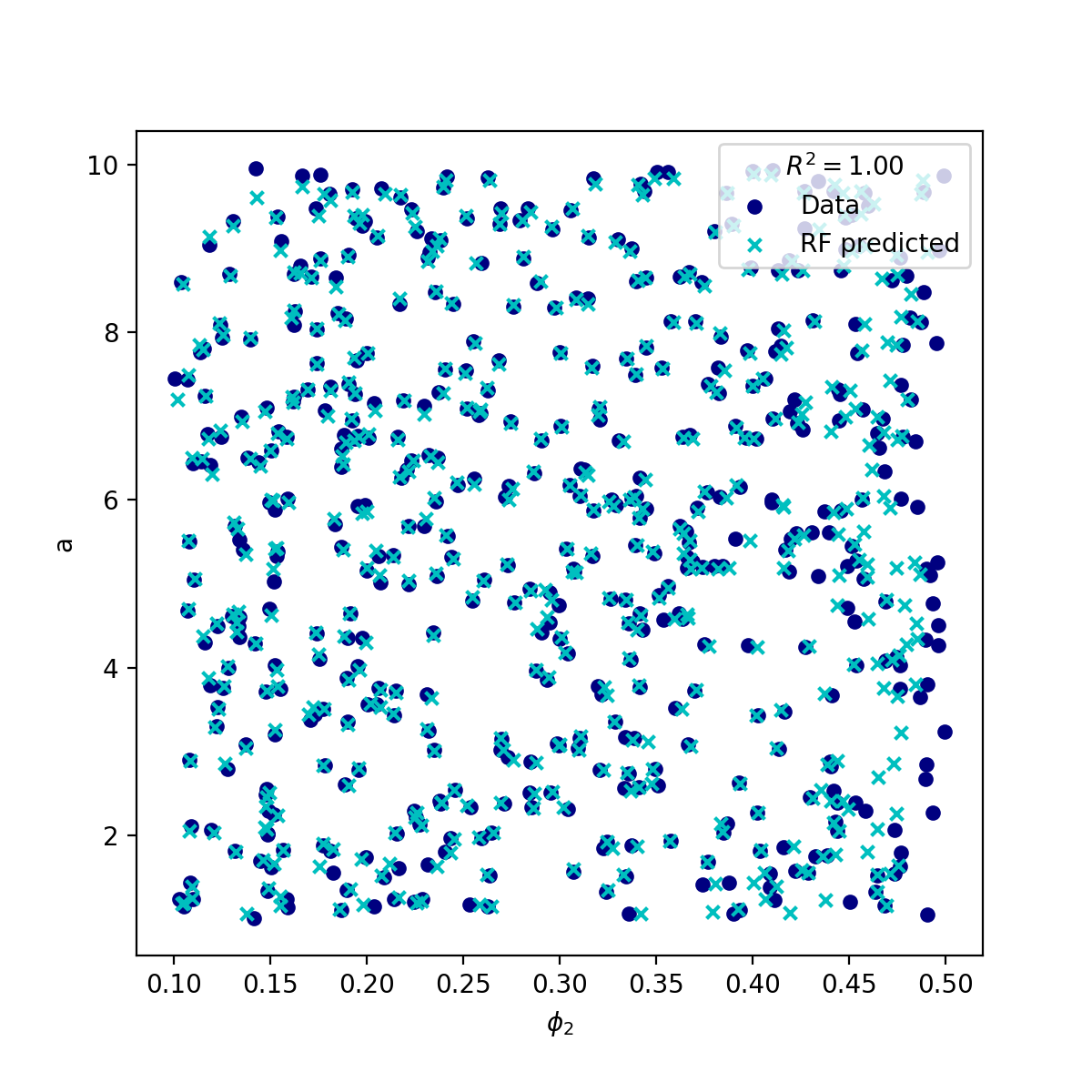}
		\caption{\label{fig:derivs}}
	\end{subfigure}
	\caption{\small Random forest regression of characteristic length scale and volume fraction, where in (\ref{fig:noderivs}) the feature space includes solely the wavespeed and attenuation and in (\ref{fig:derivs}) it additionally includes their first and second derivatives. Here we used $c_1/c_2 = A_2/A_1 = 2$ and $N_{samples} = 5000$.}
	\label{fig:withwithoutderivs}
\end{figure}

\begin{figure}[t!]
    \centering
    \begin{subfigure}[b]{0.48\textwidth}
		\centering
		\includegraphics[width=\textwidth, trim={0 0 0 0},clip]{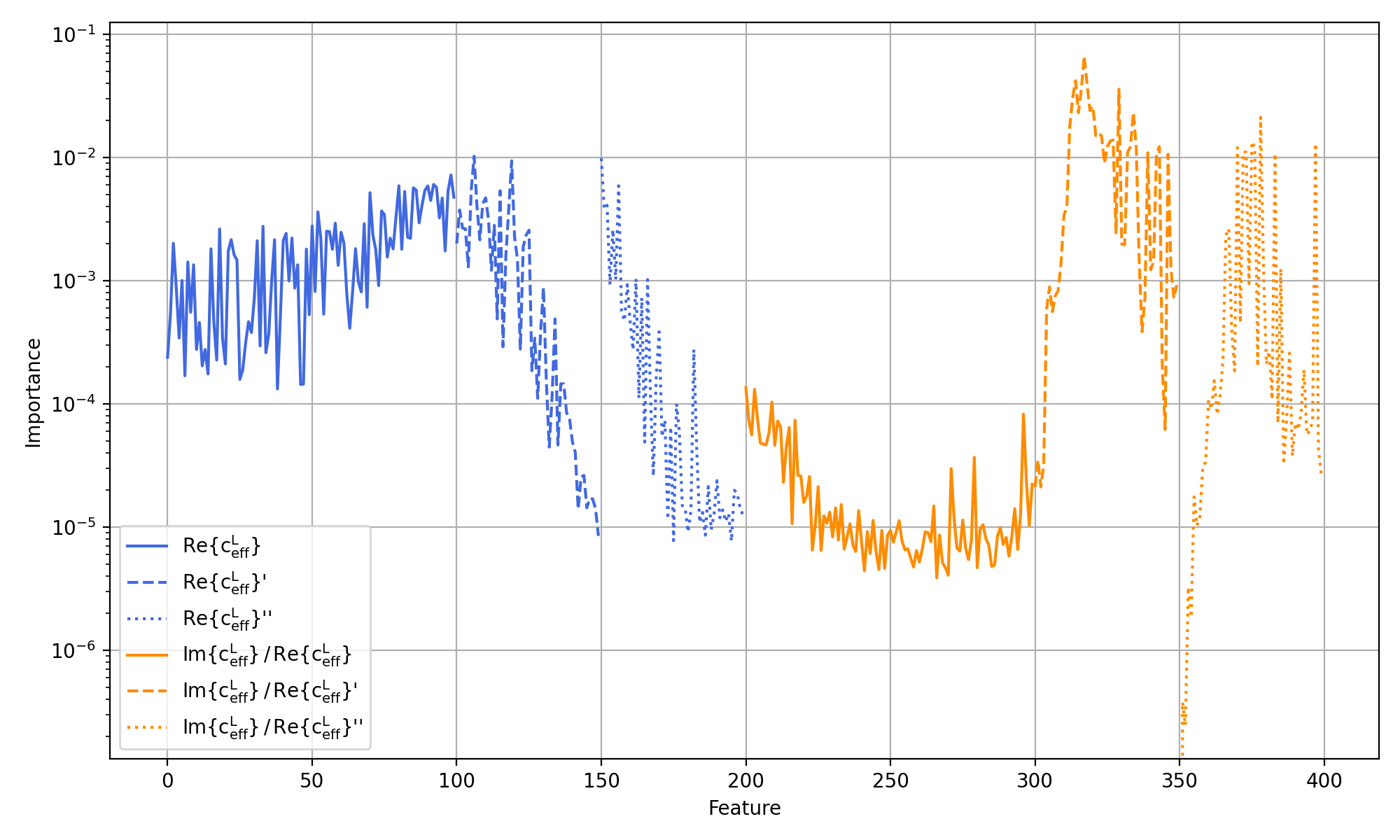}
		\caption{Debye prior acoustic \label{fig:Feature_imp_acoustic_Debye}}
	\end{subfigure}%
	\quad
	\begin{subfigure}[b]{0.48\textwidth}
		\centering
		\includegraphics[width=\textwidth, trim={0 0 0 0},clip]{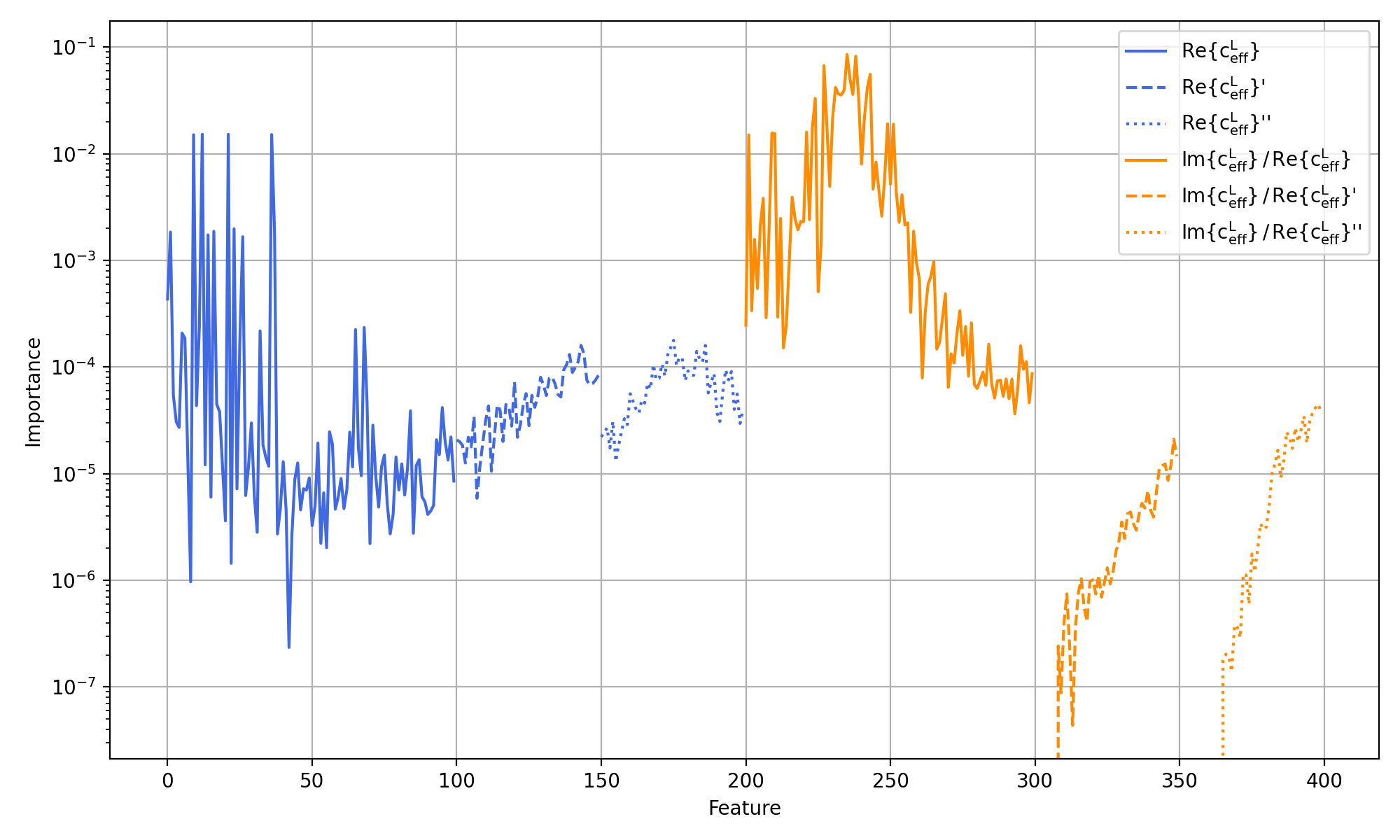}
		\caption{Complex prior acoustic\label{fig:Feature_imp_acoustic}}
	\end{subfigure}
	\begin{subfigure}[b]{0.48\textwidth}
		\centering
		\includegraphics[width=\textwidth, trim={0 0 0 0},clip]{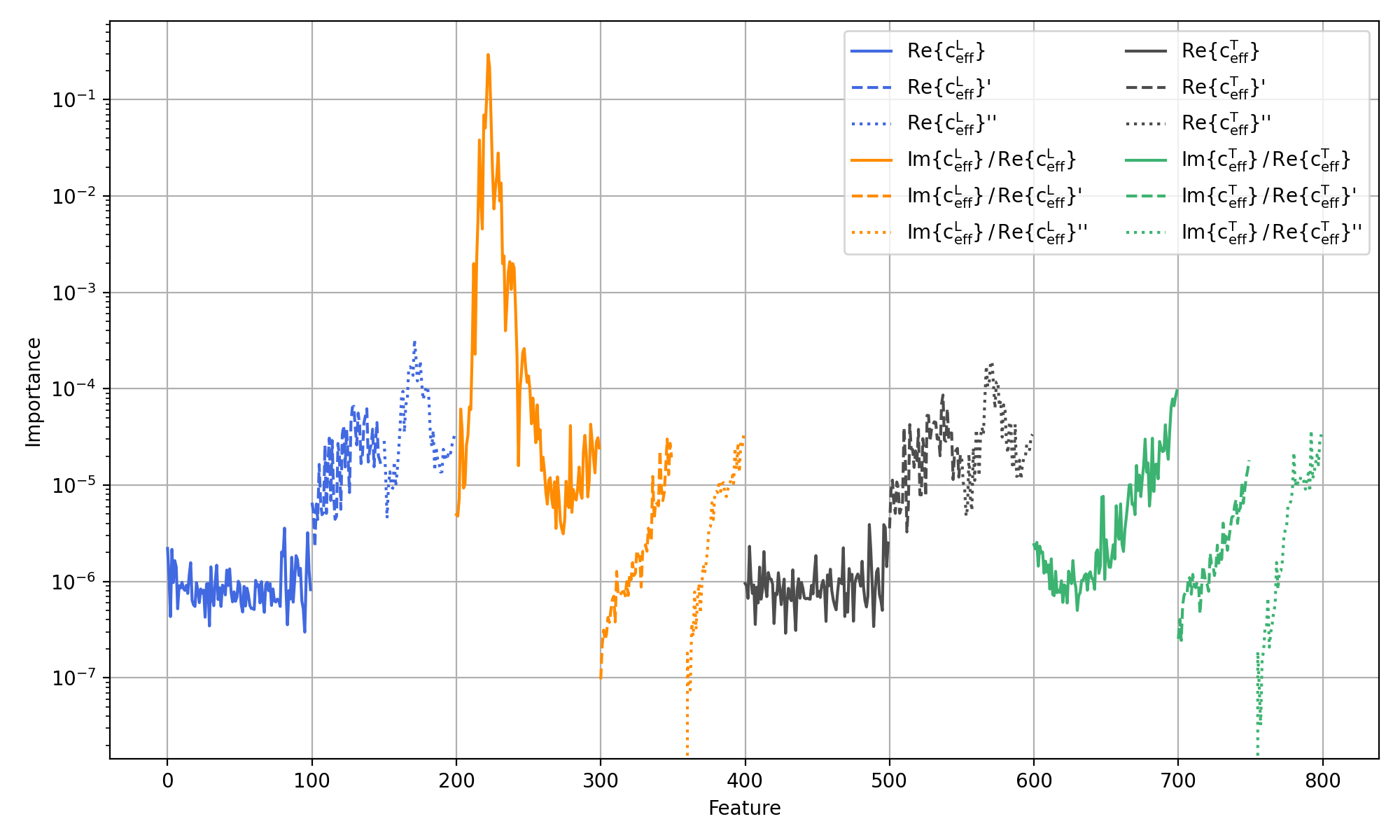}
		\caption{Complex prior local elastic\label{fig:Feature_imp_local}}
	\end{subfigure}%
	\quad
	\begin{subfigure}[b]{0.48\textwidth}
		\centering
		\includegraphics[width=\textwidth, trim={0 0 0 0},clip]{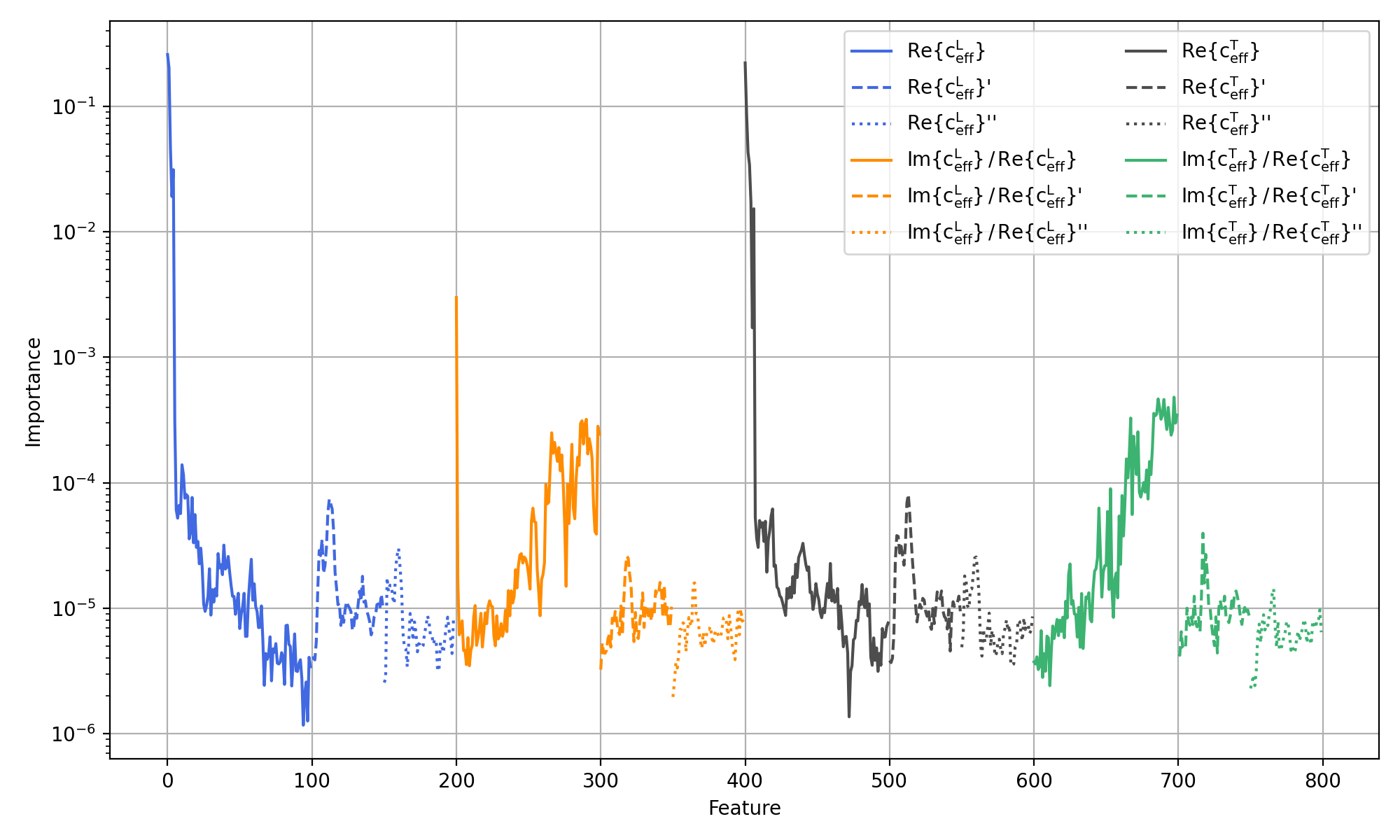}
		\caption{Complex prior nonlocal elastic\label{fig:Feature_imp_nonlocal}}
	\end{subfigure}
	\caption{\small{Feature importances of four different RF models, where for (a) the $S_2$ prior distribution in figure \ref{fig:Debyeprior} is used and for (b)-(d) that in figure \ref{fig:Broadprior}. The dashed and dotted lines represent the first and second derivatives respectively, which are truncated at half their total length as they approach zero in the long-wavelength regime. The importance per feature is calculated as the relative degree of variance reduction after a split based on that particular feature.}}
	\label{fig:Feature_imp_Debye}
\end{figure}

Secondly, after the effective wave properties are computed for each medium representation, we investigate the most efficient input and composition of our RF model. For starters it seems most trivial to include solely the scalar wavespeed and attenuation per wavenumber into the feature space. However it may be beneficial to additionally include the first and second derivative of both quantities. Figure \ref{fig:withwithoutderivs} shows a test RF model to investigate this. To this end we construct a prior with only Debye random media (figure \ref{fig:Debyeprior}), reducing our model space to the volume fraction and characteristic length $a_D$ - which controls the behaviour of $S_2$. Then, we compute their elastic wavespeed and attenuation using the acoustic theory to subsequently set up a simple 50 tree RF model. Clearly, in this case, the addition of both derivatives significantly improves the model's inference accuracy. Although the volume fraction is well resolved without the derivatives (figure \ref{fig:noderivs}), the characteristic length (i.e., microstructure geometry) is not. Looking at figure \ref{fig:MosserEffprop} we can explain this on the basis that volume fraction for a large part determines the absolute wave properties, and medium geometry the curve shapes and curvature at small wavelengths. Therefore adding derivatives to the feature space increases the model sensitivity to medium geometry. However, when we use the prior in figure \ref{fig:Broadprior} and solve for the complete $S_2$s we notice completely different results. Whereas in the previous example the derivatives play the most important role (figure \ref{fig:Feature_imp_acoustic_Debye}), the derivatives for the more complex RF regression not seem to have a significant effect looking at the feature importance (figure \ref{fig:Feature_imp_acoustic}). In this latter case the scalar attenuation values are most important, similarly to the local elastic case where longitudinal attenuation is more important than that of transverse waves (figure \ref{fig:Feature_imp_local}). However, even different behaviour occurs in the nonlocal elastic case; here both longitudinal and transverse wavespeed dominate, with some contribution of their attenuation (figure \ref{fig:Feature_imp_nonlocal}). Therefore it is impossible to determine a generally unique relation between the RF regression problem at hand and feature importance. One of the reasons for this difficulty is that the feature importance is dependent on the individual model parameter combination (volume fraction, characteristic length, or a particular $S_2$ sample). Adding derivatives adds computation time at training, but when feasible is still recommended to expand the feature space from which the RF model can learn. It is possible that additional, hereto unknown, features could further aid in the inference, but further investigation into this is the subject of future study.

\section{Microstructure inference results}
\begin{figure}[t!]
\begin{subfigure}[t!]{0.32\textwidth}
    \includegraphics[width=\linewidth]{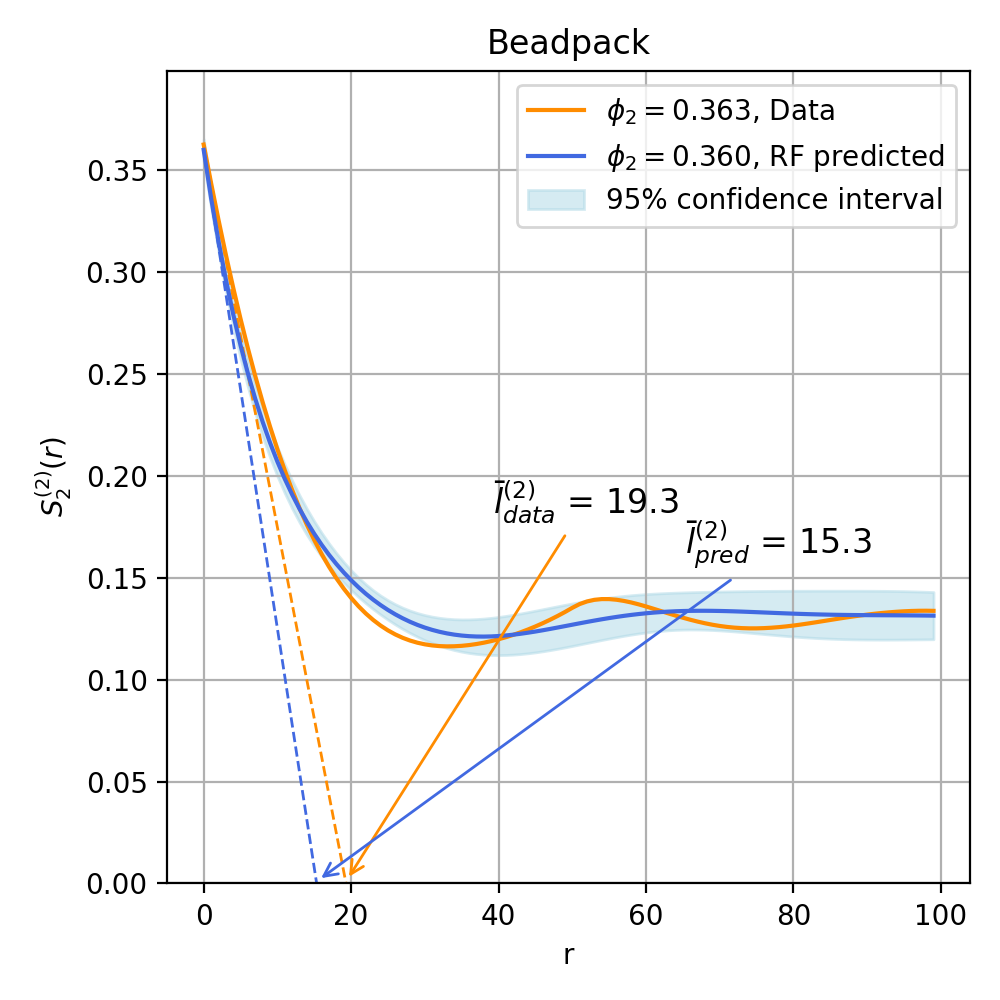}
\caption{MSE = $4.30 \cdot 10^{-5}$ \label{fig:bead_ac}}
\end{subfigure}
\begin{subfigure}[t!]{0.32\textwidth}
  \includegraphics[width=\linewidth]{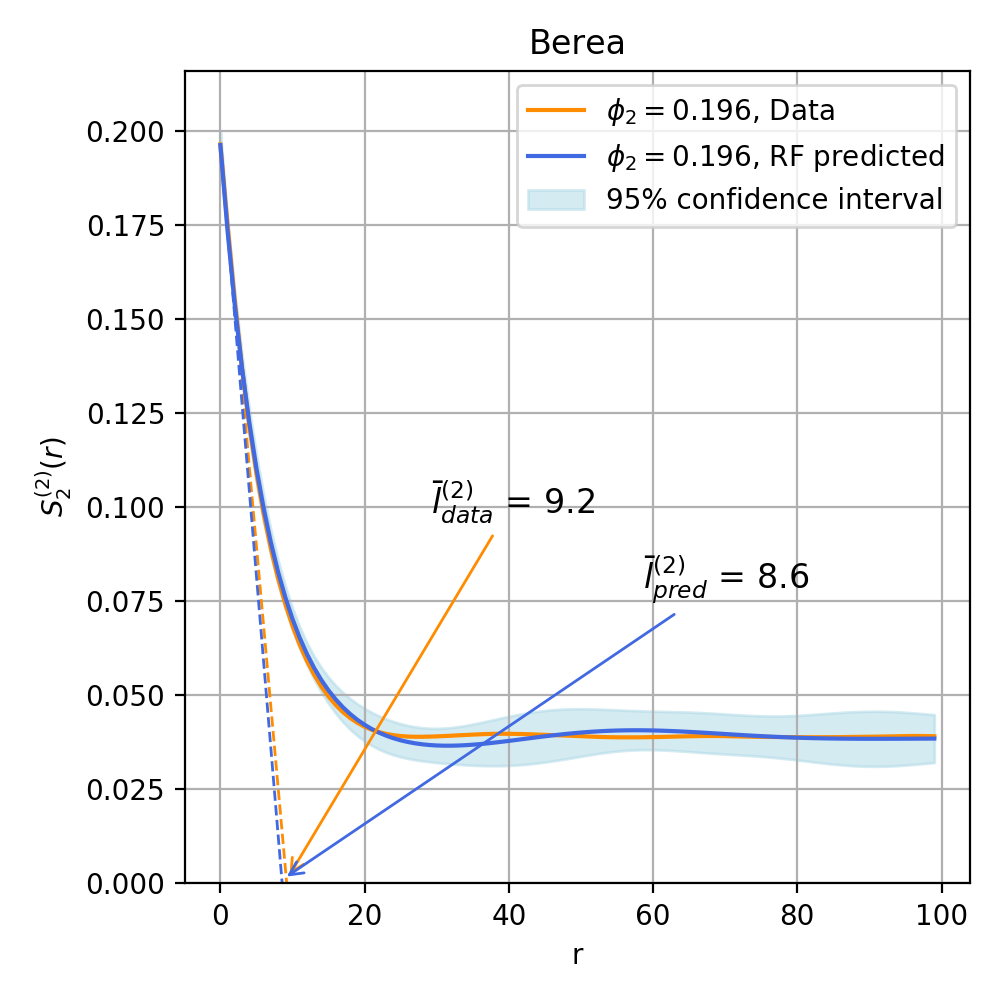}
\caption{MSE = $1.83 \cdot 10^{-6}$ \label{fig:berea_ac}}
\end{subfigure}
\begin{subfigure}[t!]{0.32\textwidth}
    \includegraphics[width=\linewidth]{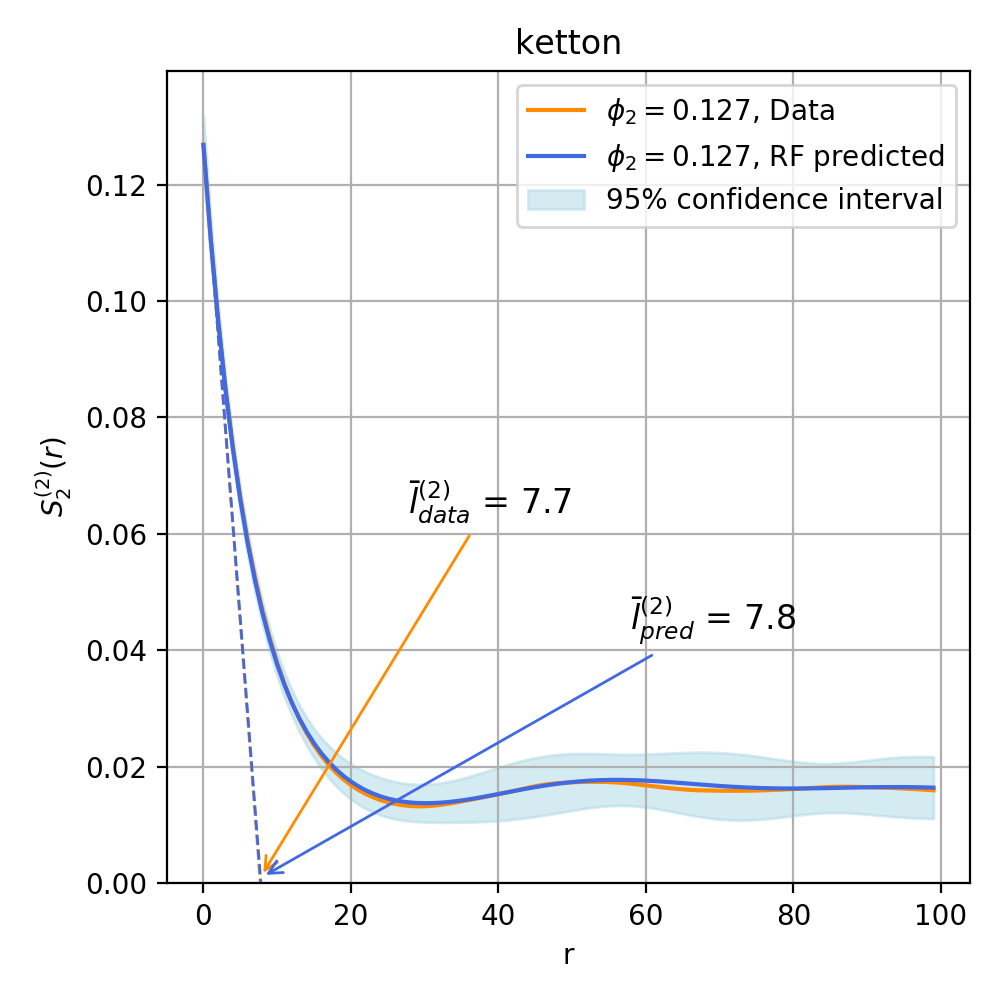}
\caption{MSE = $2.24 \cdot 10^{-7}$\label{fig:ketton_ac}}
\end{subfigure}

\begin{subfigure}[t!]{0.32\textwidth}
    \includegraphics[width=\linewidth]{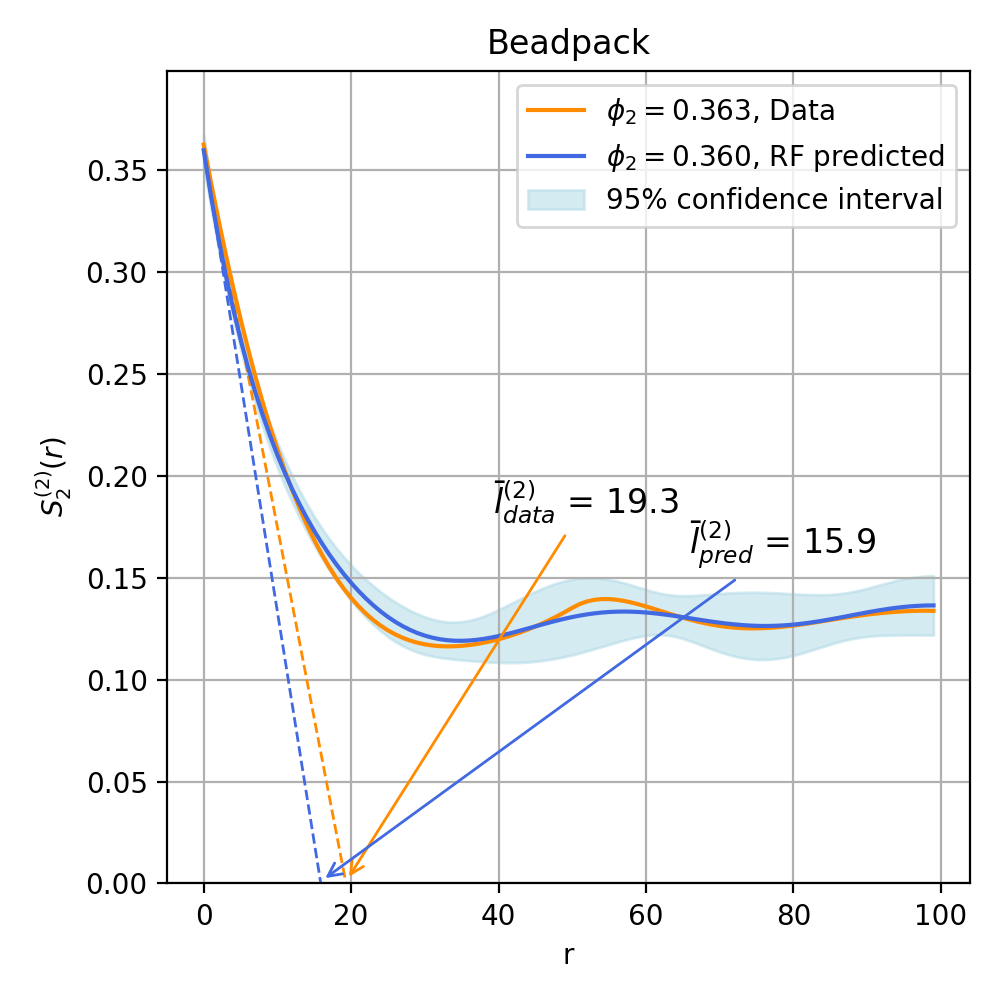}
\caption{MSE = $1.80 \cdot 10^{-5}$ \label{fig:bead_loc}}
\end{subfigure}
\begin{subfigure}[t!]{0.32\textwidth}
  \includegraphics[width=\linewidth]{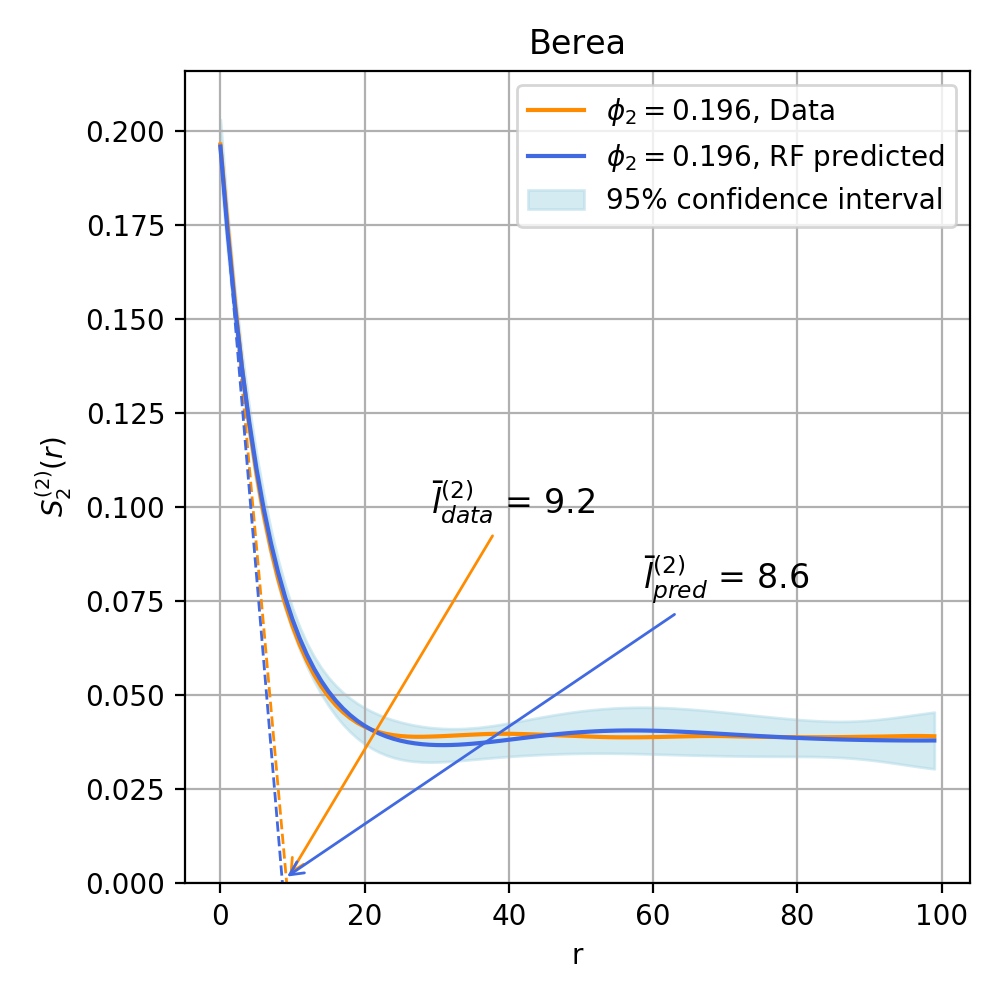}
\caption{MSE = $1.71 \cdot 10^{-6}$ \label{fig:berea_loc}}
\end{subfigure}
\begin{subfigure}[t!]{0.32\textwidth}
    \includegraphics[width=\linewidth]{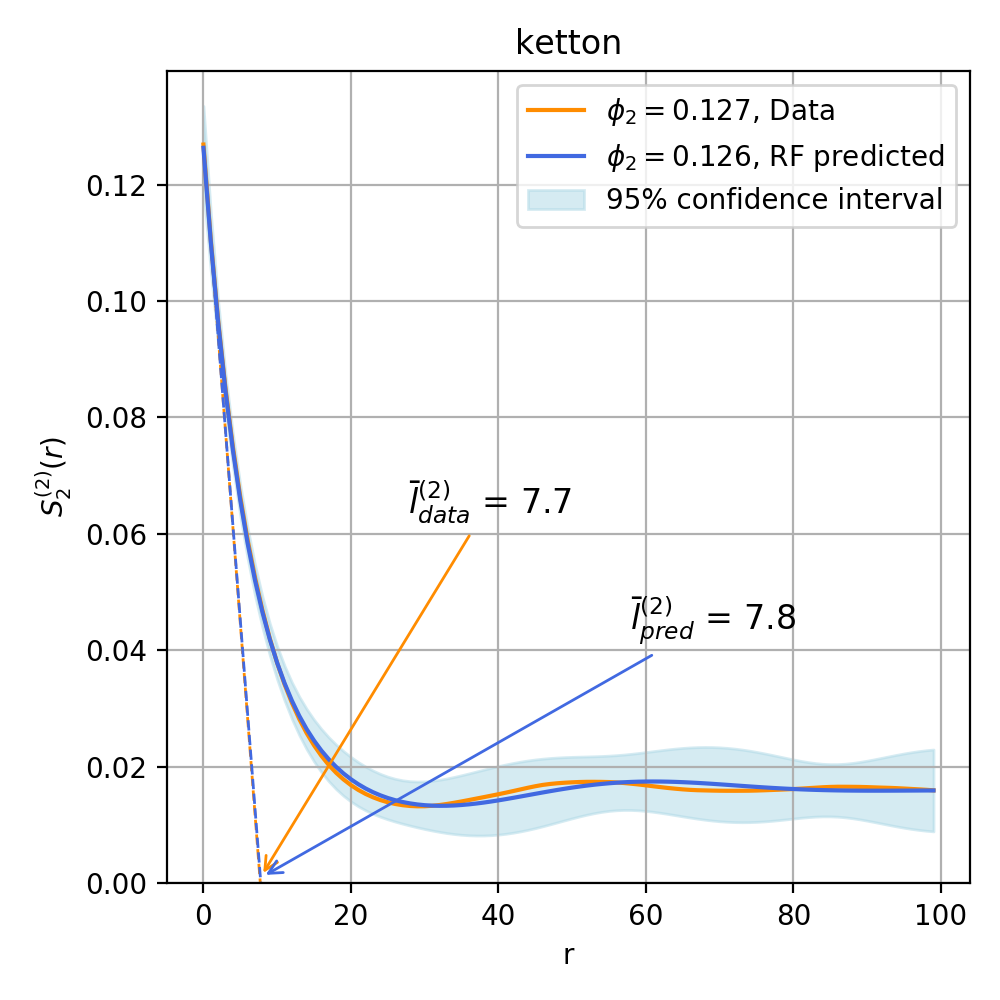}
\caption{MSE = $5.50 \cdot 10^{-7}$\label{fig:ketton_loc}}
\end{subfigure}

\begin{subfigure}[t!]{0.32\textwidth}
    \includegraphics[width=\linewidth]{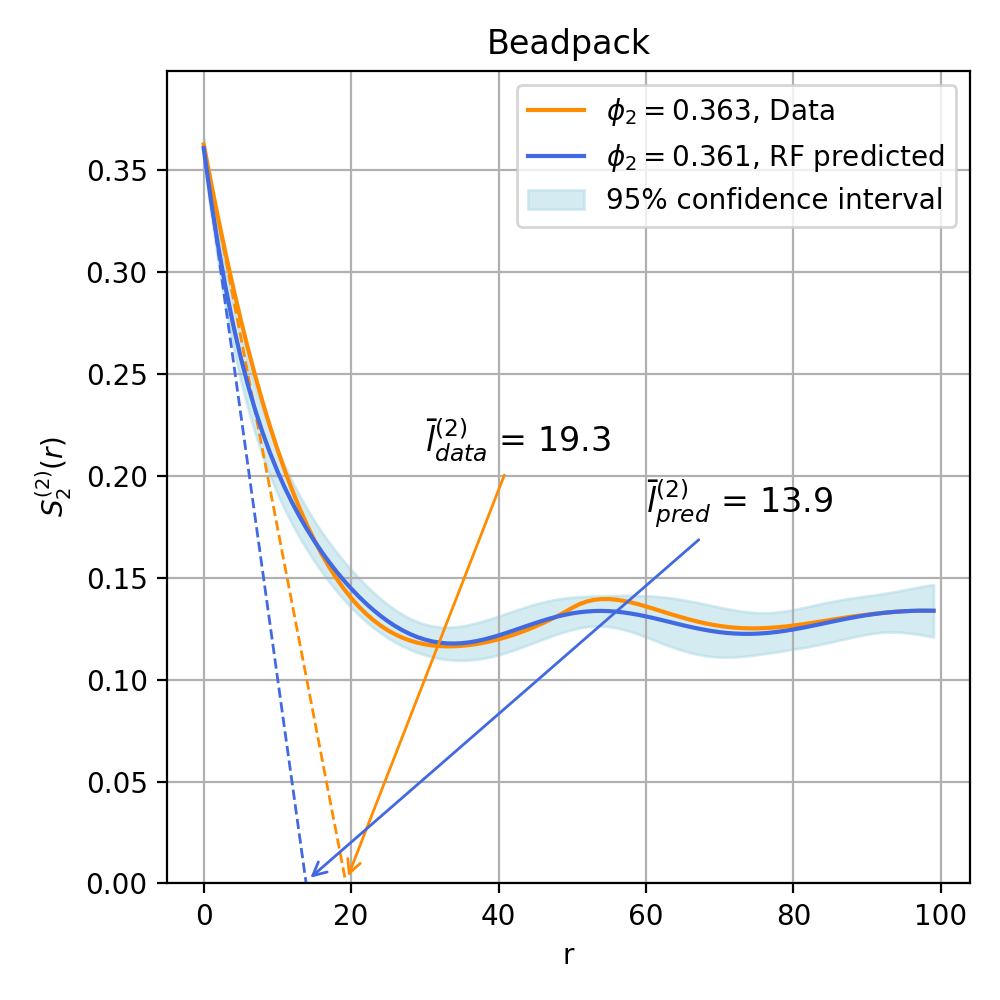}
\caption{MSE = $3.25 \cdot 10^{-5}$ \label{fig:bead_nonloc}}
\end{subfigure}
\begin{subfigure}[t!]{0.32\textwidth}
  \includegraphics[width=\linewidth]{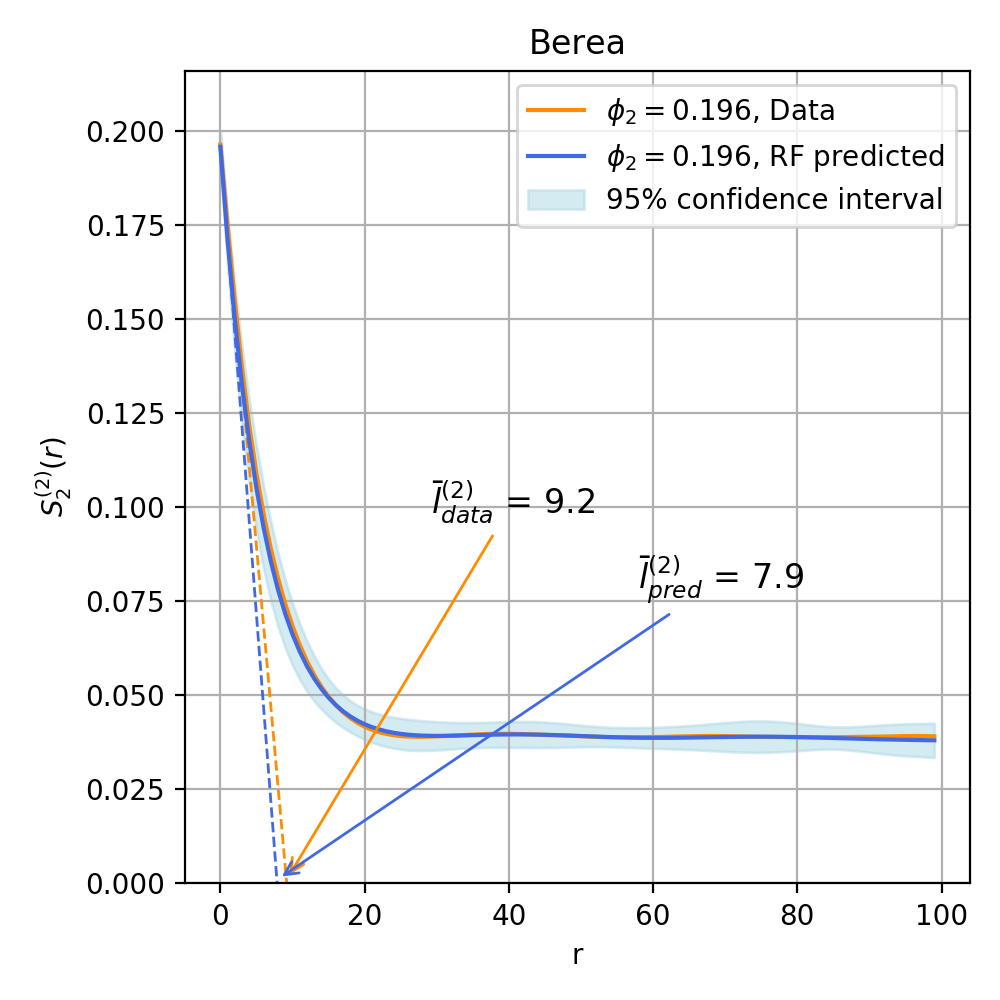}
\caption{MSE = $1.14 \cdot 10^{-6}$ \label{fig:berea_nonloc}}
\end{subfigure}
\begin{subfigure}[t!]{0.32\textwidth}
    \includegraphics[width=\linewidth]{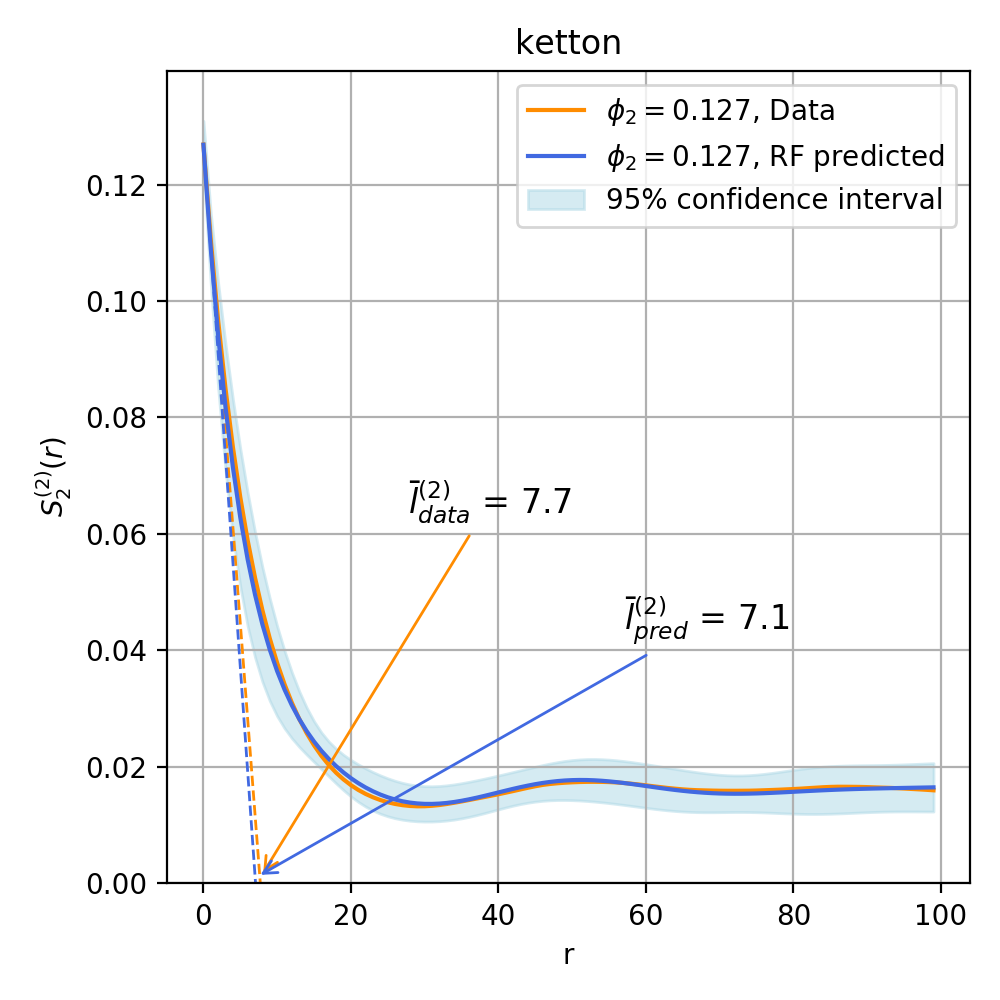}
\caption{MSE = $4.89 \cdot 10^{-7}$ \label{fig:ketton_nonloc}}
\end{subfigure}
\caption{\small RF inversion results, with prediction uncertainty showed as confidence intervals and including MSE (\ref{eqn:MSE}) and average pore size comparison (\ref{eqn:poresize}) for three SCE theories and three samples. The first row shows results from acoustic theory, the second row from local elastic theory, and the third row from the nonlocal elastic approximation. Microstructure phase properties are constant and can be found in the main text.}
\label{fig:invresults}
\end{figure}

For our inference validation, we use physical medium (i.e., individual phase) properties equal to that used in figure \ref{fig:MosserEffprop} - and the prior $S_2$ model distribution as in figure \ref{fig:Broadprior}. We built RF models with $N_{samples} = 1$ million and a hunderd trees for each of the three presented wave theories, and used these models to simultaneously predict the $S_2$s of all three samples in figure \ref{fig:Mossersamples}. The resulting nine predictions are shown in figure \ref{fig:invresults}, where the 95 \% confidence interval width gives an estimate of the posterior probability uncertainty to which a Gaussian is fitted - this is done to simplify the interpretation of our inferred posterior distributions, whose deviations from Gaussian are not extreme. In general, we see that the true model falls within confidence interval for all examples, and the estimator (i.e. the posterior mean) fits the true model well with relatively high certainty - especially for the volume fraction - however there are some differences between using different SCEs. The acoustic case, for one thing, yields a rather parsimonious solution with a significant bias to the oscillating Bead pack curve, which is expected considering that $S_2$ oscillations are not clearly pronounced in the effective acoustic properties (figure \ref{fig:MosserEff_Long}). On the other hand, it predicts small oscillations from the Berea data, which indicates that the chosen prior is biased towards oscillatory curves. The Ketton curve is nearly perfectly predicted. Looking at the local elastic model, we note a significantly better fit to the bead pack curve, telling us that the addition of transverse wave data can aid in solving for highly varying $S_2$ curves like this. Still the volume fraction is not perfectly resolved, although within respectable margins. The Berea sample is similarly solved for and the Ketton curve fits slightly worse. Then for the nonlocal elastic theory, we can conclude that all samples are fitted the best. Only the Beadpack curves deviate slightly, resulting in a different average particle size. Overall, we can thus conclude that we can fit $S_2$ curves from - especially elastic - wave data well. \\

\begin{figure}[t!]
    \centering
    \begin{subfigure}[b]{0.4\textwidth}
		\centering
		\includegraphics[width=\textwidth, trim={0 10 0 0},clip]{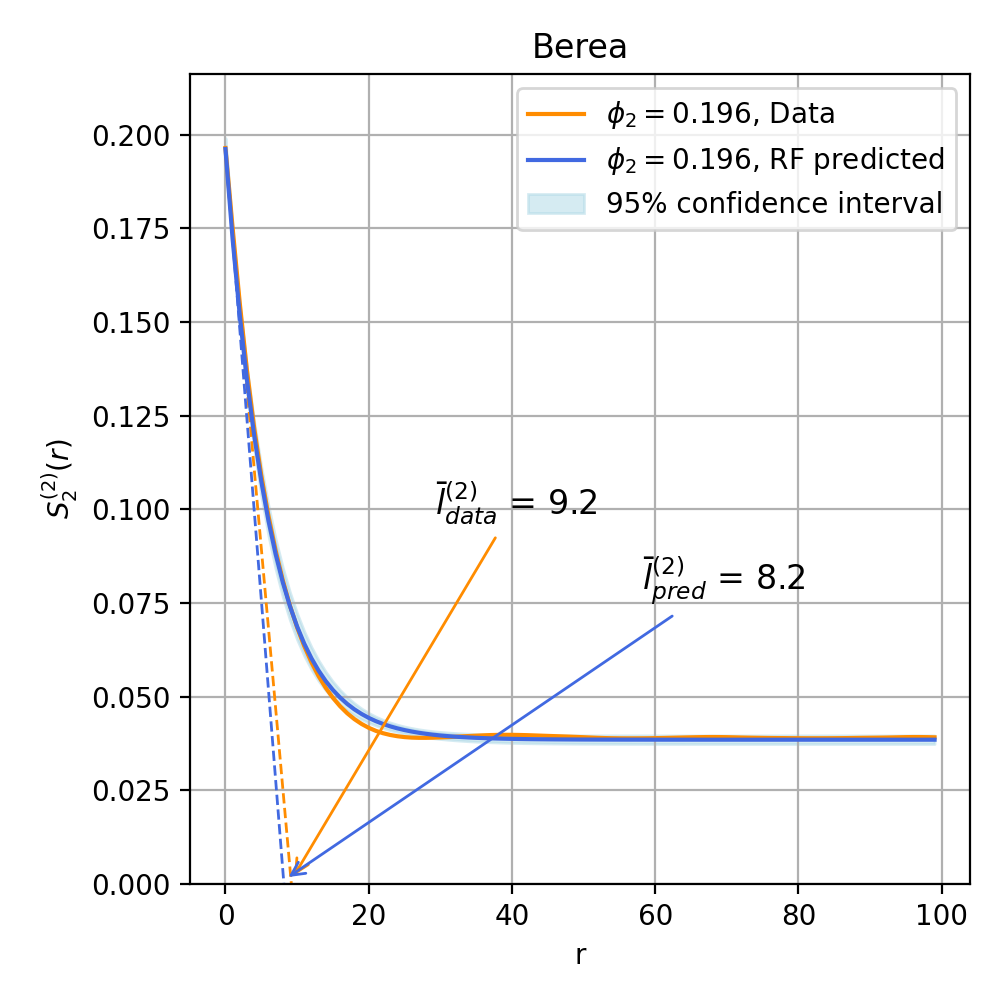}
		\caption{MSE = $1.37 \cdot 10^{-6}$ \label{fig:Berea_Debye}}
	\end{subfigure}%
	\quad
	\begin{subfigure}[b]{0.4\textwidth}
		\centering
		\includegraphics[width=\textwidth, trim={0 10 0 0},clip]{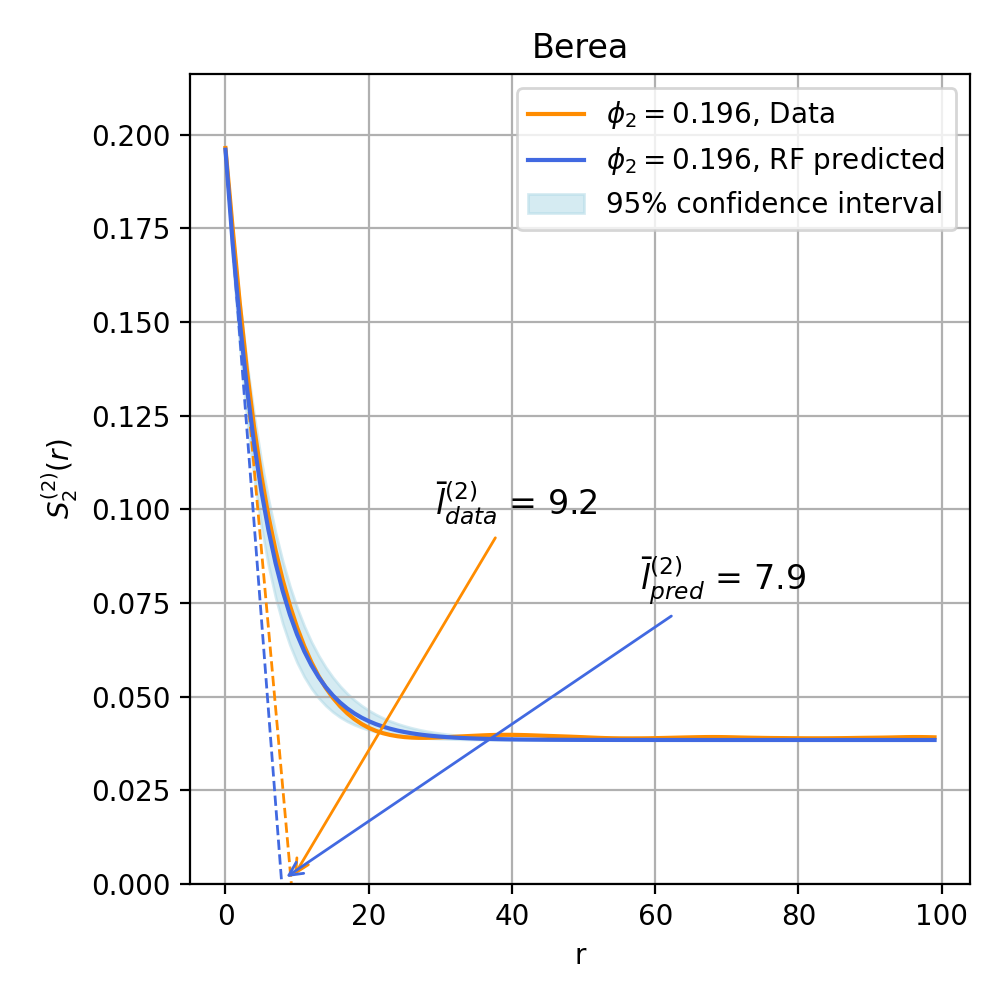}
		\caption{MSE = $2.00 \cdot 10^{-6}$ \label{fig:Berea_fixedphi}}
	\end{subfigure}
	\caption{\small{Two additional acoustic inversions of Berea data using different a priori constraints. In (a) we employ an inversion similar to figure \ref{fig:invresults}, but with the $S_2$ prior distribution shown in figure \ref{fig:Debyeprior}. In (b) we assume a fixed volume fraction and inferred both wavespeed and attenuation contrast and the scaled autocovariance function using the distribution as in (a).}}
	\label{fig:Berea_varyingprior}
\end{figure}

Secondly, we conducted two additional inference test with different prior knowledge. The first consists of assuming that the target media can be represented as a Debye random medium, which for the $S_2$ means that oscillations are negligible leaving a monotonically decaying curve. In that case, the prior can be constructed as in figure \ref{fig:Debyeprior} and the degrees of freedom are reduced enormously; again the contrast is assumed to be known a priori. From our samples the Berea sample is closest to a Debye random medium and hence is the one we use here. Independent of which wave SCE theory is used, we obtain figure \ref{fig:Berea_Debye} using $N_{samples} = 50000$, showing a good fit with much lower uncertainty than in figure \ref{fig:invresults}. The small deviation at $r \approx 20$ is due to the fact that the Berea sample can not perfectly be represented by a Debye random medium. This result shows that with a priori knowledge of the basis function we can accurately resolve for the two-point correlation function - greatly reducing model variance as a result of a more restricted prior space, at the cost of potentially introduced great bias. 

Up until now we assumed fixed physical medium properties per phase. These very quantities however have always been inference targets themselves, as they indicate, e.g., whether pores are filled by either gas, water, oil or otherwise. Therefore, we will look at constraining both contrast in longitudinal wavespeed and attenuation by similarly inverting acoustic wave properties but with fixed volume fraction (porosity) which is assumed to be known a priori. In addition, we assume the basis function is known a priori as in the previous case to minimize degrees of freedom (i.e., restricting the space of priors). Next to the $S_2$, we solve for both contrasts, where the targets values are those used in figure \ref{fig:MosserEffprop}. The reason we use the acoustic theory is that the contrast in longitudinal and transverse wave properties differ, adding two more degrees of freedom. Also, the contrast in transverse wavespeed is often much larger and difficult to benchmark. The result is shown in figure \ref{fig:Berea_fixedphi}, where we again see a close fit to the Berea sample, with slightly more uncertainty at small distances caused by uncertainties in the contrast prediction. The contrasts are predicted accurately: the MSE for wavespeed ($c_1^L / c_2^L = 3$) and attenuation ($Q_1^L / Q_2^L = 5$) are $6.2 \cdot 10^{-5}$ and $5.7 \cdot 10^{-4}$, respectively. 

\section{Discussion}

In our results section we show that we can accurately infer two-point correlation functions from effective wave data with known a priori medium phase properties, especially when relying on the nonlocal elastic wave SCE. Here, we discuss how closely the data is fitted, how sensitive the RF model is to hyperparameter changes, and limitations.

\subsection{Data fit and feature importance}

\begin{figure}[t!]
    \centering
    \begin{subfigure}[b]{0.475\textwidth}
		\includegraphics[width=\textwidth, trim={0 0 0 0},clip]{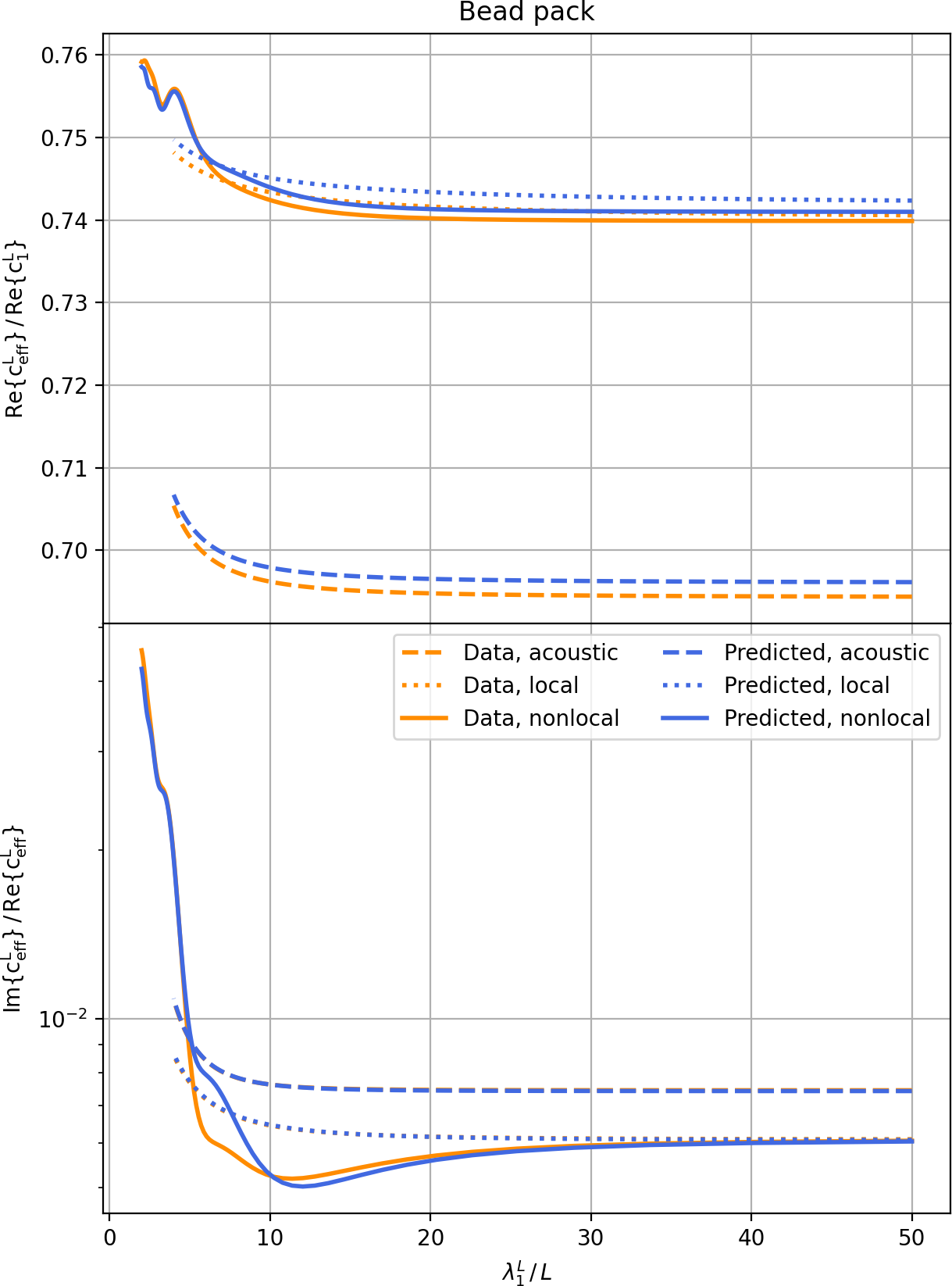}
		\caption{Longitudinal waves \label{fig:Bead_datafit_P}}
	\end{subfigure}%
	\quad
	\begin{subfigure}[b]{0.48\textwidth}
		\includegraphics[width=\textwidth, trim={0 0 0 0},clip]{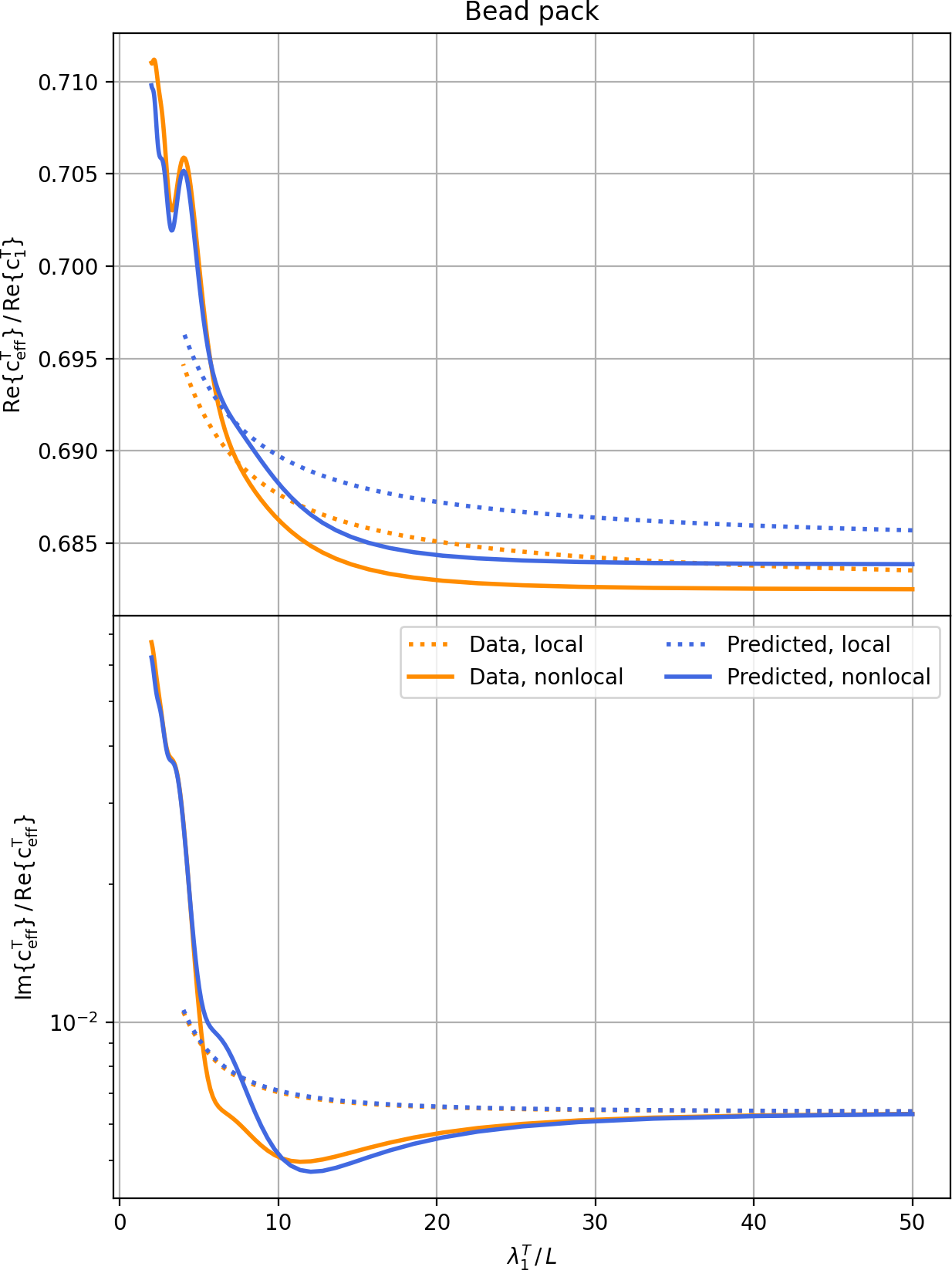}
		\caption{Transverse waves\label{fig:Bead_datafit_S}}
	\end{subfigure}
	\caption{\small{Longitudinal (a) and Transverse (b) data fit of the $S_2$ models in figure \ref{fig:invresults} for the bead pack sample.}}
	\label{fig:Beadpack_datafit}
\end{figure}

In figure \ref{fig:invresults} we observe different fits to the Bead pack $S_2$ depending on the choice of SCE, which are then reflected in the predicted and true models in figure \ref{fig:Beadpack_datafit}. Interestingly, the attenuation data is systematically fitted better than  wavespeed. This makes sense, considering at the feature importances for the local acoustic and elastic inversions (figures \ref{fig:Feature_imp_acoustic} and \ref{fig:Feature_imp_local}), where predictions are predominantly controlled by the longitudinal wave attenuation. For the nonlocal elastic inversion, however,  wavespeeds are significantly more important (figure \ref{fig:Feature_imp_nonlocal}) which as such are fitted closely at small wavelengths where a pronounced oscillation is present - indicating that wave dispersion at the near-coherent scattering regime is an important factor in infering microstructure information. Also, the attenuation is relatively well fitted, with some exceptions caused by the difference in scaled autocovariance visible in figure \ref{fig:invresults}. Most noticeable here however, is the fact that the wavespeeds at longer wavelengths are inferred with relatively larger errors, highlighting the small imperfections on predicted volume fraction. Most likely, this is due to limitations of our chosen prior $S_2$ distribution - but this requires futher investigation. Moreover, it seems that both the volume fraction and scaled autocovariance curve of the bead pack sample are not simultaneously occuring in our prior, and that fitting the latter is prioritised. Indeed, volume fraction influences effective wavespeeds more strongly relatively speaking (figure \ref{fig:MosserEffprop}), which we observe in figure \ref{fig:Beadpack_datafit}. Transverse wave data are fitted likewise, although play less important role in the inversion of local elastic data. The Berea and Ketton data are shown in Appendix \ref{sec:AppC}, showing similar behaviour, but in general are better fitted. 

\subsection{RF performance}

\begin{figure}[t!]
\begin{subfigure}[t!]{0.3\textwidth}
    \includegraphics[width=\linewidth]{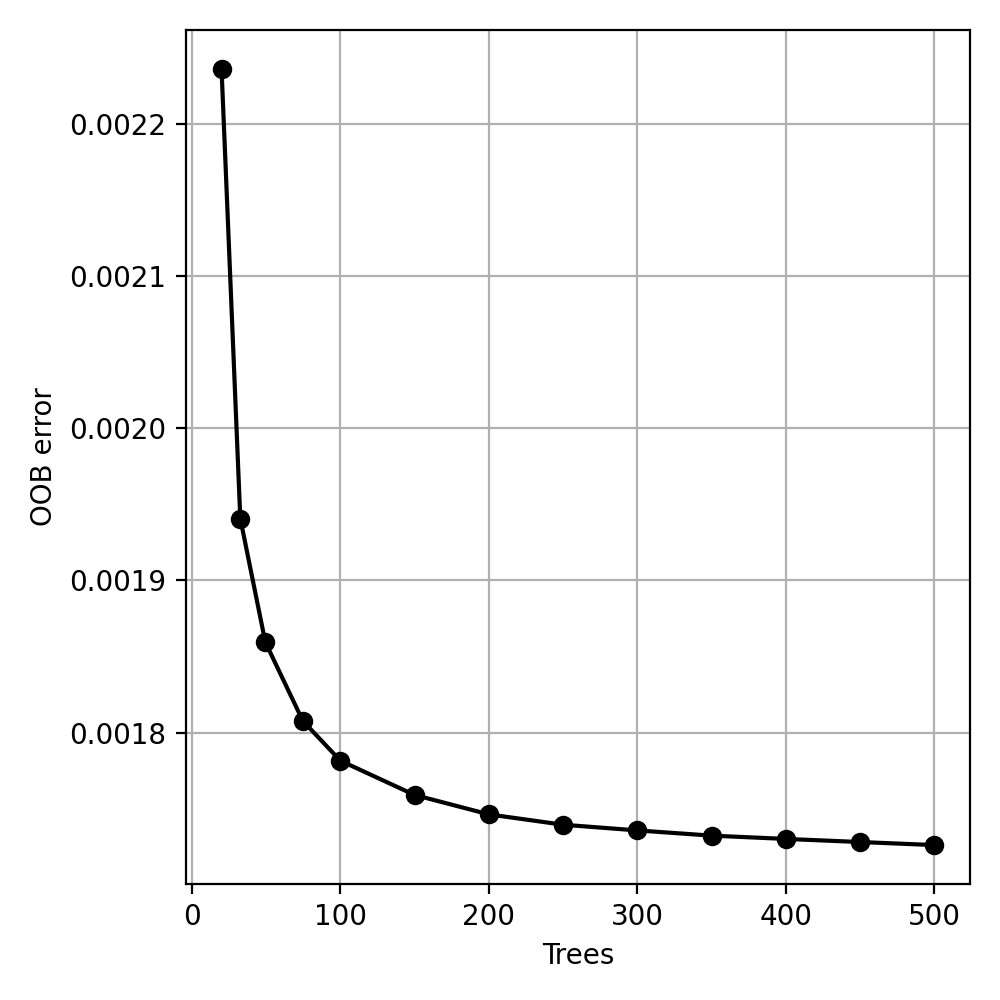}
\caption{Trees \label{fig:OOB_trees}}
\end{subfigure}\hfill
\begin{subfigure}[t!]{0.3\textwidth}
  \includegraphics[width=\linewidth]{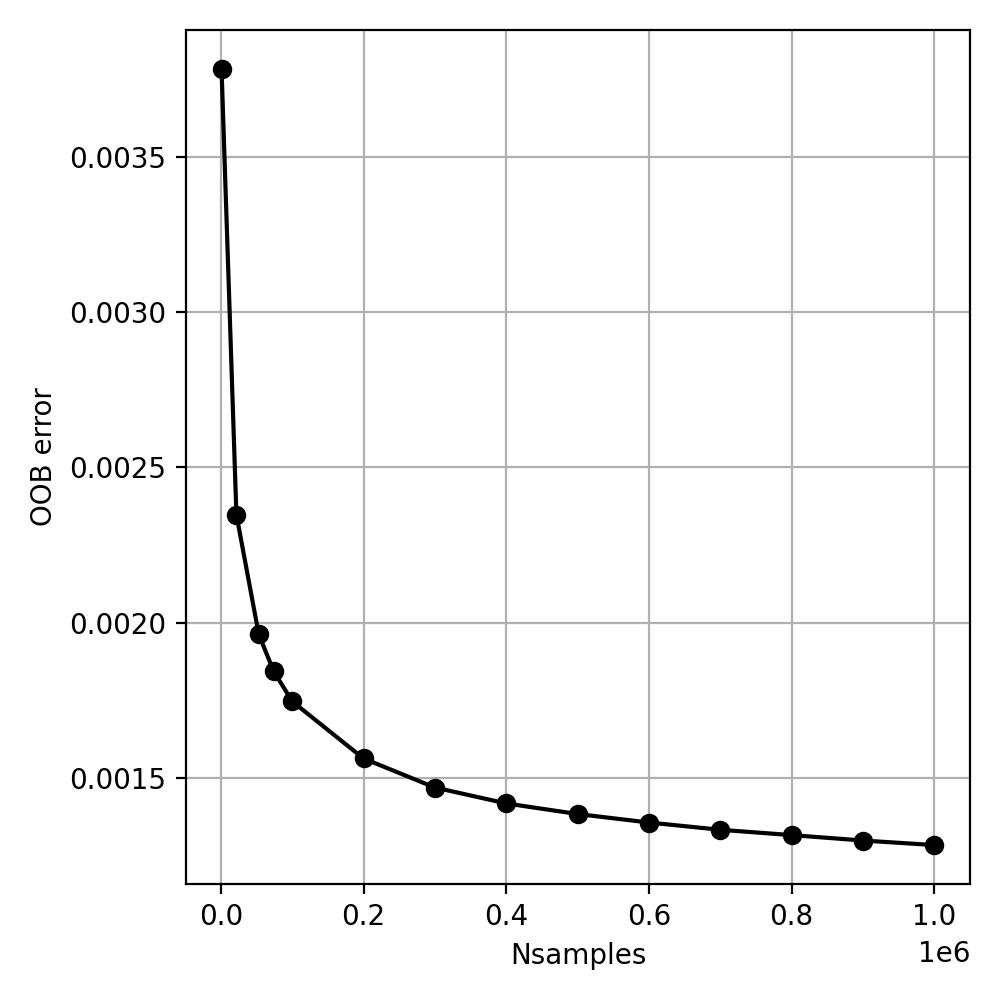}
\caption{Training samples\label{fig:OOB_Nsamples}}
\end{subfigure}\hfill
\begin{subfigure}[t!]{0.3\textwidth}
    \includegraphics[width=\linewidth]{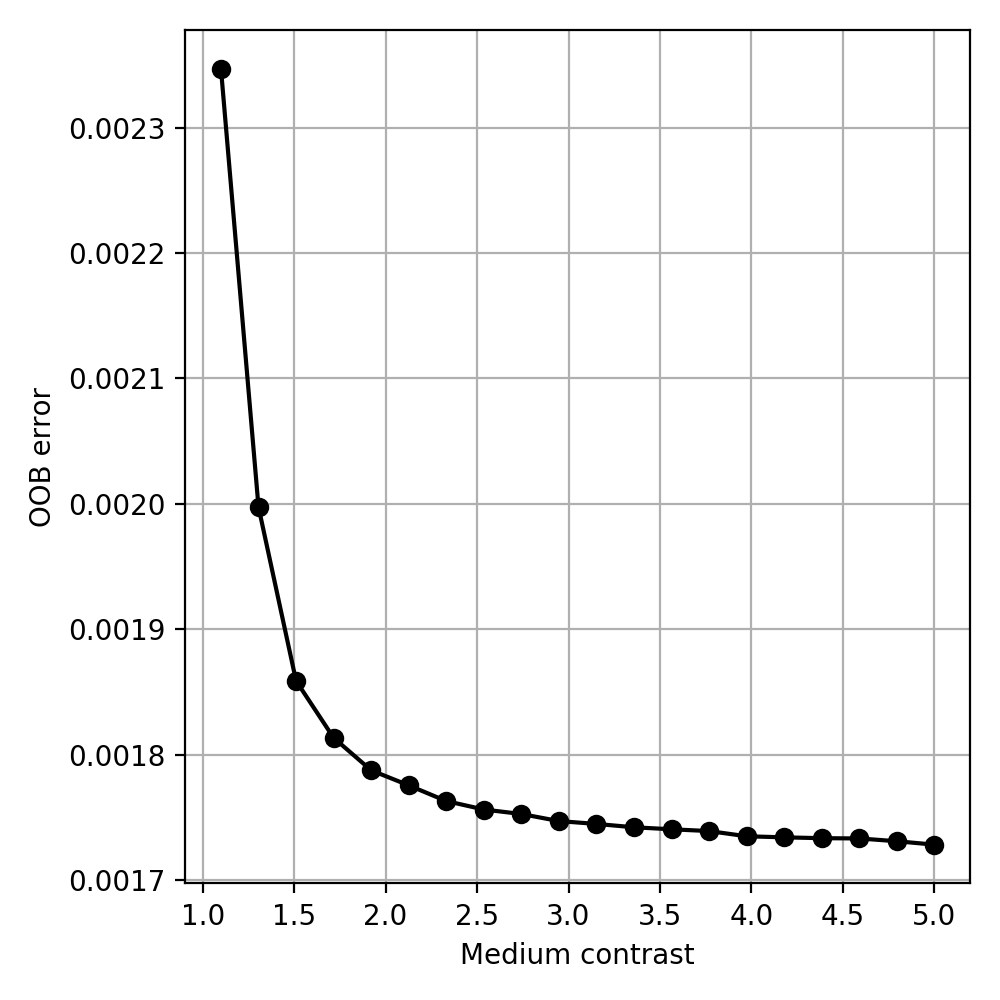}
\caption{Medium contrast \label{fig:OOB_contrast}}
\end{subfigure}
\caption{ \small Out-of-bag error for different variables. In (c) the attenuation and wavespeed contrast is assumed to be equal.}
\label{fig:OOB}
\end{figure}

Generally RF models perform better when more estimators (trees) are used since overfitting - or variance - can then suppressed. We can measure this using OOB samples that give an estimate of the generalisation error. Figure \ref{fig:OOB_trees} shows the OOB error as a function of the amount of trees used in the RF model. We clearly observe a near-exponential decay with increasing tree count, which is due to the reduction of overfitting and thus variance (equation (\ref{eqn:MSE})). In our problem adding more trees could help reducing the generalisation error slightly, however the increase in computation time was found to not be worth the small difference in performance. Figure \ref{fig:OOB_Nsamples} shows the importance of the prior space size. Since the posterior completely depends on the prior (\ref{eqn:bayes}) that is built from chosen $S_2$ functions from which we analytically compute effective wave properties, the prior will converge when its size is large enough to systematically represent new data samples - or OOB samples. Here we notice a similar error decay, continuing up until 1 million, used in our inference. Using 1 CPU per RF model the inversions can be performed within 24 hours. However adding more than 1 million samples did not significantly improve the results, which tells us that the use of distinct basis functions with random perturbations (see previous section on the joint fit of volume fraction and scaled autocovariance) rather than the prior size hindered closer fits especially to the bead pack sample. Therefore not only the quantity but also the quality - in terms of appropriately sampling the function space of priors -  of the prior matters. In a third OOB analysis we computed the influence of medium contrast on the RF model performance, because larger contrasts cause larger difference between the effective and reference wave properties. In addition, differences of attenuation due to varying microstructure geometry, and in particular when considering hyperuniformity, are sensitive to medium contrast (see \cite{rechtsman2008effective} and \cite{kim2020effective}). Hence our choice of $c_1^L / c_2^L = 3$ and $Q_1^L / Q_2^L = 5$ might bias the RF model as it benefits from these larger differences when splitting data and making inferences. From figure \ref{fig:OOB_contrast}, where we used equal contrast for wavespeed and attenuation for simplicity, we find that only for contrasts near 1 (i.e. near homogeneity) the model is significantly worse. Therefore for most geophysical purposes where the medium contrast is significant we assume the inversion to have similar results as in this work.  

\subsection{Limitations and future work}

The accuracy of the SCE theories depends on the assumptions made, which among others are constant phase densities, and macroscopic and elastic isotropy. Allowing density variations in the SCE is not straightforward (\cite{kim2020effective}), and since elastic moduli generally differ more strongly between phases we somewhat justify this assumption. Although the acoustic theory only allows macroscopic isotropy due to the scalar equations, the elastic theory can be extended to more complex cases. Namely, even though elastic isotropy is necessary to decouple longitudinal and transverse waves, macroscopic anisotropy could be accounted for and would yield the fourth-rank stiffness tensor $\bm C_e$. Indeed, all theories allow, in principle, for macroscopic isotropy and statistical anisotropy, i.e. $\bm C_e$ would reduce to the effective elastic moduli as in our problem and the n-point correlation functions over which is integrated (\ref{eqn:A2int}, \ref{eqn:F_nonlocal}) depend on three directions instead of on the radial distance $r$. This is because all statistically isotropic media yield macroscopic isotropy, whereas the converse is not necessarily true (\cite{kim2020effective}). 

An additional limitation is the use of only two-point correlations, without the addition of higher-order descriptors. It has been shown that two-point statistics alone are not sufficient to describe complex micro-heterogeneities (\cite{gommes2012density, gommes2012microstructural}). For example, media with similar two-point correlation functions can have significantly different effective permeability, which can be of large importance in for instance CO$_2$ sequestration (\cite{mosser2017reconstruction}). It is noteworthy that the inclusion of higher order statistics would likely yield different inversion results. Although the inversion ill-posedness of geometry for low-order descriptors may decrease due to a more accurate description of multiple scattering, the results in figures \ref{fig:Berea_Debye} and \ref{fig:Berea_fixedphi} can be different. On the other hand, the inclusion of higher-order descriptors will increase the model-parameter space for inference thus potentially adding ill-posedness of a different kind than that present in our current examples. Moreover, one could argue that the $S_2$ degeneracy works in our favour in inference, as we can now restrict the prior space to only include e.g., Debye autocovariance functions. Describing the Berea sample with higher-order correlations would possibly not allow such a restriction since these describe the samples more uniquely. Apart from that, the extension to a more complete set of SMDs would benefit most geophysical applications. Only when the numerical computation of correlation functions with $n \geq 3$ will become feasible this would be possible. Alternatively the use of two-point cluster functions as an addition to two-point correlation functions for reconstruction purpoes can help to overcome degeneracy (\cite{jiao2009superior}). The recent proposal of n-point polytope functions can aid to constrain anisotropy in microstructures (\cite{chen2019hierarchical}). However there are no SCE formulations available for these SMDs as of the time of this study. 
Lastly, the correction term in the nonlocal elastic approximation is far from perfect, since the scattered wave mode and the interaction between modes is not taken into account (\cite{kim2020multifunctional, kim2020effective}). The most efficient way to benchmark this theory is by numerically simulating elastic waves propagating through two-phase microstructures, which therefore is the most crucial future step forward from this work. It is expected that such simulations more accurately account for mode interaction and therefore yield more distinct effective wave characteristics per mode. This would add more unique data for the RF inversion, and therefore can only improve the results found in this study.

One very interesting application of this work is to classify the degree of hyperuniformity in materials, which seems feasible based on this work. While the samples evaluated here are nonhyperuniform, they did show relative degrees of hyperuniformity which are captured in the RF predictions (\ref{fig:invresults}). In connection to material integrity,  \cite{xu2017microstructure} showed that stress accumulation is reduced in hyperuniform heterogeneous materials due to the exceptional ordering of the brittle/weak phase, resulting in a larger overall brittle strength. One suggested workflow therefore is aiming at distinguishing hyperuniform from nonhyperuniform materials through classification based on AI procedures as that shown here. In that way, we might get an estimate of which materials are weaker than others using remote effective wave observations, aiding e.g, in non-destructive testing for material failure, or in detecting potential earthquake nucleation sites.

\section{Conclusions}

In this study, we infer complex microstructure information in the form of two-point correlation functions from frequency-dependent effective wave data using a Random-Forest (RF) Bayesian approach. We analytically compute the effective wavespeed and attenuation - our input data - for pure-mode longitudinal waves and elastic waves in the long-wavelength regime and additionally approximations for elastic waves in the scattering regime using recent strong-contrast expansions (SCEs). We find that with a priori knowledge of medium phase properties (i.e., contrast) the microstructure geometry descriptors of a bead pack, Berea sandstone and Ketton limestone can be accurately determined. We also show that this is especially true when using (nonlocal) elastic wave theory. In addition, for well-known samples, we can predict the $S_2$ using a restriction on the prior used in the probabilistic inversion, and when the volume fraction instead of the medium contrast is known a priori. Further improvements could be achieved via numerical elastic wave simulations in the scattering regime and the inclusion of higher order statistical microstructure descriptors to account for more complex heterogeneities. Our approach paves the way to the promise of sub-wavelength microstructure characterisation and monitoring from wave data in applications such as acoustics, geophysics, medical imaging, non-destructive testing and meta-material design.

%\section{Acknowledgements}

\bibliographystyle{abbrvnat}
\bibliography{main}

\newpage

\section*{Appendix A. Statistical (an)isotropy samples}
\label{sec:AppA}
\addcontentsline{toc}{section}{Appendix A. Statistical (an)isotropy samples}

\begin{figure}[t!]
\begin{subfigure}[t!]{0.32\textwidth}
    \includegraphics[width=\linewidth]{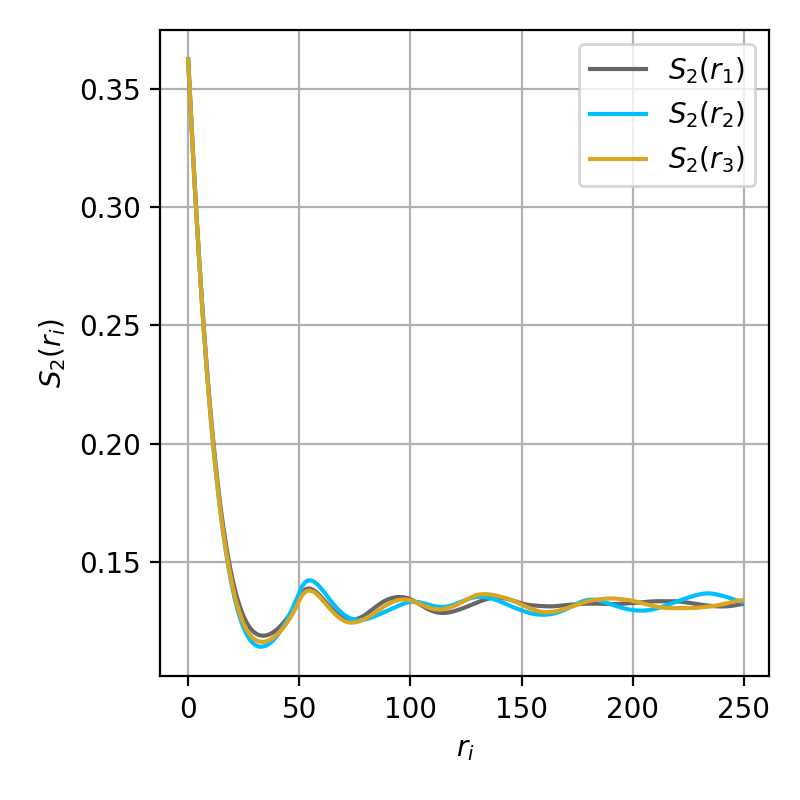}
\caption{\small Bead pack \label{fig:bead_3dims}}
\end{subfigure}\hfill
\begin{subfigure}[t!]{0.32\textwidth}
  \includegraphics[width=\linewidth]{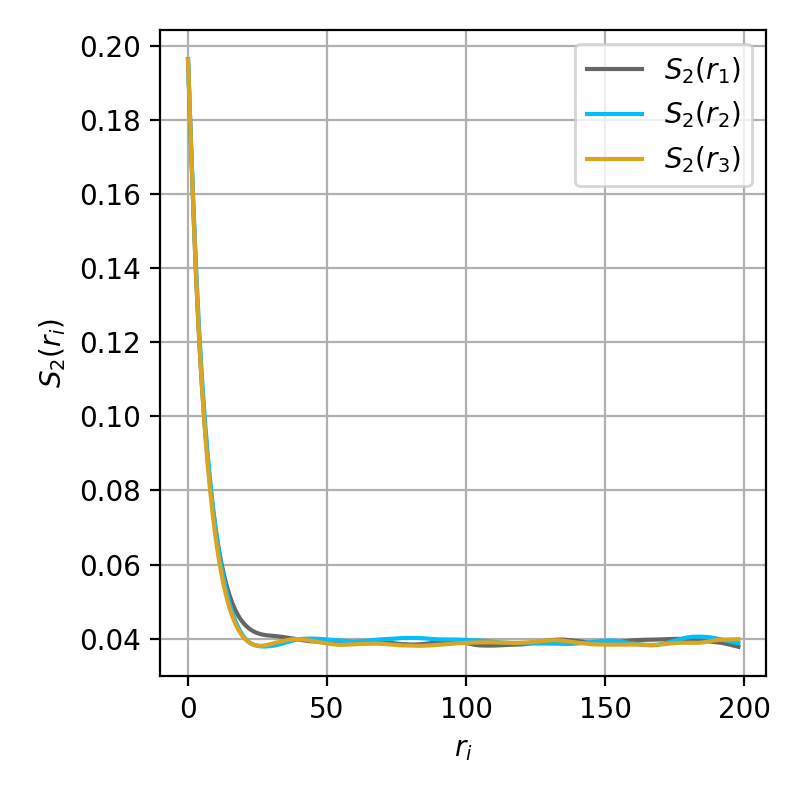}
\caption{\small Berea \label{fig:berea_3dims}}
\end{subfigure}\hfill
\begin{subfigure}[t!]{0.32\textwidth}
    \includegraphics[width=\linewidth]{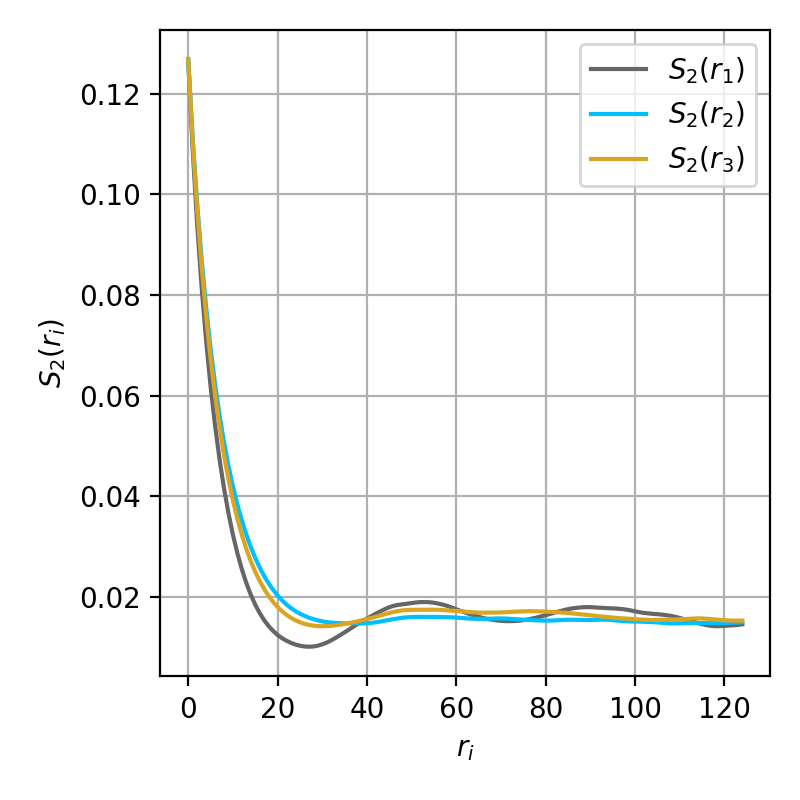}
\caption{\small Ketton \label{fig:ketton_3dims}}
\end{subfigure}
\caption{\small $S_2(r)$ in three directions for the samples in figure \ref{fig:Mossersamples}.}
\label{fig:samples_S2_3dims}
\end{figure}

Figure \ref{fig:samples_S2_3dims} shows the $S_2$ for the bead pack, Berea and Ketton samples in the x-, y- and z-direction ($r_1, r_2$ and $r_3$, respectively). Logically, these curves overlap in the case of statistical isotropy. Even though there will always be small differences, we can assume that both the bead pack and Berea samples satisfy this condition. In contrast we observe significant anisotropy in the Ketton sample. While there is little variation in the y- and z-direction the curve sampled in the x-direction oscillates strongly. Therefore the Ketton sample resembles a combination of impenetrable and hard spheres, depending on the direction.
% When looking at figure \ref{fig:kettonim} we can explain this difference by observing that the grains are somewhat layered in the y-direction. Therefore the resulting $S_2$ in the y-direction resembles that of impenetrable spheres, whereas that in the x-direction indicates the pore space separating each `layer'.

\section*{Appendix B. Strong-contrast expansion derivation}
\label{sec:AppB}
\addcontentsline{toc}{section}{Appendix B. Strong-contrast expansion derivation}
Here we derive the exact strong-contrast expansions discussed in the main text.
\subsection*{B.1. Pure-mode longitudinal waves}
\label{subsec:AppB_1}
\addcontentsline{toc}{subsection}{B.1. Pure-mode longitudinal waves}

Imagine a plane longitudinal wave $u_0^L(\boldsymbol{r})$ with angular frequency $\omega$ and velocity $c(\boldsymbol{r})$ (which might be complex) travelling through a two-phase microstructure, obeying the scalar Helmholtz equation:
\begin{equation}
    \label{eqn:Helmholtz}
    \left[ \grad^2 + \left( \frac{\omega}{c(\boldsymbol{r})}\right)^2  \right] \, u(\boldsymbol{r}) = 0,
\end{equation}
where the superscript L referring to the longitudinal nature of waves is neglected for simplicity, as will be done in the rest of the derivation. Also, we will use the wavenumber formulation $k^2(\boldsymbol{x}) =  \sigma(\boldsymbol{r}) = ( \omega / c(\boldsymbol{r}) )^2$.
This can be rewritten, when we acknowledge the fact that medium contrast can be interpreted as a source term, as:
\begin{equation}
    \label{eqn:Helmholtz_source}
    \left[ \grad^2 - \sigma_q \right] \, u(\boldsymbol{r}) =  (\sigma(\boldsymbol{r}) - \sigma_q)  \, u(\boldsymbol{r}).
\end{equation}
The solution is the Lippmann-Schwinger equation (\cite{lippmann1950variational}):
\begin{equation}
\label{eqn:Lippmann}
    u(\boldsymbol{r}) = u_0(\boldsymbol{r}) + \int d\boldsymbol{r}' G(\boldsymbol{r}, \boldsymbol{r}') \left[ \sigma(\boldsymbol{r}) - \sigma_q \right] u(\boldsymbol{r}'),
\end{equation}
where $G(\bm{r}, \bm{r}')$ is the Green's function to the Helmholtz operator in (\ref{eqn:Helmholtz_source}), given by:

\begin{equation}
        G(\bm{r}, \bm{r}') = - \frac{1}{3 \sigma_q}\delta (\bm{r} - \bm{r}') - \rho \frac{e^{i k_q r}}{4 \pi r}
\end{equation}

Equation (\ref{eqn:Lippmann}) can be written in linear operator form as:
\begin{equation}
     \mathrm{U} =  \mathrm{U_0} + \mathrm{G} \Phi,
\end{equation}
where we call $\Phi$ the contrast source, given by:
\begin{equation}
    \Phi = ( \sigma(\boldsymbol{r}) - \sigma_q ) \, \mathrm{U}.
\end{equation}
Next we will extract the static part of the Green's function, which yields the following field:
\begin{equation}
\label{eqn:cavityfield}
     \mathrm{C} = \mathrm{U_0} + \mathrm{H} \Phi,
\end{equation}
in which in three dimensions, $\mathrm{H}$ is the dynamic part of the Green's function, and equals:
\begin{equation}
    \mathrm{H} (\bm r - \bm r') = - \frac{\rho_q \, e^{i \, k_q \, r}}{4 \, \pi \, r},
\end{equation}
with $r = |\bm r - \bm r'|$. Now we find an expression for the contrast source:
\begin{equation}
    \Phi (\bm r) = \frac{\sigma (\bm r) - \sigma_q}{1 + \frac{1}{3\sigma_q}(\sigma (\bm r) - \sigma_q)} \mathrm{C}(\bm r).
\end{equation}
Note that this quantity is local in space. Now we take the spatial average of both fields and assume that both are related via a constant (i.e. homogenized):
\begin{equation}
    \langle \Phi (\bm r) \rangle = L_e \langle \mathrm{C}(\bm r) \rangle, 
\end{equation}
where the subscript $e$ stands for effective, and:
\begin{equation}
    L_e = \frac{\sigma_e - \sigma_q}{1+\frac{1}{3\sigma_q}(\sigma_e - \sigma_q)},
\end{equation}
where $\sigma_e = (\omega / c_e)^2$, with $c_e$ the effective velocity. 
Solving equation (\ref{eqn:cavityfield}) iteratively yields the following exact strong-contrast expansion for statistically homogeneous and isotropic media:
\begin{equation}
    \beta^2_{pq} \, \phi^2_p \, \beta^{-1}_{eq} = \phi_p \, \beta_{pq} - \sum_{n = 2}^{\infty} A_n^{(p)} \, \beta^n_{pq}.
\end{equation}
Here, $\beta_{ab} = (\sigma_a - \sigma_b)/(\sigma_a + 2 \sigma_b)$ in three dimensions and $A_n^{(p)}$ are integrals over n-point correlation functions, which for $n = 2$ is given in equation (\ref{eqn:A2int}).

\subsection*{B.2. Deriving the effective stiffness tensor}
\label{subsec:AppB_2}
\addcontentsline{toc}{subsection}{B.2. Deriving the effective stiffness tensor}
Let us assume an incident plane wave $\epsilon_0(\boldsymbol{x})$ with angular frequency $\omega$ propagating through a two-phase composite. In this case we consider the full stiffness tensor - which locally equals $\boldsymbol{C}(\boldsymbol{x}) = \boldsymbol{C}_q \mathcal{I}^{(q)}(\boldsymbol{x}) + \boldsymbol{C}_p \mathcal{I}^{(p)}(\boldsymbol{x})$, using the indicator function (\ref{eqn:indic}) - for elastically isotropic phases, given by
\begin{equation}
    \boldsymbol{C}_i = dK_i \boldsymbol{\Lambda}_h + 2G_i \boldsymbol{\Lambda}_s, (i = p, q).
\end{equation}
The hydrostatic projection tensor $\boldsymbol{\Lambda}_h$ and shear projection tensor $\boldsymbol{\Lambda}_s$ are given by
\begin{equation}
    (\Lambda_h)_{ijkl} \equiv \frac{1}{3}\delta_{ij}\delta_{kl}
\end{equation}
\begin{equation}
    (\Lambda_s)_{ijkl} \equiv \frac{1}{2}(\delta_{ik}\delta_{jl} + \delta_{il}\delta_{jk}) - \frac{1}{3}\delta_{ij}\delta_{kl}
\end{equation}
with $\delta_{ij}$ the Kronecker delta. The local displacement field $\boldsymbol{u(x)}$ is the solution to the following wave equation where again $q$ is the reference phase (\cite{kim2020effective}):
\begin{equation}
    \omega^2 \boldsymbol{u}(\boldsymbol{x}) + ({c_q^L}^2 - {c_q^T}^2) \grad (\div{\boldsymbol{u}(\boldsymbol{x})}) + {c_q^T}^2 \grad^2 \boldsymbol{u}(\boldsymbol{x}) = \div{\boldsymbol{P}(\boldsymbol{x})},
\end{equation}
When the displacement dependence on time is sinusoidally oscillating with frequency $\omega$: $\boldsymbol{u}(\boldsymbol{x}, t) = \boldsymbol{u}(\boldsymbol{x}) \exp (-i\omega t)$. $\boldsymbol{P}(\boldsymbol{x})$ is the stress polarisation field, induced by stiffness contrasts in the medium:
\begin{equation}
    \boldsymbol{P}(\boldsymbol{x}) \equiv \frac{1}{\rho} (\boldsymbol{C}(\boldsymbol{x}) - \boldsymbol{C}_q) : \boldsymbol{\epsilon}(\boldsymbol{x}),
\end{equation}
and clearly is only non-zero in the polarised phase $p$. The displacement field can then be expressed by the following Green's function formalism: 
\begin{equation}
    \boldsymbol{u}(\boldsymbol{x}) = \boldsymbol{u}_0(\boldsymbol{x}) + \int \boldsymbol{g}^{(q)}(\boldsymbol{x} - \boldsymbol{x}')\cdot (\div{\boldsymbol{P}(\boldsymbol{x}')})d\boldsymbol{x}',
\end{equation}
which can be rewritten in terms of strain when we take the symmetric part of its gradient:
\begin{equation}
\label{eqn:strain}
    \boldsymbol{\epsilon}(\boldsymbol{x}) = \boldsymbol{\epsilon}_0(\boldsymbol{x}) + \int \boldsymbol{G}^{(q)}(\boldsymbol{x} - \boldsymbol{x}') : (\div{\boldsymbol{P}(\boldsymbol{x}')})d\boldsymbol{x}'.
\end{equation}
Here $\boldsymbol{G}^{(q)}(\boldsymbol{r})$ with $\boldsymbol{r} \equiv \boldsymbol{x} - \boldsymbol{x}'$ is the fourth rank Green's function and has a singularity at $\boldsymbol{x}' = \boldsymbol{x}$. Therefore it is necessary to exclude a small region around this location and integrate equation (\ref{eqn:strain}) in the limit that its volume goes to zero. Therefore the total Green's function can be decomposed into
\begin{equation}
    \boldsymbol{G}^{(q)}(\boldsymbol{r}) = - \boldsymbol{D}^{(q)}\delta(\boldsymbol{r}) + \boldsymbol{H}^{(q)}(\boldsymbol{r}),
\end{equation}
where the delta function is the dipole source, the constant tensor $\boldsymbol{D}^{(q)}$ determines the `exclusion-zone' shape around it (chosen to be spherical) and $\boldsymbol{H}^{(q)}(\boldsymbol{r})$ contains the contribution outside this region. The exact explicit expression of both tensors can be found in \cite{kim2020effective}. The linear operator form of equation (\ref{eqn:strain}) reads
\begin{equation}
    \boldsymbol{\epsilon} = \boldsymbol{\epsilon}_0 + \boldsymbol{G} \boldsymbol{P}
\end{equation}
Which reduces to the following equation for the `cavity strain field' when we exclude contributions from the exclusion zone:
\begin{equation}
    \boldsymbol{f} = \boldsymbol{\epsilon}_0 + \boldsymbol{H} \boldsymbol{P}
\end{equation}
We can now express $\boldsymbol{P}$ as a function of this strain tensor:
\begin{equation}
    \boldsymbol{P}(\boldsymbol{x}) = (\boldsymbol{L}^{(q)} \, \mathcal{I}^{(p)} (\boldsymbol{x})) : \boldsymbol{f}(\boldsymbol{x})
\end{equation}
With 
\begin{equation}
    \boldsymbol{L}^{(q)} \equiv (\boldsymbol{C}_p - \boldsymbol{C}_q) \, / \, \rho \, : \, (\boldsymbol{I} + \boldsymbol{D}^{(q)}:(\boldsymbol{C}_p - \boldsymbol{C}_q)/\rho)^{-1},
\end{equation}
Which is the constitutive relation arising from the influence of the polarized phase $p$, and zero in phase $q$.  
Note the influence of the exclusion zone shape $\boldsymbol{D}^{(q)}$ in this expression. Now we can postulate the following relation by taking ensemble averages and assuming a constant constitutive relation in space ($\langle \cdot \rangle$):
\begin{equation}
    \langle \boldsymbol{P} \rangle (\boldsymbol{x}) = \boldsymbol{L}_e^{(q)}(k_q^L) : \langle \boldsymbol{f} \rangle (\boldsymbol{x}),
\end{equation}
with the effective constant tensor equal to
\begin{equation}
    \boldsymbol{L}_e^{(q)}(k_q^L) = (\boldsymbol{C}_e(k_q^L) - \boldsymbol{C}_q)/\rho : (\boldsymbol{I} + \boldsymbol{D}^{(q)}:(\boldsymbol{C}_e(k_q^L) - \boldsymbol{C}_q)/\rho)^{-1}.
\end{equation}
 Then, for macroscopically isotropic media we use this equation to obtain the following using series expansion:
\begin{align}
\label{eqn:SCE}
    \phi_p^2 \boldsymbol{L}^{(q)} : (\boldsymbol{L}_e^{(q)}(k_q^L))^{-1} &= \phi_p^2 \left( \frac{\kappa_{pq} }{\kappa_{eq}(k_q^L)} \boldsymbol{\Lambda}_h + \frac{\mu_{pq} }{\mu_{eq}(k_q^L)} \boldsymbol{\Lambda}_s \right) \\
    &= \phi_p \boldsymbol{I} - \sum_{n = 2}^\infty \boldsymbol{B}_n^{(p)} (k_q^L)
\end{align}
where $\kappa_{pq}$ is given in the main text and $\kappa_{eq} \equiv (K_e - K_q)/(K_e + 4 G_q/3)$. Similarly, $\mu_{pq}$ is given in the main text and $\mu_{eq} \equiv (G_e - G_q)/(G_e + (3K_q / 2 + 4G_q/3)G_q/(K_q + 2G_q))$. We can isolate both the effective bulk and shear modulus from equation (\ref{eqn:SCE}) by contracting both $\boldsymbol{\Lambda}_h$ and $\boldsymbol{\Lambda}_s$ using quadruple inner products with the other (\cite{torquato1997effective, kim2020effective}. When we subsequently truncate the series after $n = 2$, i.e. at the two-point correlation level, we obtain equations (\ref{eqn:K_eff}) and (\ref{eqn:G_eff}).

\section*{Appendix C. Datafit Berea and Ketton}
\label{sec:AppC}
\addcontentsline{toc}{section}{Appendix C. Datafit Berea and Ketton}

\begin{figure}[t!]
    \centering
    \begin{subfigure}[b]{0.48\textwidth}
		\includegraphics[width=\textwidth, trim={0 0 0 0},clip]{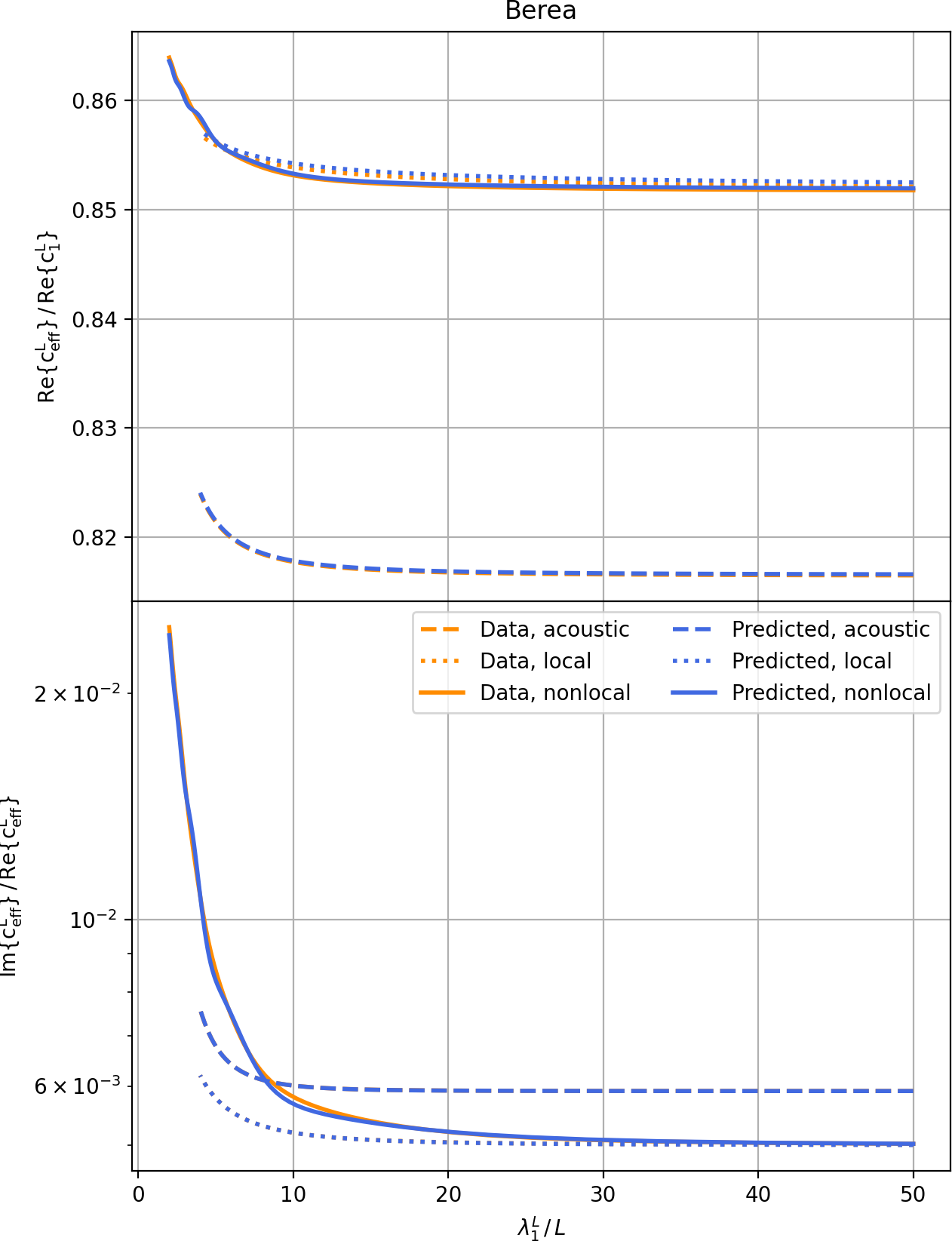}
		\caption{\label{fig:Berea_datafit_P}}
	\end{subfigure}%
	\quad
	\begin{subfigure}[b]{0.48\textwidth}
		\includegraphics[width=\textwidth, trim={0 0 0 0},clip]{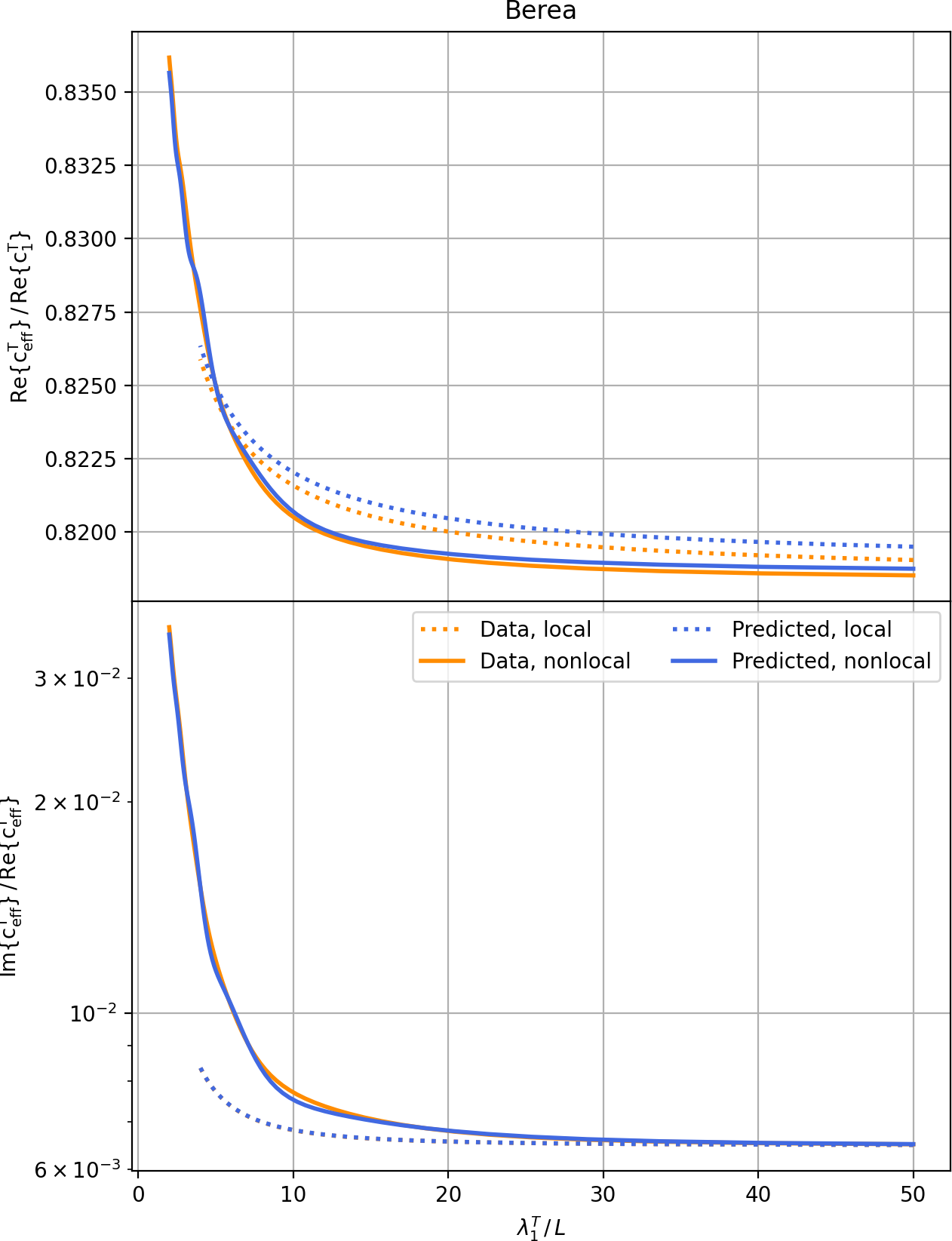}
		\caption{\label{fig:Berea_datafit_S}}
	\end{subfigure}
	\caption{\small{Longitudinal (a) and Transverse (b) data fit of the $S_2$ models in figure \ref{fig:invresults} for the Berea sample.}}
	\label{fig:Berea_datafit}
\end{figure}

\begin{figure}[t!]
    \centering
    \begin{subfigure}[b]{0.48\textwidth}
		\includegraphics[width=\textwidth, trim={0 0 0 0},clip]{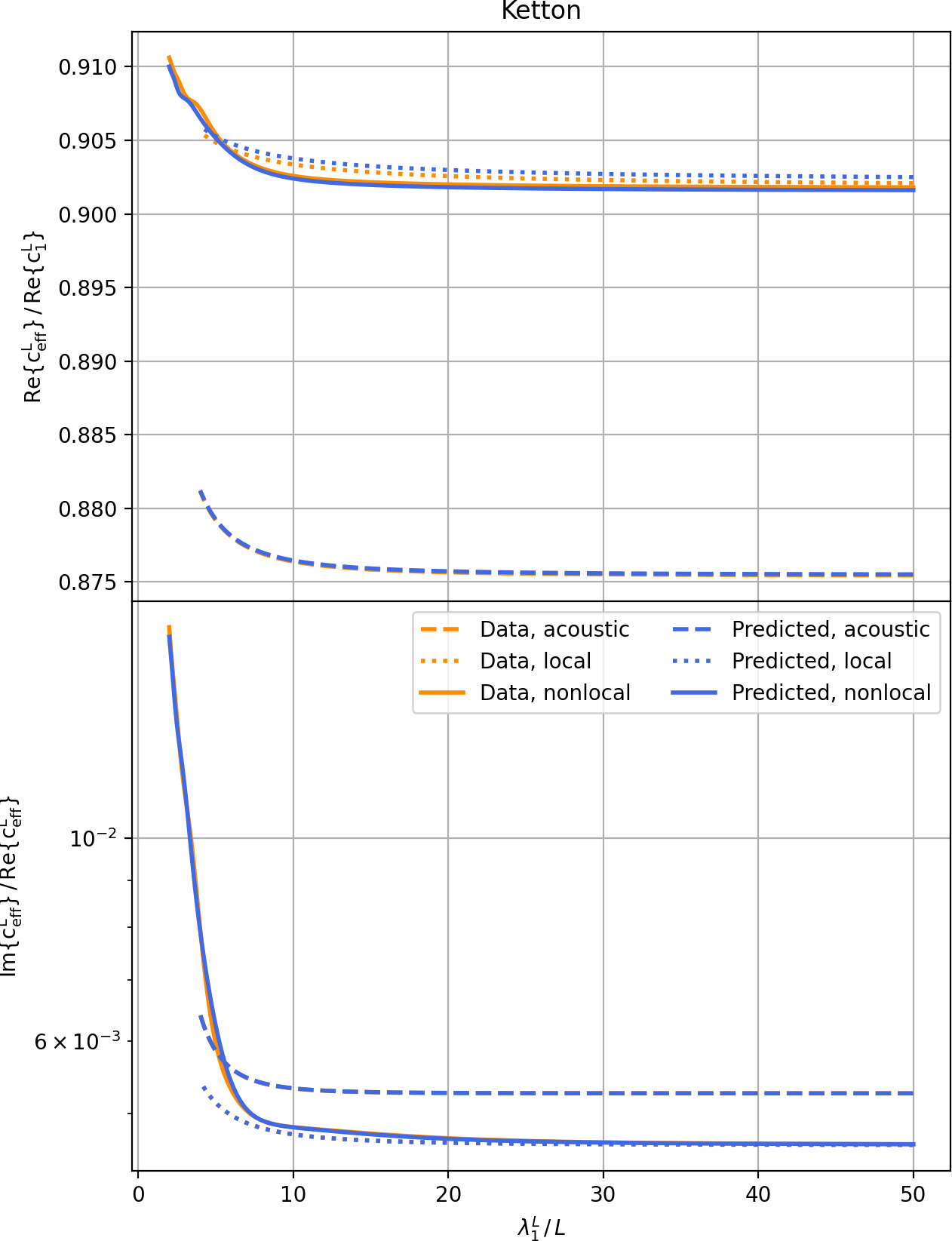}
		\caption{\label{fig:Ketton_datafit_P}}
	\end{subfigure}%
	\quad
	\begin{subfigure}[b]{0.48\textwidth}
		\includegraphics[width=\textwidth, trim={0 0 0 0},clip]{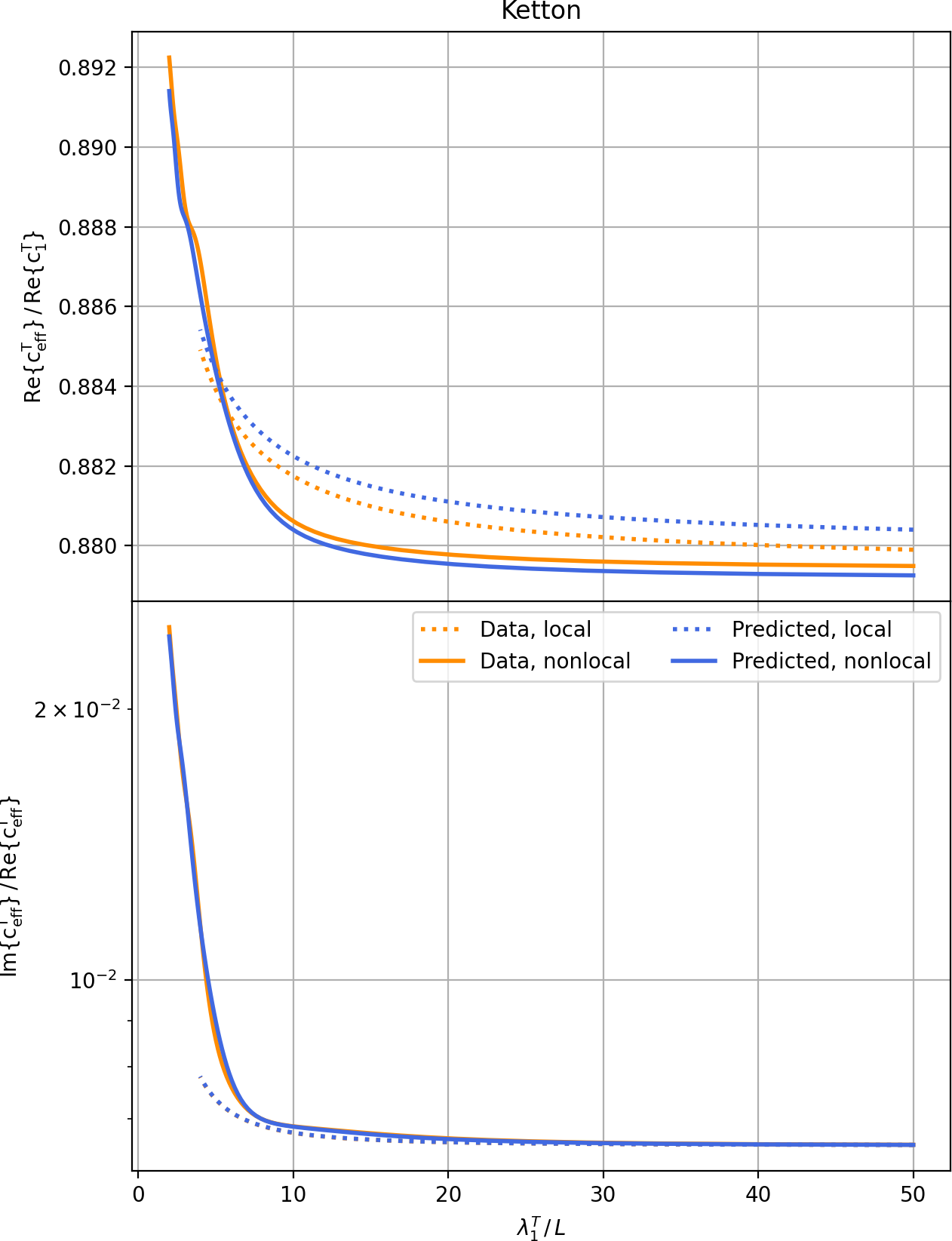}
		\caption{\label{fig:Ketton_datafit_S}}
	\end{subfigure}
	\caption{\small{Longitudinal (a) and Transverse (b) data fit of the $S_2$ models in figure \ref{fig:invresults} for the Ketton sample.}}
	\label{fig:Ketton_datafit}
\end{figure}

\end{document}